\documentclass[12pt,twoside,openright]{book}
\usepackage[pdftex]{graphicx}
\usepackage{pdfpages}
\usepackage[a4paper, tmargin=3.5cm, bmargin=2.5cm, lmargin=3.75cm, rmargin=2.5cm, twoside]{geometry}
\usepackage{latexsym,amsmath,amssymb,amsfonts,bm}             
\usepackage{rotating}                    
\usepackage{natbib}          
\usepackage{soul}                                         

\usepackage{fancyhdr}                    
\usepackage{setspace}
\usepackage[english]{babel}
\doublespace          
\usepackage{tocloft}
\usepackage{lscape}

\usepackage[pdfpagelabels=true,plainpages=false,colorlinks=true,linkcolor=black,citecolor=black,urlcolor=blue]{hyperref}
\usepackage[small,hang]{caption2}
\usepackage[compact]{titlesec}
\numberwithin{equation}{section}
\usepackage[T1]{fontenc}                    
\makeatletter
\def\cleardoublepage{\clearpage\if@twoside \ifodd\c@page\else%
	\hbox{}%
	\thispagestyle{empty}
	\newpage%
	\if@twocolumn\hbox{}\newpage\fi\fi\fi}
\makeatother
\clearpage{\pagestyle{plain}\cleardoublepage}
\newcommand{\be}{\begin{equation}}
	\newcommand{\ee}{\end{equation}}
\newcommand{\bea}{\begin{eqnarray}}
	\newcommand{\eea}{\end{eqnarray}}
\newcommand{\ba}{\begin{array}}
	\newcommand{\ea}{\end{array}}
\newcommand{\bi}{\begin{itemize}}
	\newcommand{\ei}{\end{itemize}}
\newcommand{\bc}{\begin{center}}
	\newcommand{\ec}{\end{center}}
\newcommand{\bfr}{\begin{flushright}}
	\newcommand{\efr}{\end{flushright}}

\begin{document}
\thispagestyle{empty}

{ \renewcommand{\baselinestretch}{1.5}
\begin{center}
\begin{spacing}{2}
\noindent{\Large \bf PROBING LATTICE DYNAMICS IN \\ 
REAL-SPACE AND REAL-TIME}
\end{spacing}
\end{center}

\vspace{18cm}
\begin{flushright}
{ {\LARGE\em  \textbf{NAVDEEP RANA} ~~}}
\end{flushright}}

	\cleardoublepage
	\frontmatter
	\newpage
\thispagestyle{empty}

\begin{titlepage}
\begin{center}
{
%


\textbf{\Large PROBING LATTICE DYNAMICS IN  REAL-SPACE AND  REAL-TIME } \vskip1.0cm 
{\large\emph{Submitted in partial fulfillment of the requirements}} \vskip 0.03cm 
{\large\emph{of the degree of}}
\singlespacing 
\textbf {\large Doctor of Philosophy}\\
\singlespacing
 {\large\emph{by}}\\
\singlespacing \textbf{\large NAVDEEP RANA} \vskip0.2cm
\vskip1cm
{\large Supervisor:}\\
\singlespacing \textbf{\large{Prof. Gopal Dixit}} \vskip1cm

\includegraphics[height=42mm]{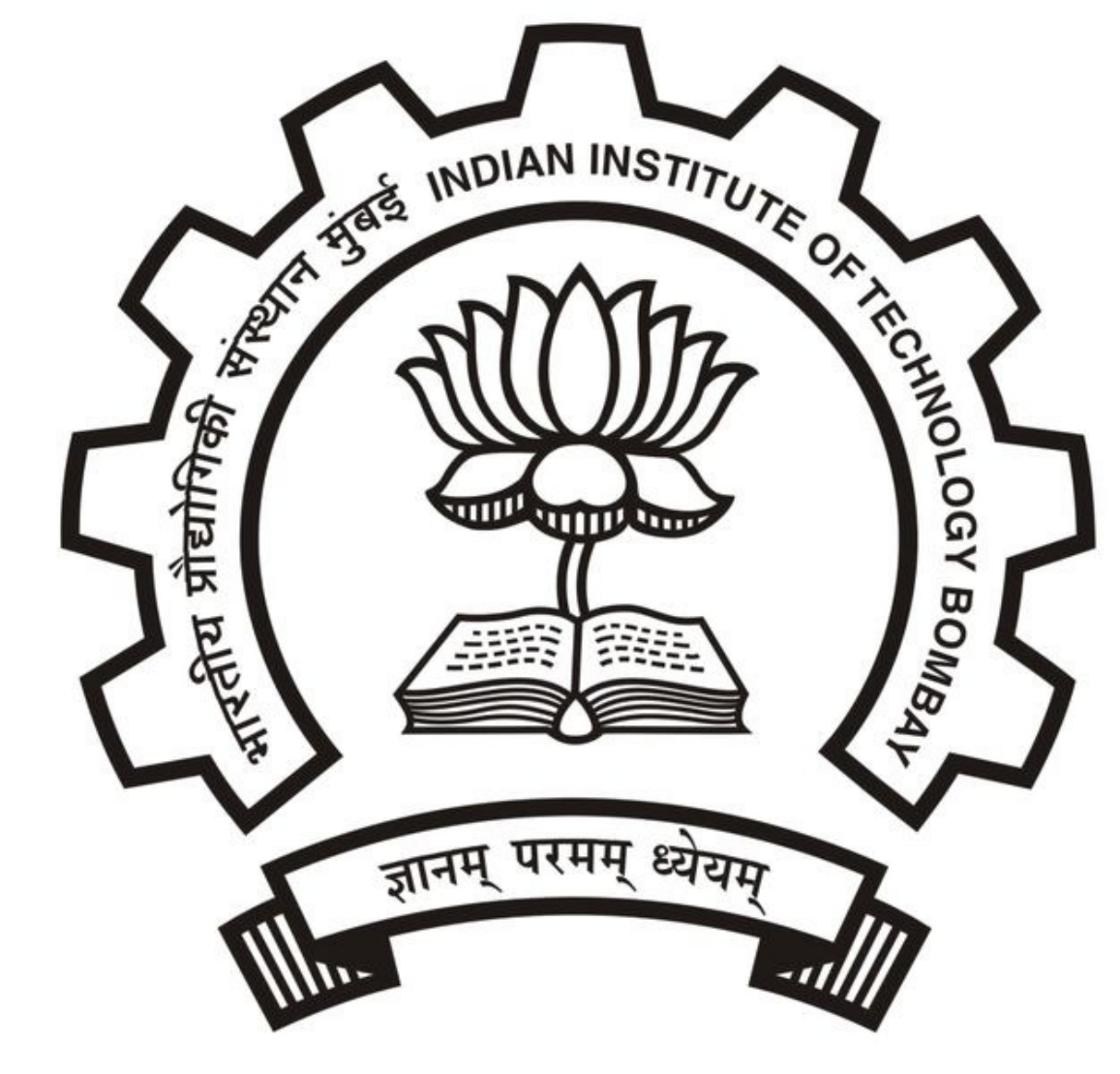}
\doublespacing
\begin{center}
\textbf{\large{DEPARTMENT OF PHYSICS\\
INDIAN INSTITUTE OF TECHNOLOGY BOMBAY\\\singlespacing
2023}}\\\singlespacing \copyright~2023 NAVDEEP RANA All rights reserved.
\end{center}

\fboxsep6mm
\fboxrule1.3pt
}
\end{center}
\end{titlepage}
\pagestyle{fancy}

	\cleardoublepage
	\newpage
\thispagestyle{empty}
\setlength{\baselineskip}{32pt}
\bigskip
\bigskip
\bigskip
\bigskip
\bigskip
\vspace*{5cm}
\begin{center}
\vspace*{0.5cm}
{\Large  \bf {Dedicated to my family} } \\
\end{center}
%

\setlength{\baselineskip}{18pt}

	\cleardoublepage
	\addcontentsline{toc}{chapter}{Acknowledgements}
	\newpage
\thispagestyle{empty}
\begin{center}
\vspace*{-0.4cm}
{\LARGE {\textbf{Acknowledgements}}}
\end{center}

{\setlength{\baselineskip}{8pt} \setlength{\parskip}{2pt}
\begin{spacing}{1.5}

Foremost, I would like to express my sincere gratitude to my supervisor, \textbf{Prof. Gopal Dixit}, Department of Physics, IITB, for his unwavering guidance, expertise, and support throughout my doctoral research journey. Your insightful feedback, patient mentorship, and constant encouragement have been invaluable in shaping this work and helping me navigate through challenges. I extend my heartfelt thanks to him for introducing me into the field of ultrafast optics. Especially, I am thankful to him for all the opportunities I have received, and the skills I have developed under his supervision during my PhD.

I want to express my sincere thanks to my RPC members, \textbf{Prof. Dipanshu Bansal} and \textbf{Prof. Hridis Kumar Pal}  for their valuable comments and insightful questions, which helped to shape my research.

I extend my heartfelt thanks to my labmates \textbf{Irfana N Ansari}, \textbf{M S Mrudul}, \textbf{Adhip Pattanayak}, \textbf{Sucharita Giri}, and \textbf{Amar Bharti}, for their constant support and cooperation. I will always remember our coffee breaks and social events which have brought spirit to our lab and team work. We have created great memories and memorable moments which I will cherish forever. Also, I want to thank all my friends inside and outside IIT campus. Furthermore, I would like to record my thanks to all the staff members of IITB for making life inside IITB much easier. Finally, I would like to thank my institution for providing the resources and infrastructure.

To my beloved parents and sisters, I owe a debt of gratitude for your continuous encouragement, understanding, and belief in my abilities. Your unconditional support, both emotionally and morally, and sacrifices has given me the strength to persevere even when faced with difficulties.

Last but certainly not least, I want to express my deepest appreciation to my dear wife. Your unwavering love, patience, and constant encouragement have been my anchor throughout this journey. Your sacrifices and understanding during long hours of work are deeply cherished, and I am truly grateful for your presence in my life.

This thesis is a culmination of the collective efforts and contributions of these exceptional individuals, each of whom has played a unique role in shaping my academic and personal growth. I am profoundly grateful for their presence and support, without which this achievement would not have been possible.

\vskip2cm
%
%

\begin{flushright}
Navdeep Rana\\
Department of Physics\\
IITB
\end{flushright}
\begin{flushleft}
	Date: 21/08/2023\\
\end{flushleft}
\end{spacing}

	\cleardoublepage
	\addcontentsline{toc}{chapter}{Abstract}
	\newpage
\thispagestyle{empty}
\begin{center}
\vspace*{-0.4cm}
{\LARGE {\textbf{Abstract}}}
\end{center}

{\setlength{\baselineskip}{8pt} \setlength{\parskip}{2pt}
\begin{spacing}{1.5}

The coherent lattice vibrations significantly impact a wide array of physical and chemical processes in solids, such as  heat transfer, displacive phase transition, thermal conductivity, to name but a few. 
Thus, probing lattice dynamics in real-space and real-time is essential for understanding numerous ubiquitous  phenomena in solids.  
High-harmonic spectroscopy  has emerged as a preferred technique for investigating numerous  
static and dynamic properties  of  solids on ultrafast timescale. 
Yet, despite these accomplishments,  the applicability of high-harmonic spectroscopy to probe 
the influence of coherent lattice vibrations on electronic responses has remained an unexplored territory. 

In this thesis, we have explored  the impact of coherent lattice dynamics on attosecond electronic response in solids using high-harmonic spectroscopy.
It is observed that the 
coherent excitation of the in-plane phonon mode in graphene results in the appearance of sidebands in the harmonic spectrum, and the subsequent sidebands are separated by the frequency of the excited phonon mode. 
Additionally, we have demonstrated  the capability of  high-harmonic spectroscopy to characterize energy, polarization, phase difference as well as ``chirality'' of the phonon modes. 
This thesis offers an avenue to probe the phonon-driven processes in solids with sub-cycle temporal resolution.

In the later segment of this thesis, our focus has shifted toward probing coherent lattice dynamics in real-space and real-time.
We have demonstrated  that the inelastic scattering techniques, combined with theoretical analysis, can yield comparable results to those obtained from time-resolved diffraction and imaging measurements within pump-probe configurations. 
Our findings exhibit excellent agreement with the results obtained from a time-resolved diffuse 
x-ray  scattering experiment. 
Our proposed method can serve as  an alternative to time-resolved diffraction and imaging methods in probing  lattice dynamics  in real-space and real-time with atomic-scale spatiotemporal resolution. 


\end{spacing}

	\cleardoublepage
	\renewcommand{\contentsname}{Contents}  
	\begin{spacing}{1.2}  
		\tableofcontents
		\cleardoublepage
		\addcontentsline{toc}{chapter}{List of Figures} 
		\listoffigures
		\cleardoublepage
		\addcontentsline{toc}{chapter}{List of Symbols and Abbreviations} 
		\markboth{List of Symbols}{List of Symbols and Abbreviations}  
		\chapter*{List of Symbols and Abbreviations}
\noindent {\bf Symbols}
\begin{tabbing}
aaaaaaaaaaaa \= abababababab \kill
$\mathcal{A}(t)$\> Vector potential\\
$\mathcal{E}(t)$\> Electric field\\
$\mathcal{H}_{0}$\> Field-free Hamiltonian \\
$\mathcal{H}_{e}$\> Electronic Hamiltonian \\
$\mathcal{H}_{t}$\> Time-dependent tight-binding Hamiltonian \\ 
$\mathcal{H}_{L}$\> Lattice Hamiltonian \\
$\mathcal{H}(t)$\> Time-dependent Hamiltonian for light-matter interaction\\
$\rho_{mn}$ \> Density matrix element \\
$\textbf{d}_{mn}$ \> Dipole matrix element \\
$\textbf{p}_{mn}$ \> Momentum matrix element \\
$\textrm{T}_2$ \> Dephasing time \\
$\textbf{J}(t)$ \> Total current \\
$\mathcal{I}(\omega)$ \> Intensity of HHG \\
$\omega_{\textrm{ph}}$ \> Frequency of phonon mode  \\
$\mathsf{S}(\mathbf{k}, \omega)$ \> Dynamical structure factor \\ 
$\chi$ \> Response function \\ 

\end{tabbing}
\noindent {\bf Abbreviations}
 \begin{tabbing}
aaaaaaaaaaaa \= abababababab  \kill
HHG \> High-harmonic generation\\
2D \> Two-dimensional \\
TDSE \> Time-dependent Schr\"odinger equation\\
SBE \> Semiconductor-Bloch equations\\
$\textsf{iLO}$ \> In-plane Longitudinal optical \\
$\textsf{iTO}$ \> In-plane Transverse optical \\
DS  \> Dynamical symmetry \\
4D \> Four-dimensional \\
IX(N)S \> Inelastic x-ray(neutron) scattering \\
EELS \> Electron-energy loss spectroscopy 
\end{tabbing}
%

		\cleardoublepage
		
	\end{spacing}
\mainmatter
\begin{spacing}{1.5}
\chapter{Introduction}\label{Chaper1}

Throughout the course of history, humanity has been captivated by the fascinating nature of light, leaving scientists perplexed for millennia. 
Mysterious observations and associated properties of light 
have been described by both wave as well as particle nature. 
In contemporary physics, marked by heightened knowledge and understanding, it is widely accepted that light exhibits a dual nature, consisting of oscillating electric and magnetic fields that propagate at a constant speed within our cosmos  -- a fundamental constant in the universe.

Light plays a crucial role in enabling human perception of the material world. We rely on light to comprehend and interact with our surroundings. 
In everyday scenarios, we observe visual stimuli through transparent mediums like glass, 
which allows the transmission of light, enabling us to perceive various scenes. 
In social gatherings, we recognize acquaintances by observing the illumination of their facial features. In essence, our acquisition of knowledge about the world is made possible through the interaction between light and matter. 
This phenomenon is not an exception in the realm of science either. The process of scientific research typically involves four essential stages: (1) observing a phenomenon, (2) forming a hypothesis to explain the observed occurrence, (3) testing and refining the hypothesis to develop a theory, and (4) generating predictions based on the established theory. 
The initial stage of scientific inquiry, observation, is often facilitated through the interplay between light and matter. Scientists employ light as a means to investigate and observe particular phenomena of interest. Therefore, understanding light-matter interaction is of utmost significance to discern the phenomena effectively perceived during the observation process, ultimately paving the way for the continued scientific quest.

In measuring physical processes occurring within a specific timescale, 
it is necessary to utilize a device with a response time shorter than the duration of the physical process.
In the early days of time-resolved measurements, the aim was to resolve dynamics that are barely  outside of what the human eye can resolve.  
In the late 19$^{\textrm{th}}$ century, a notable example was Eadweard Muybridge's ground breaking work, where a series of cameras was used to record the gait of a horse and  
finally settling the long-debated question of whether all four hooves of a horse leave the ground during trotting and galloping. 
Over time, numerous stroboscopic techniques have been developed  to capture faster motion, but they do have limitations in achieving adequate temporal resolution. 
The introduction of the laser marked a groundbreaking  breakthrough in improving the time resolution capabilities of stroboscopic measurements.  

The advent of the laser in 1960 brought a revolutionary advancement in the study of matter's structure and its properties with unprecedented detail.
Beyond its remarkable impact on precision, the laser also unveiled a hitherto inconceivable ability in conventional optics: the power to generate light beams with electric field so intense that it 
could significantly alter the fundamental electronic and optical characteristics of materials. 
This pioneering  revelation gave rise to an extensive range of novel optical phenomena, collectively known as nonlinear optics. 

For many years, physicists had established a framework to elucidate light-matter interaction, relying on the perturbation theory within the quantum mechanics framework. 
In the realm of linear optics, the response of a semiconductor and dielectric materials to light 
is elegantly expressed through electric permittivity. 
This implies that the polarization induced in the material by the incident light is directly proportional to the electric field of the light, with a scaling constant governing the relationship.  
However, about 70 years ago, it was discovered that this linearity is not always the case. 
Merely a year after the laser's invention, Franken and co-workers made a remarkable discovery  
in which they demonstrated the doubling of the laser's frequency by focusing it on a quartz crystal~\citep{franken1961hill}. 
Simultaneously, Kaiser and Garrett reported an observation of  two-photon absorption in doped CaF${_2}$~\citep{kaiser1961two}. 
Both these phenomena were promptly understood as second-order nonlinear effects, 
facilitated by the substantial photon flux achievable with the newly created laser~\citep{franken1961hill,kaiser1961two,armstrong1962interactions}.
Subsequently, the rapid advancement of perturbative nonlinear optics has resulted in numerous applications in various fields, such as consumer electronics, communication, and medicine~\citep{garmire2013nonlinear}, significantly revolutionizing spectroscopy and microscopy~\citep{bloembergen1982nonlinear,zipfel2003nonlinear}.  

The introduction of chirped pulse amplification by Strickland and Mourou in 1985, 
for which they were awarded the Nobel Prize in 2018, enabled the amplification of laser pulses to mJ-level energies with femtosecond  pulse durations (1 fs = 10$^{-15}$ s) \citep{strickland1985compression}. 
This led to a significant enhancement in the achievable peak powers of laser pulses. 
By compressing these pulses to sufficiently short durations, even micro-joules of energy could generate electric field strengths comparable to intra-atomic electric fields. 
Shortly after the advent of chirped pulse amplification, 
it was observed that higher-order harmonics could be generated when a strong laser was 
focused on gaseous medium~\citep{mcpherson1987studies}.  
To understand the generation of higher-order harmonics, 
physicists  ventured into the realm of nonlinear optics, 
where the induced polarization is expanded further to incorporate higher-order functions of the light's electric field. 
In both linear and nonlinear optics, the light-matter interaction is conventionally approached as a small perturbation, assuming the influence of light on materials to be insignificantly small. 
To be more precise, it is assumed that the electric field (intensity) of the light remains small in comparison to the  binding energy of the electrons within the materials. 
In numerous cases, this assumption holds true, leading to a well-understood description of light-matter interaction via linear and nonlinear optics.
However, the conventional theory of perturbative optical harmonic generation proved inadequate  in explaining how a laser pulse could produce the observed plateau-like structure  of higher-order harmonics extending into the extreme ultraviolet (XUV) region of the electromagnetic spectrum [see Fig.~\ref{fig:hhggas}(a)]. 
Over the years, high-harmonic generation (HHG) has facilitated a route to generate 
 a coherent light source with attosecond pulse durations as well as used to interrogate ultrafast electronic motion in matter.  

Ultrashort laser pulses,  synthesized via HHG, have become vital  for investigating the ultrafast electron dynamics in atoms, molecules, and solids. 
These pulses are well-suited for studying various processes occurring on the femtosecond timescale, such as molecular vibrations, chemical bond breaking and formation. 
On the other hand,  electron dynamics in matter presents an even more intriguing aspect, 
taking place on an even faster  timescale - attosecond (1 as = 10$^{-18}$ s )~\citep{krausz2009attosecond}. 
An illustrative example of this phenomenon is the orbital period of an electron in a hydrogen atom within the Bohr model, which lasts a mere 152 attoseconds~\citep{chang2016fundamentals}. 
To delve into these ultrafast  processes and unravel the intricate structures and behaviors of atoms and electrons within matter, attosecond pulses have proven to be indispensable. 
These  laser pulses allow us to initiate and probe  various ultrafast electronic processes in matter, 
providing an unprecedented opportunity to improve our understanding of the world surrounding us~\citep{rading2016intense}. 
By probing events on the attosecond timescale, 
scientists gain valuable insights into the fundamental processes that govern the behavior of matter, 
leading to groundbreaking discoveries and advancements in numerous scientific fields.
In the subsequent section, our focus will be on strong-field-driven HHG. 
We will delve into the fundamental principle and mechanism underlying this intriguing phenomenon. Subsequently, we will explore HHG from solids, and delve into both interband and intraband electron dynamics in great detail. 
These discussions will shed light on the fascinating processes occurring within solids and provide a comprehensive understanding of HHG processes in different contexts.

\section{High-Harmonic Generation}
High-harmonic generation is a non-perturbative frequency up-conversion process in which 
higher integer multiple frequencies of the incident laser frequency are emitted. 
HHG from  gases was first observed by Ferray $\it et~al.$ using intense laser pulse~\citep{ferray1988multiple}. 
The HHG spectrum shown in Fig.~\ref{fig:hhggas}(a) displays an exponential drop in low-order harmonics, followed by a plateau that extends over several harmonic-orders with a relatively constant strength of harmonics. Subsequently, a cutoff  was observed where the harmonic yield  experiences a drastic  decrease.

\begin{figure}
\centering
\includegraphics[width= 0.8 \linewidth]{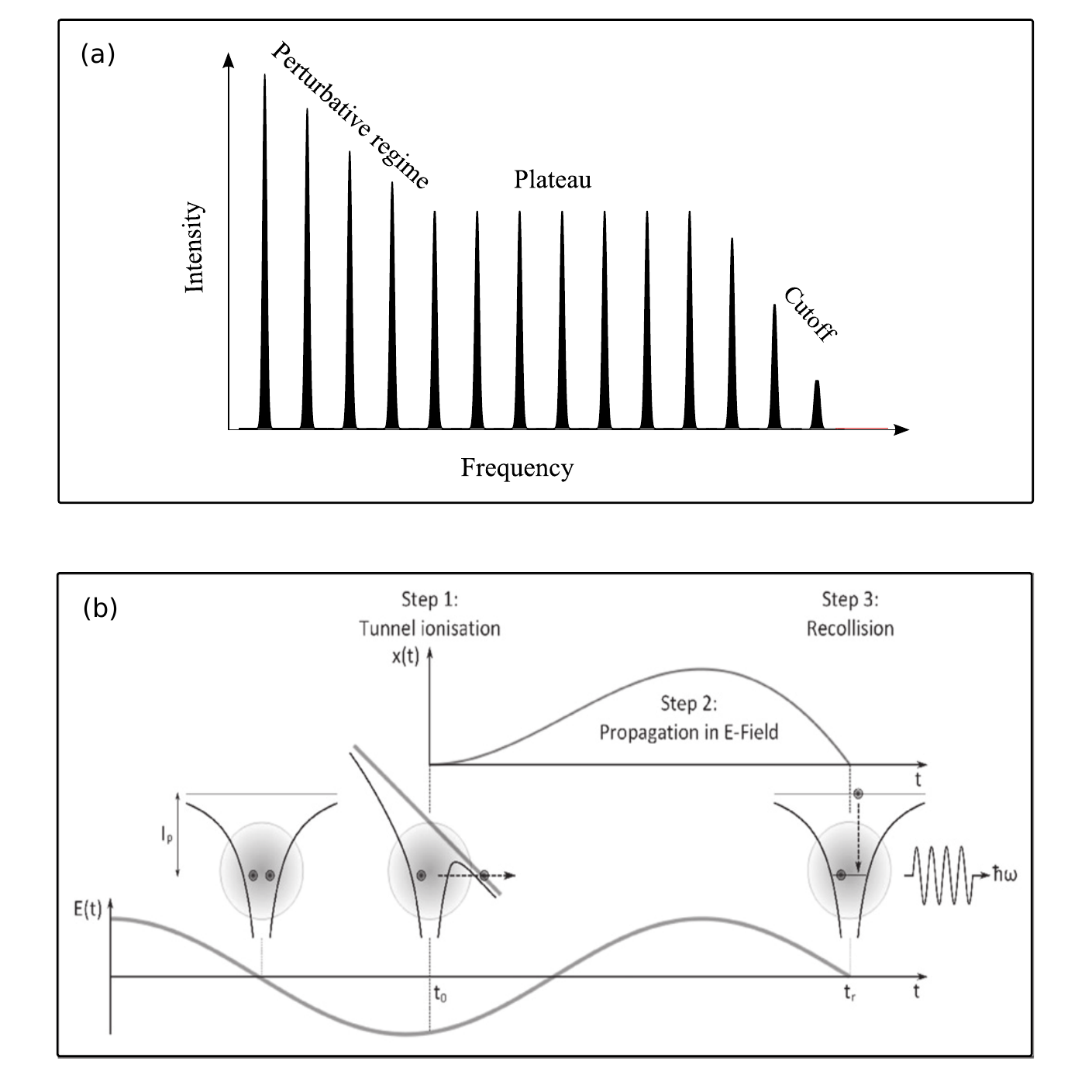}
\caption{(a) A representative  high-harmonic spectrum from gaseous medium. 
(b) The three-step recollision mechanism of HHG from gases in real-space, adapted from Ref.~\citep{thomson2013ultrafast}.}
\label{fig:hhggas}
\end{figure}

The underlying mechanism to understand HHG for gaseous targets was proposed by Paul Corkum in 
1993, which  is commonly referred as the semiclassical three-step (recollision) model~\citep{corkum1993plasma}. 
The strong incident laser pulse tunnel ionizes an electron as a  first step within this model. 
During the second step, the tunneled electron  propagates in the presence of the strong laser field
 and gains kinetic energy. 
Once the direction of the incident laser field reverses, the liberated electron accelerated back towards its parent ion, allowing it to eventually recombine as a third step, i.e., recombination. 
A photon, comprising the ionization energy as well as the total kinetic energy of the tunneled electron garnered from the laser field, is emitted as a result of the recombination as shown in Fig.~\ref{fig:hhggas}(b). 
A distinct theoretical approach, beyond the known perturbative framework, becomes necessary in the strong-field regime where the field-free potential acts as a perturbation.
Around a year after Corkum's semiclassical model, Lewenstein and colleagues proposed a quantum mechanical model of HHG based on the strong-field approximation, known as the Lewenstein model~\citep{lewenstein1994theory}.

The success of HHG from gases over the past decades has also stimulated attempts to generate strong-field-driven high-harmonics from solids.
Prior to extending extreme non-perturbative nonlinear optics to solids, substantial advances in laser technology were required to circumvent the material degradation caused by 
the atomic-scale electric field of the order of  1 V~\AA$^{-1}$.  
In the last two decades, significant  advancement in laser technologies in the mid-infrared and terahertz  frequency regimes  enabled the study of strong-field-driven processes in solids \citep{andriukaitis201190, schmidt2014frequency, manzoni2016design, sell2008phase}. 
These mid-infrared and terahertz  laser pulses made it possible to expose dielectric and semiconductors to laser fields exceeding their static dielectric strength without causing damage. 

\subsection{High-Harmonic Generation in Solids}

Ghimire $\it et~al.$ in 2011 demonstrated a pioneering work in which an intense mid-infrared pulse generated  coherent radiation at energies beyond the band gap of  ZnO~\citep{ghimire2011observation}. 
Subsequent experiments employing terahertz and mid-infrared  driving fields stretched the limits of solid-state nonlinear optics, gaining access to previously inaccessible spectral ranges in terahertz and extreme ultraviolet energy regimes~\citep{schubert2014sub, luu2015extreme}. 
Currently, HHG is being investigated in a variety of solid-state systems ranging from  semiconductors~\citep{ghimire2011observation, lanin_mapping_2017, schubert2014sub, gholam2017high}, dielectrics~\citep{luu2015extreme, garg2016multi, ndabashimiye2016solid, you2017anisotropic} to two-dimensional semiconductors~\citep{liu2017high}, gapless semimetals~\citep{yoshikawa2017high}, metasurfaces~\citep{liu2018driving}, and nanostructured solids~\citep{sivis2017tailored,vampa2017merge}.
Thus, HHG is observed when a solid interacts with an intense mid or near-infrared laser pulse. 
In the following, we will discuss the underlying mechanism responsible for HHG from solids. 

\begin{figure}
\centering
\includegraphics[width= \linewidth]{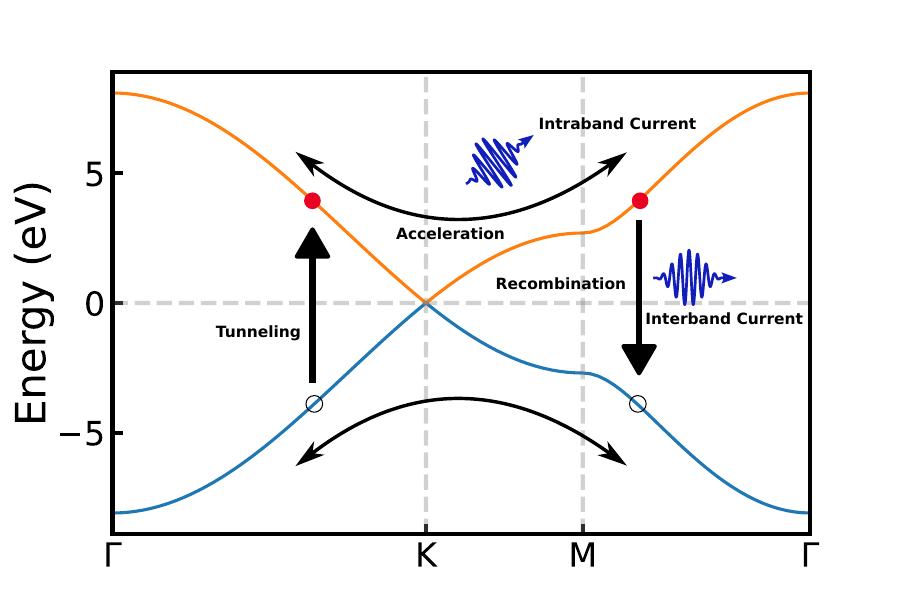}
\caption{The underlying mechanism of HHG in solids. The energy band structure of monolayer graphene along high-symmetry directions within two band picture in which the conduction (valence) band is shown by blue (orange) in momentum space.}
\label{fig:hhgmechanism}
\end{figure}
Let us draw a parallel between the three-step model of the HHG in gases and solids in order to provide a comprehensive explanation as shown in Fig.~\ref{fig:hhgmechanism}.
An intense laser pulse  promotes electrons from the valence to  conduction bands via tunnel ionization during the first step. 
The liberated electron (created hole) in the  conduction (valence) band  experiences acceleration due to the influence of the oscillating laser field  and undergoes oscillation in the second step. 
These oscillations, referred to as intraband oscillations, result in the generation of an intraband current.
Once the direction of the incident laser field changes its direction,  there is a probability for the 
electron in the conduction band to recombine with the hole in the valence band as a third step. 
This process, known as an interband transition, gives rise to an interband polarization.
The three-step model shown in Fig.~\ref{fig:hhgmechanism} provides a qualitative understanding of the HHG in solids.  

As discussed above, interband and intraband electron dynamics are two major processes 
that contribute to the high-harmonic spectrum of solids. 
Theoretical investigations have revealed that both types of dynamics lead  to high harmonics, with energies below and above the minimum  band gap of the solid, aligning with the experimental findings discussed in the preceding section. 
Depending on the specific choice of solid under study, the laser parameters employed, and the energy of emitted harmonics, either intraband or interband mechanism may dominate the generation process of higher harmonics in an experiment. 
In the following, we will provide a brief overview of the distinct characteristics of these  interband and intraband mechanisms.

\subsubsection{Intraband Mechanism:}

Let us introduce a semiclassical treatment to describe the intraband process, aiming to offer a simple and intuitive understanding of the HHG process. 
A more intricate and  complex mathematical model for the intraband process will be discussed  
in the later part of this thesis.
The phenomenon of intraband-driven  high-harmonic emission is linked to the nonlinear characteristics of the current, as indicated by the black two-sided arrows in Fig.~\ref{fig:hhgmechanism}. 
This nonlinearity arises when charge carriers are driven by the intense laser field in the anharmonic bands.  The strong anharmonic motion of the charge carrier leads to an overall nonlinear current that results in the emission of higher-order harmonics.  
It is worth mentioning that  intense terahertz-driven harmonic generation is 
dominated by the intraband current~\citep{feise1999semiclassical}.  

The group velocity of an electronic wave packet  in the conduction band can be written as
\begin{equation}\label{eq:grpvel}
v_{g}(t) = \frac{d \varepsilon(\mathbf{k})}{d\mathbf{k}}. 
\end{equation}
Here, the electronic wave packet is consist of a superposition of Bloch waves with different crystal momenta $\mathbf{k}$ and $\varepsilon (\mathbf{k})$ is the energy dispersion of the band structure in a solid.  
Let us express the band structure's dispersion in terms of higher-order cosine as 
\begin{equation}\label{eq:highcos}
\varepsilon(\mathbf{k}) = \sum_{l = 0}^{\infty} c_{l} \cos{(l \mathbf{k} a)}, 
\end{equation}
where $a$ is the lattice constant and $c_{l}$ are the expansion coefficient 
that can be obtained by fitting Eq.~\eqref{eq:highcos}.
We can end the series expansion at $l_m$, i.e., the maximum order required to fit the band structure. 
This means the highest characteristic distance involved is $l_{m}a$ in the solid~\citep{luu2015extreme}. On substituting Eq.~\eqref{eq:highcos} into Eq.~\eqref{eq:grpvel}, the expression of the group velocity reads as 
\begin{equation}
v_{g}(t) = \sum_{l = 0}^{l_{m}} l a c_{l} \sum_{m = 1} J_{2m-1}\left(\frac{la E_{0}}{\omega_{0}}\right) \sin[(2m-1)\omega_{0}t], 
\end{equation}
where $E_{0}$ and $\omega_{0}$ are the peak strength and frequency of the laser's electric field, respectively.  
In the above expression, Jacobi-Anger expansion is used to express sin term in 
the form of the Bessel function of the first kind of order $m$, $J_{m}$~\citep{golde2008high}.
The current due to the intraband dynamics is obtained from the group velocity as
\begin{equation}
\textbf{J}_{intra}(t) 
\propto \int_{BZ} v_{g}(t)\eta_{g}(\mathbf{k},t)d\mathbf{k},
\end{equation}
where  $\eta_{g}(\mathbf{k},t)$ is the time-dependent distribution  function of the wave packet, i.e.,
population distribution function~\citep{feise1999semiclassical, luu2015extreme}. 
The yield of  the $n^{\textrm{th}}$ harmonic can be obtained 
from the time-derivative of the intraband current as~\citep{mucke2011isolated}
\begin{equation} \label{eq:intraharm}
\textbf{H}_{n, intra}(\omega) \propto \left|n \omega_{0}\sum_{l=1}^{l_{m}} l a c_{l}~J_{n}\left(\dfrac{l\omega_{B}}{\omega_{0}}\right) \right|.
\end{equation}
Here,  $\omega_{B} = a E_{0}$ is Bloch frequency. 
The dependency of the intraband contribution to HHG yield on Bloch frequency relies  upon the context of Bloch oscillations, which 
occur  when electrons in a periodic solid are subjected to a continuous electric field~\citep{bloch1929quantenmechanik}.
In the present context, an electric field associated with an ultrashort laser pulse exhibits  sinusoidal behavior rather than being constant. 
This provides the notion of ``dynamical Bloch oscillations''~\citep{foldi2013effect,ghimire2014strong,yoshikawa2017high,silva2018high}.
In addition, analysis of  Eq.~\eqref{eq:intraharm} reveals nonlinear relationship between the
intraband-driven harmonic yield and the electric field strength, which can be understood by considering the amplitude and Fourier components of the effective band. 

As discussed above, 
HHG in solid was first observed in ZnO and the intraband mechanism was used to explain the observed features in the experiment, such as 
plateau-like structure in the harmonic spectrum and linear dependence of the energy cutoff on the 
laser's electric field~\citep{ghimire2011observation}. 
In addition, the intraband mechanism has reproduced the limited sensitivity to the carrier-envelope phase 
 of high harmonic emission in solids, while also providing a reasonable estimation of the spatial characteristics of the macroscopic harmonic beams~\citep{luu2015extreme,garg2016multi, garg2018ultimate}.  
Moreover, intraband dynamics  
further anticipates the production of high harmonics without any chirping, demonstrating synchronization with the peak field~\citep{garg2016multi,langer2016lightwave}.  
However, subsequent studies revealed that the interband mechanism provides the more  accurate explanation for the temporally chirped high harmonic emission~\citep{vampa2015linking, vampa2015all}.  
Thus, let us discuss the interband mechanism responsible for HHG.
 
\subsubsection{Interband Mechanism:}
An analogous  three-step model, resembling HHG from gases,  is developed 
to explain the interband-driven harmonics with energy above  the minimum  band gap in solids. 
An interband mechanism involves an electronic transition from the conduction band to the valence band, which  can also be comprehended by considering its polarization current. 
The applied laser field drives  electrons and holes in their respective bands in opposite directions, which leads to an accumulation of polarization and subsequently emission of harmonics~\citep{vampa2015semiclassical}. 
The interband current can be expressed in terms of  the time derivative of the polarization as
\begin{equation}
\textbf{J}_{inter}(t) = \dfrac{\partial \textbf{P}(t)}{\partial t},
\end{equation}
where $\textbf{P}(t)$ is the polarization. 
The yield of the higher-order harmonics can be 
obtained  by the Fourier transform of the time-derivative of $\textbf{J}_{inter}(t)$ as
\begin{equation}
\textbf{H}_{n, inter}(\omega) \propto \left| \omega ^{2} \tilde{\textbf{P}}(\omega) \right|.
\end{equation}
Owing to the polarization accumulation,  the interband-driven generalized recollision model of HHG indicates the occurrence of a chirped emission of harmonics  and 
not synchronized with the crests of the driving field. 
This implies that the emission time of the harmonics,  during the recollision process,   
exhibits variation depending on the order of the harmonic.  
The chirped emission  has been experimentally verified by analyzing HHG from  ZnO~\citep{vampa2015linking} and MgO~\citep{vampa2020attosecond}. 
Moreover, the importance  of the classical trajectories  during the recombination 
has been established through their comparison with windowed Fourier transforms of full quantum calculations~\citep{vampa2014theoretical}. 
The simple quasi-classical model  also successfully explains the linear dependence of the harmonic cutoff on the electric field strength. 
An intricate interplay between interband and intraband harmonics has been discussed 
in the realm of gapless semimetal, such as graphene~\citep{mrudul2021high}.

In recent times, HHG in solids has emerged as a powerful method to probe numerous static and dynamic 
properties of solids, such as  band dispersion~\citep{vampa2015all, luu2015extreme, lanin2017mapping, pattanayak2019direct}, the density of states~\citep{tancogne2017impact},  band defects~\citep{mrudul2020high, pattanayak2020influence}, valley pseudospin~\citep{mrudul2021light, langer2018lightwave, mrudul2021controlling, rana2023all, rana2022generation}, 
Bloch oscillations~\citep{schubert2014sub},
topology and light-driven phase transitions, 
including strongly correlated systems ~\citep{bauer2018high, bai2021high, imai2020high, borsch2020super, baykusheva2021all, bharti2022high, pattanayak2022role, shao2022high}, and even combine attosecond temporal with picometer spatial resolution of the electron trajectories in the lattices~\citep{lakhotia2020laser}. 
On the technological side, HHG from solids offers an attractive possibility of fashioning
all-solid-state compact optical devices to generate bright, coherent attosecond pulses in 
extreme ultraviolet energy range.
These studies have provided  exciting insights into the behavior of electrons and the associated properties of solids. 
The applications mentioned have significant implications for the advancement of ultrafast science and the exploration of fundamental phenomena in solids.

\section{Motivation}
Since its inception in 2011, high-harmonic spectroscopy  has become a method of choice to probe various 
static and dynamic aspects of solids on ultrafast timescale. 
In spite of the major advances, it is still not known how the coherent lattice dynamics affect HHG from solids: remarkably, the role of lattice vibrations in solid HHG remains uncharted territory. 
This stands in stark contrast to the HHG from a gaseous medium, where high-harmonic spectroscopy has been employed to probe nuclear motion in various molecules~\citep{patchkovskii2009nuclear, wagner2006monitoring, le2012theory, baker2006probing, lein2005attosecond, worner2011conical}. 
Thus, there is an exciting opportunity to leverage high-harmonic spectroscopy as a sensitive probe of  coherent lattice dynamics in solids, offering a plethora of potential applications in diverse fields, such as studying lattice-induced symmetry breaking~\citep{nova2017effective}, observing phase transitions~\citep{shin2018phonon}, exploring energy transfer through nonradiative processes~\citep{lin2017ultrafast}, to name but a few. 
This thesis focuses on probing lattice dynamics and the interplay of lattice vibrations with electron dynamics in solids via HHG~\citep{rana2022high, rana2022probing}. 

Coherent  lattice dynamics influence a diverse range of physical and chemical processes in solids. Thus, probing  lattice dynamics is essential to comprehend several ubiquitous phenomena in solids. 
By elucidating the vibrational characteristics of crystalline solids and the behavior of constituent atoms  undergoing coherent oscillatory motion, lattice dynamics facilitate a comprehensive understanding of diverse  aspects of solids. 
These properties encompass thermal attributes, such as heat capacity and thermal conductivity~\citep{Fultz2010,lindsay2020thermal}; structural aspects 
including crystalline arrangement characterization and the manifestation of phase transitions arising from external stimulus~\citep{Dove1993}.


Over the years, various experimental techniques are developed  to  interrogate  lattice dynamics and unravel the associated  vibrational modes of the solids.  
Raman spectroscopy, characterized by its ability to discern vibrational modes based on frequency shifts, stands as a cornerstone in this pursuit~\citep{smith2019modern}. 
Infrared spectroscopy, by measuring light absorption due to lattice vibrations, complements this endeavor by revealing lattice vibrations and the nature of involved chemical bonds~\citep{socrates2004infrared}.
Neutron scattering, encompassing both inelastic and elastic scattering, emerges as a formidable probe, unveiling phonon dispersion relations and highlighting defects and impurities within solids~\citep{squires1996introduction,Willis2009}. 
Additionally, spectroscopic and imaging methods based on terahertz, nuclear magnetic resonance, ultrasonic, and x-rays  collectively enrich the landscape of lattice dynamics exploration~\citep{averitt2002ultrafast}.

Typically, these above-mentioned measurements are employed  within the domains of energy and momentum, and thus  they do not provide temporal information about lattice dynamics.  
Time-resolved spectroscopic methods within pump-probe configuration are a standard approach to overcome the issue of temporal dynamics. 
In this context, a crucial consideration is the duration of the probe pulse, which  must be shorter than the characteristic timescale of the lattice dynamics under investigation. 
The timescale associated with inter-atomic  lattice dynamics, responsible for chemical bond breaking and formation and phase transitions, lies on the femtosecond regime.   
Concurrently, dynamics intertwined with electron correlation and exchange interactions can transpire even more expeditiously. 
Thus, time-resolved mapping of the interplay of the coherent lattice vibration with electronic motion on the electronic timescale is challenging. 
Moreover, light-driven lattice dynamics may lead to changes in the symmetry, which 
could alter  the electronic structure of solids.
How do coherent lattice vibrations and associated alterations in symmetries affect the electronic properties of a system, and how can these changes be probed with all-optical approaches?
By delving into these questions, we aim to gain valuable insights into the interplay between lattice and electron dynamics and advance our understanding of solid's behavior at a fundamental level.

Owing to large coherent bandwidth, high-harmonic spectroscopy offers a very promising all-optical approach as it is sensitive to the sub-cycle electron dynamics, the transient changes in symmetry, and the 
electronic  properties of solids. Moreover, if we aim to employ light-driven electron dynamics in solids for technological applications, such as valleytronics and petahertz electronics, we have to understand 
how lattice vibrations affect electron dynamics in solids. 
The present thesis aims to address these crucial issues.
Another essential aspect of this thesis is to introduce a method to image coherent lattice 
dynamics in real-space and real-time  without using the pump-probe approach. 
Our proposed method is based on inelastic scattering, which provides an alternative approach to the conventional pump-probe method. 

Throughout the majority of this thesis, our focus will be on probing coherent lattice dynamics and their interplay with electron dynamics in  monolayer graphene. 
The study of two-dimensional (2D) materials, such as graphene transition metal dichalcogenides,  has emerged as a rapidly growing research field  in the last few decades~\citep{neto2009electronic, xiao2010berry}. 
These atomically-thin 2D materials exhibit a wide range of unique electronic, optical, and mechanical properties that have attracted significant interest from researchers in diverse fields, including materials science, physics, chemistry, and engineering~\citep{novoselov20162d, zhang2014two,zhu2014semiconducting}. 
One of the key features of 2D materials is the confinement of electrons and phonons within a thin layer, leading to quantum mechanical effects and enhanced interactions between these particles.  Understanding the behavior of electrons and phonons in 2D materials is essential for designing and optimizing the performance of electronic and optoelectronic devices based on 2D materials.

\section{Thesis Overview}
This thesis consists of six chapters and is organized as mentioned below. 

We present the fascinating world of ultrafast spectroscopy and the intricate interaction between light and matter  in \textbf{Chapter 1}.
We  discuss the three-step model of HHG in solids, including a comprehensive analysis of both  intraband and interband   mechanisms that govern this phenomenon.  
Moreover, we highlight the significance of probing coherent lattice dynamics in solids, 
encouraging readers to delve deeper into this captivating field. 
By shedding light on these fundamental aspects, we set the stage for a comprehensive understanding of the intricate interplay among light, electrons, and lattice vibrations in solids. 

\textbf{Chapter 2} delves into an extensive discussion of the theory of electron and lattice dynamics. 
We start by introducing  a tight-binding approach for the electronic properties of graphene.  The light-matter interaction is addressed by solving the time-dependent Schrodinger equation in length gauge. Density-matrix-based semiconductor Bloch equations in the Houston basis  are introduced to numerically  simulate HHG. 
In addition, we present the Born model of lattice dynamics, which aids in obtaining eigenvectors corresponding to coherent phonon modes and phonon dispersion of graphene. 
The culmination of this chapter involves a detailed description of simulating HHG with lattice dynamics using  a time-dependent tight-binding model wherein the  coherent lattice dynamics is treated classically.

\textbf{Chapter 3} presents insights into the nonlinear optical response of graphene with coherent lattice dynamics, specifically  focusing on the in-plane longitudinal and  transverse  phonon modes. The investigation includes impacts  of transient-symmetry breaking and the dynamical symmetries (DSs)
on HHG. 
Graphene with coherent lattice dynamics results sidebands generation in the high-harmonic spectrum.  Appearances of  even- and odd-order sidebands depend on the symmetry of the coherently excited phonon mode and the polarization of the HHG probe pulse. The observed findings are explained using DSs within the Floquet framework. 
Finally, we thoroughly discuss the sensitivity of the HHG spectra with respect to the properties of transiently-evolving graphene and various laser parameters.

The capability of high-harmonic spectroscopy,  by retrieving the phase between two coherently excited phonon modes,  is demonstrated in \textbf{Chapter 4}. 
This chapter introduces a recipe to  probe the ``chirality'' of circular phonons via high harmonic spectroscopy.  
Towards the end of this chapter, we  show that the coherent phonon dynamics alters the dynamical symmetry of graphene with the probe pulse and leads to the generation of symmetry-forbidden harmonics. 
The symmetry of the in-plane phonon mode can be characterized by analyzing involved 
DSs and polarization analysis of the sidebands associated with the main harmonic peaks. 

\textbf{Chapter 5} introduces a robust approach for four-dimensional imaging of  
lattice dynamics in real-space and  real-time with atomic-scale spatiotemporal resolution.  
Our approach is based on analyzing experimental data in the momentum-frequency domain, obtained from inelastic scattering; and subsequently  transforming the results in the cartesian space-time domain. 
The fluctuation-dissipation theorem is used to obtain the complete response function from the dynamical structure factor in the momentum-frequency domain. 
An excellent agreement is shown by comparing the results of coherent lattice dynamics in germanium obtained by our proposed approach and time-resolved diffuse x-ray scattering within the pump-probe configuration. 
Additionally, we also show that our methodology is capable of four-dimensional imaging of coherent lattice dynamics, whether emanating from a point-like nucleation site or an extended defect. 

\textbf{Chapter 6} provides the conclusion and the future scope of this thesis.

\cleardoublepage
\chapter{Theory of Electron and Lattice Dynamics}\label{chp:Chapter2}


To gain a comprehensive understanding 
about various facades of solids, it is essential  to consider  an intricate interplay between the valence electrons behavior, and the combined effects of both internal forces stemming from the lattice structure and external stimuli applied to the solids. 
This scientific problem generally  entails addressing a non-equilibrium many-body problem, given the entangled interactions among multiple degrees of freedom.  
However, the majority of the phenomena of our interest can be effectively elucidated using a one-electron description.
Our objective is to understand the behavior of valence electrons in a solid interacting with an intense  laser pulse. 
Subsequently, we aim to investigate 
how coherent  lattice dynamics affect laser-driven electron dynamics in solids. 
For this purpose, we can start our discussion by writing the total Hamiltonian of a system. 
We can separate the lattice and electronic degree of freedoms as the nuclei are heavier  than the electrons\footnote[1]{Using Born-Oppenheimer and Adiabatic approximation.}. The total time-independent Hamiltonian can be written as 
\begin{equation}
 \mathcal{H}_{\textrm{total}} = \underbrace{\mathcal{T}_{\textrm{e}}+\mathcal{V}_{\textrm{e}}}_{\mathcal{H}_{\textrm{e}}}+\underbrace{\mathcal{T}_{\textrm{L}}+\mathcal{V}_{\textrm{L}}}_{\mathcal{H}_{\textrm{L}}}+\mathcal{H}_{\textrm{eL}}.
\end{equation}
Here, $\mathcal{H}_{\textrm{e (L)}}$ represents the electronic (lattice) Hamiltonian consists of kinetic ($\mathcal{T}_{\textrm{e (L)}}$) and potential ($\mathcal{V}_{\textrm{e (L)}}$) energies, 
and $\mathcal{H}_{\textrm{eL}}$ stands for the 
interaction between the electron and lattice. 

As a starting point, we will discuss the tight-binding approach to solve the electronic Hamiltonian  corresponding to the equilibrium lattice structure of the monolayer graphene. 
After obtaining eigenstates and eigenenegies of graphene within the tight-binding picture,  
a density-matrix-based theory to describe laser-driven electron dynamics will be developed. 
The developed density-matrix formalism will be calibrated with the available  theoretical   and experimental results. 
Born approach will be adopted to solve the lattice Hamiltonian and obtain the phonon dispersion for graphene. 
We will update our density-matrix formalism to incorporate the coherent lattice dynamics and its impact on 
the laser-driven electron dynamics. 

This chapter is designed as follows: Section~\ref{section:2.1}  will discuss the tight-binding 
model of graphene, which can be extended to gapped graphene by adding a mass term in the electronic Hamiltonian. 
We will recap light-matter interaction and present the derivation of density-matrix-based semiconductor Bloch equation approach in section~\ref{section:2.2}. 
We will also analyze and discuss the high-harmonic spectrum for monolayer graphene. 
Section~\ref{section:2.3} will be devoted to the Born model, which is employed to derive the phonon dispersion. 
Furthermore, we will also discuss the electron-phonon interaction by extending the tight-binding approach  
in the time domain.  Atomic units are used  throughout the entirety of this thesis unless stated otherwise. 

\section{Tight-Binding Approach}\label{section:2.1}
The carbon atoms in graphene are arranged in a honeycomb lattice structure  as illustrated in Fig.~\ref{fig2.1}(a). 
The unit cell is composed of two carbon atoms and is defined by the primitive lattice vectors:  $\textbf{a}_{1}$ = $a_{0}$($-1/2$,$\sqrt{3}/2$) and $\textbf{a}_{2}$ = $a_{0}$($1/2$,$\sqrt{3}/2$)
with  $a_{0} = 2.46$~\AA~being the lattice parameter.  
These lattice vectors  provide  reciprocal lattice vectors as $\textbf{b}_{1}$ = 2$\pi$/$a_{0}$($-1$,$1/\sqrt{3}$) and  $\textbf{b}_{2}$ = 2$\pi$/$a_{0}$($1$,$1/\sqrt{3}$). 
The high symmetry points in graphene are 
$\mathsf{\Gamma}$, $\mathsf{\bm{M}}$, $\mathsf{\bm{K}}$, and $\mathsf{\bm{K}^{\prime}}$ as shown in Fig.~\ref{fig2.1}(b). 

\begin{figure}[t!]
\centering
\includegraphics[width=0.8\linewidth]{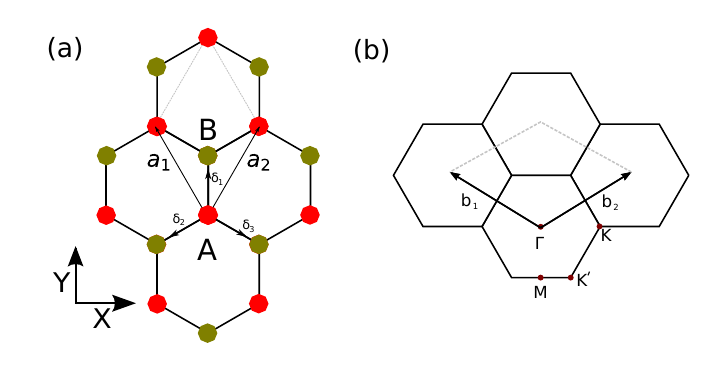}
\caption{Hexagonal honeycomb lattice structure of graphene. 
(a) Real-space structure of graphene with $\textbf{a}_{1}$ and $\textbf{a}_{2}$ as primitive lattice vectors. The A and B represent two non-equivalent atoms of graphene arranged in the honeycomb lattice. (b)  Brillouin zone in the momentum space with 
$\mathsf{\Gamma}$, $\mathsf{\bm{M}}$, $\mathsf{\bm{K}}$, and $\mathsf{\bm{K}^{\prime}}$  as the high symmetry points.} \label{fig2.1}
\end{figure}

Each of the carbon atoms in graphene has a total of four valence electrons, two of which occupy the 2$s$ orbital, while the other two occupy the 2$p$ orbital. 
During $sp^{2}$ hybridization, one of the 2$s$ electrons is promoted to the 2$p$ orbital, 
and the three 2$p$ electrons participate in hybridization to form three $sp^{2}$ hybrid orbitals.
The three $sp^{2}$ hybrid orbitals of each carbon atom overlap with those of neighboring carbon atoms to form three \textit{$\sigma$} bonds, which create the planar hexagonal lattice structure of graphene. 
The remaining unhybridized $p$ orbital, specifically the $p_{z}$ orbital, 
is oriented perpendicular to the plane of the hexagonal lattice and contains one electron.
The $p_{z}$ orbitals of neighboring carbon atoms overlap to form the delocalized \textit{$\pi$} bonding network that gives rise to the \textit{$\pi$}  and \textit{$\pi^{*}$} bands in the graphene's electronic structure. 
These bands are important for writing the tight-binding model of graphene as they dominate the electronic properties near the Fermi level. The \textit{$\pi$} and \textit{$\pi^{*}$} band dispersion determines the electronic band structure and the density of states, and it plays a crucial role in determining various  electronic properties of graphene.
Moreover, the \textit{$\pi$} and \textit{$\pi^{*}$} electrons in graphene exhibit a high degree of spatial delocalization, which means that they are relatively insensitive to local perturbations, 
and are less likely to be affected by defects or impurities in the lattice. 
However, the \textit{$\sigma$} bands, arise from the overlap of the $s$, $p_{x}$, and $p_{y}$ orbitals 
of the carbon atoms, are less important for the electronic properties as they are ``more''  localized.
Therefore, the tight-binding model of graphene focuses only on the \textit{$\pi$} and \textit{$\pi^{*}$} bands as they are the most relevant for the electronic properties of graphene~\citep{dresselhaus1998physical}.

Owing to the delocalization of the $p_{z}$ orbitals, the electronic wave function can be estimated using a tight-binding approximation. 
Moreover, Bloch functions can be derived from the atomic orbitals of two non-equivalent carbon atoms. 
The electronic ground state of graphene can be  best described by the corresponding Hamiltonian within  tight-binding approximation as~\citep{dresselhaus1998physical}
\begin{equation}\label{tb:1}
\mathcal{H}_{\textrm{e}} = - \gamma_0 \sum_{i,j} (\hat{a}_{i}^{\dagger} \hat{b}_{j} + \hat{b}_{j}^{\dagger} \hat{a}_{i}). 
\end{equation}
Here, $i~(j)$ is the label for the site in the sublattice A (B), $\hat{a}_{i}^{\dagger}$ ($\hat{a}_{i}$) is the fermionic operator associated with creation (annihilation) of an electron with position \textbf{r$_{i}$} at the site A, and similarly for $\hat{b}_{j}^{\dagger} ~(\hat{b}_{j})$. $\gamma_0$ is the nearest-neighbor hopping energy, which is chosen to be 2.7 eV~\citep{reich2002tight,moon2012energy}. 

\begin{figure}[t!]
\centering
\includegraphics[width=0.9\linewidth]{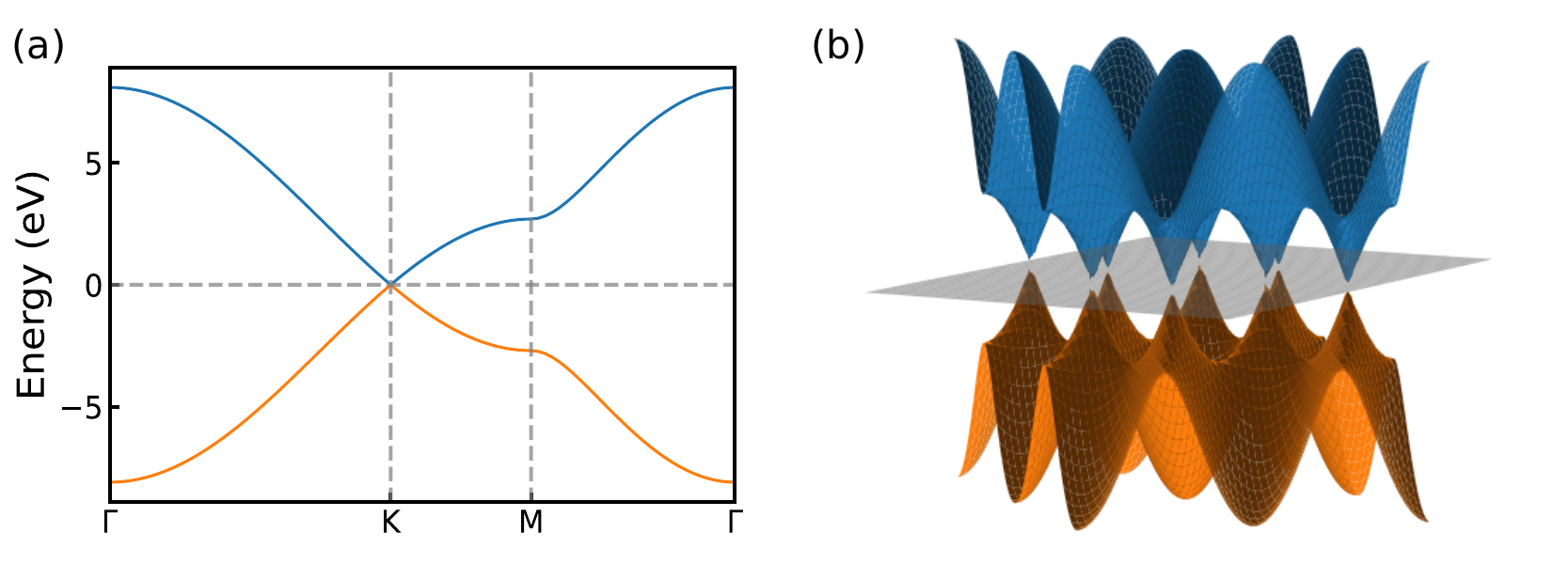}
\caption{Energy dispersion in graphene obtained by simulating the tight-binding Hamiltonian in  Eq.~(\ref{tb:4}). (a) Valence and conduction bands along the high-symmetry points in  one dimension, and 
(b) energy band structure  in three dimensional.  }\label{fig2.2}
\end{figure}

The tight-binding Hamiltonian in  Eq.~(\ref{tb:1}) can be further simplified  within nearest-neighbor interaction as 
\begin{equation}\label{tb:2}
    \mathcal{H}_{nn} = - \gamma_0 \sum_{i} \sum_{\bm{\delta}} (\hat{a}_{i}^{\dagger} \hat{b}_{i+\bm{\delta}} + \hat{b}_{i+\bm{\delta}}^{\dagger} \hat{a}_{i}), 
\end{equation}
where summation $\bm{\delta}$ is performed over nearest-neighbor vectors as shown in Fig.~\ref{fig2.1}(a). 
The creation and annihilation operators in the momentum space can be written as 
\begin{equation}\label{tb:3}
 \hat{a}_{i}^{\dagger} = \dfrac{1}{\sqrt{N/2}} \sum_{\textbf{k}}~e^{i\textbf{k}\cdot \textbf{r}_i} \hat{a}_{\textbf{k}}^{\dagger}, ~~~~\textrm{and}~~~~
 \hat{a}_{i} = \dfrac{1}{\sqrt{N/2}} \sum_{\textbf{k}}~e^{- i\textbf{k}\cdot \textbf{r}_i} \hat{a}_{\textbf{k}}.   
\end{equation}
Here, $\hat{a}_{\textbf{k}}^{\dagger}$ ($\hat{a}_{\textbf{k}}$) is the creation (annihilation) operator for atom A in the unit cell, $\textbf{k}$ is the crystal momentum and $N/2$ is the number of sites for the sublattice A. 
Similar expressions can be written for the sublattice B. 
Using the definition of the creation and annihilation operators from Eq.~(\ref{tb:3}), the tight-binding Hamiltonian for graphene in momentum space within nearest-neighbor interaction can be written as 
\begin{equation}\label{tb:4}
\mathcal{H}_{\textbf{k}} = - \gamma_0 \sum_{\bm{\delta}} ~e^{i\textbf{k}\cdot \bm{\delta}} \hat{a}_{\textbf{k}}^{\dagger} \hat{b}_{\textbf{k}} + \textrm{H.c.}
\end{equation}
Here, \textrm{H.c.} stands for Hermitian conjugate. 
The energy eigenvalues corresponding to the above Hamiltonian can be written as 
\begin{equation}\label{tb:5}
    E_{\pm}(\textbf{k}) = \pm \gamma_0 \sqrt{3+2|f(\textbf{k})|},
\end{equation}
where $|f(\textbf{k})|$ = $\exp\left(\dfrac{i\text{k}_{\text{y}}a_{0}}{\sqrt{3}}\right) + 2\cos\left(\dfrac{\text{k}_{\text{x}}a_{0}}{2}\right) \exp\left(\dfrac{-i\text{k}_{\text{y}}a_{0}}{2\sqrt{3}}\right)$.
Energy dispersion  along the high-symmetry points in one dimension is shown in 
Fig.~\ref{fig2.2}(a). 
The energy gap is zero and maximum at $\mathsf{\bm{K}}$ and $\mathsf{\bm{\Gamma}}$ points, respectively.
The three-dimensional view reveals that the valence and conduction bands touch at six points when Eq.~(\ref{tb:5}) approaches to zero. 
These points are commonly known as Dirac points, and it can be demonstrated with ease that only two of them are independent, and referred as  $\mathsf{\bm{K}}$ and $\mathsf{\bm{K}^{\prime}}$ points as shown in Fig.~\ref{fig2.1}(b).

\section{Light-Matter Interaction}\label{section:2.2}
Light-matter interaction is a process in which electromagnetic radiation (light) interacts with matter, resulting  in various phenomena, such as absorption, reflection, and transmission of light. 
These phenomena play 
an important role in many areas of science and technology, including optics, spectroscopy, and electronics
~\citep{hecht2017optics,boyd2020nonlinear}. 
Moreover, understanding the nature of light-matter interaction is crucial for the design and development of many  devices,  such as lasers, solar cells, and sensors to name 
but a few~\citep{yariv2007photonics,saleh2019fundamentals}.
At the fundamental level, light-matter interaction is governed by quantum mechanics and involves the exchange of energy between photons (quanta of electromagnetic radiation) and electrons or other particles in matter. 
The interaction is typically described by the Maxwell equation, which governs the propagation of electromagnetic waves, and the time-dependent Schr\"{o}dinger equation (TDSE)  describing the behavior of quantum mechanical particles, such as electrons in the presence of light. 
These equations can be combined to yield the so-called Maxwell-Schr\"{o}dinger equations, which provide a complete description of light-matter interaction in a given system~\citep{saleh2019fundamentals,klingshirn2007semiconductor}. 

Interaction of a solid with a laser pulse may lead to  a variety of processes, such as the generation of electron-hole pairs, the excitation of electrons to higher energy states, and the recombination of electrons and holes~\citep{gorbachev2014detecting}.
In the following, we will discuss a theoretical framework to describe laser-driven coherent electron dynamics in solids including the behavior of electrons in the conduction and valence bands.
Specifically, we will focus on the electric field component of the laser pulse as it is typically much stronger than the magnetic field component, which is weaker by a factor of $1/c$.

\subsection{Laser-Driven Electron Dynamics}
Let us solve  TDSE in the Houston basis  to describe laser-driven  electron dynamics in a  solid~\citep{wu2015high}. 
The Houston states are the instantaneous eigenstates of the time-dependent Hamiltonian $\mathcal{H}(t)$ as
\begin{equation}\label{SBE:1}
\mathcal{H}(t)\left|\widetilde{\phi}_{n,\textbf{k}} \right\rangle = \mathcal{E}_n^{\textbf{k}_{t}}\left|\widetilde{\phi}_{n,\textbf{k}} \right\rangle.
\end{equation}
Here, $\textbf{k}_{t} = \textbf{k} + \mathcal{A}(t)$ represents the canonical momentum 
in the presence of a vector potential $\mathcal{A}(t)$, $\left|\widetilde{\phi}_{n,\textbf{k}} \right\rangle$ is an
instantaneous eigenstate of $\mathcal{H}(t)$, $\mathcal{E}_n^{\textbf{k}_{t}}$ is the energy eigenvalue 
corresponding to the $n^{\textrm{th}}$ eigenstate.
The Houston state basis is related to the Bloch state basis with crystal momentum by $\textbf{k}_{t}$ as 
\begin{equation}\label{SBE:6}
	\left|\widetilde{\phi}_{n,\textbf{k}}(t)\right\rangle = e^{-i\mathcal{A}(t)\cdot\hat{\textbf{r}}}\left|\phi_{n,\textbf{k}_{t}}\right\rangle,
\end{equation}
where $\left|\phi_{n,\textbf{k}_{t}}\right\rangle$ is the eigenstates of a field-free Hamiltonian and  
known as Bloch state basis. The Bloch state can be written as the product of a plane wave and a periodic function of a lattice as $\left\langle \hat{\textbf{r}} \right. \left|\phi_{n,k_{t}}\right\rangle = e^{i\textbf{k}_{t}\cdot\textbf{r}}u_{n,\textbf{k}_{t}}(\textbf{r})$.

Let us assume time-dependent wave function $\left| \psi(t)\right\rangle$ satisfies TDSE as
\begin{equation}\label{SBE:3}
i\frac{ d}{dt}\left|\psi(t)\right\rangle = \mathcal{H}(t) \left|\psi(t)\right\rangle.
\end{equation}
In the present scenario, $\left|\psi(t)\right\rangle$ can be expanded in the Houston basis as 
\begin{equation}\label{SBE:4}
	\left|\psi(t)\right\rangle = \sum_{n,\textbf{k}} c_{n,\textbf{k}}(t)\left|\widetilde{\phi}_{n,\textbf{k}}(t)\right\rangle.
\end{equation}
By substituting Eq.~(\ref{SBE:4}) in Eq.~(\ref{SBE:3}), TDSE can be written as 
\begin{equation}\label{SBE:5}
	i \sum_{n,\textbf{k}} \left[\dot{c}_{n,\textbf{k}}(t) \left|\widetilde{\phi}_{n,\textbf{k}}(t) \right\rangle 
	+ c_{n,\textbf{k}}(t) \partial_t  \left|\widetilde{\phi}_{n,\textbf{k}}(t) \right\rangle \right] = 
	\sum_{n,\textbf{k}} c_{n,\textbf{k}}\mathcal{H}(t) \left|\widetilde{\phi}_{n,\textbf{k}}(t) \right\rangle.
\end{equation}

The partial derivative of the Houston state,  i.e., $\partial_{t}\left|\widetilde{\phi}_{n,\textbf{k}}(t)\right\rangle $ can be simplified using Eq.~(\ref{SBE:6}) as 
\begin{equation}\label{SBE:7}
\begin{aligned}
    \partial_{t}\left|\widetilde{\phi}_{n,\textbf{k}}(t)\right\rangle & = \partial_{t} \left[ 
 e^{-i\mathcal{A}\cdot\hat{\textbf{r}}} \left|\phi_{n,\textbf{k}_{t}} \right\rangle \right] \\ 
  & =  \partial_{t} \left[ e^{-i\mathcal{A}\cdot\hat{\textbf{r}}} e^{i\textbf{k}_{t}\cdot\textbf{r}} \left| u_{n,\textbf{k}_{t}} \right\rangle \right] \\ 
  & = \partial_{t} \left[ e^{i\textbf{k}\cdot\textbf{r}} \left| u_{n,\textbf{k}_{t}} \right\rangle \right] \\ 
  & =  e^{i\textbf{k}\cdot\textbf{r}}\left[ \nabla_{\textbf{k}_{t}}\left| u_{n,\textbf{k}_{t}} \right\rangle \right]  \partial_{t} \textbf{k}_{t} \\ 
  & = - \textbf{E}(t) e^{i\textbf{k}\cdot\textbf{r}} \nabla_{\textbf{k}_{t}}\left| u_{n,\textbf{k}_{t}} \right\rangle. 
\end{aligned}
\end{equation}
Here, $\textbf{E}(t)$ is the electric field of the laser pulse. 
After substituting the result from the above equation and taking an inner product with $\left\langle \widetilde{\phi}_{m,\textbf{k}}\right|$,  Eq.~(\ref{SBE:5})  reduces as 
\begin{multline}\label{SBE:8}
     i \sum_{n,\textbf{k}}\left[\dot{c}_{n,\textbf{k}}(t) \underbrace{\left\langle \widetilde{\phi}_{m,\textbf{k}}(t) |\widetilde{\phi}_{n,\textbf{k}}(t) \right\rangle}_{\delta_{nm}} -  c_{n,\textbf{k}}(t) \textbf{E}(t) \cdot \left\langle\widetilde{\phi}_{m,\textbf{k}}\right|\nabla_{\textbf{k}_{t}}  \left| u_{n,\textbf{k}_{t}} \right\rangle \right] \\ = \sum_{n,\textbf{k}} c_{n,\textbf{k}} \mathcal{E}_n^{\textbf{k}_{t}} \underbrace{\left\langle \widetilde{\phi}_{m,\textbf{k}}(t) | \widetilde{\phi}_{n,\textbf{k}}(t) \right\rangle}_{\delta_{nm}}.   
\end{multline}
The above equation can be expressed in compact form as
\begin{equation}\label{SBE:10}
i \dot{c}_{m,\textbf{k}}(t)  = \mathcal{E}_n^{\textbf{k}_{t}}c_{m,\textbf{k}} -\textbf{E}(t)\cdot\sum_{n}
\textbf{d}_{mn}^{\textbf{k}_{t}} c_{n,\textbf{k}}, 
\end{equation}
where $\textbf{d}_{mn}^{\textbf{k}_{t}} = i\left\langle u_{m,\textbf{k}}\right|\nabla_{\textbf{k}} \left| u_{n,\textbf{k}} \right\rangle$ is dipole matrix element.
The electronic population of a specific energy state can be obtained by the squared modulus of the
time-dependent coefficients, $c_{i,\textbf{k}}(t)$. 
Density-matrix formalism is required to estimate the coherence between different energy states 
as well as to account for various decoherence effects. 
Thus, let us  define the density matrix elements as $ \rho_{mn}^{\textbf{k}} = c_{m,\textbf{k}} c{^*}_{n,\textbf{k}}$, and its time derivative can be  expressed as
\begin{equation}\label{SBE:11}
    \dot{\rho}_{mn}^{\textbf{k}} =  \dot{c}_{m,\textbf{k}}(t)c{^*}_{n,\textbf{k}}(t) + c_{m,\textbf{k}}(t)\dot{c}{^*}_{n,\textbf{k}}(t).
\end{equation}
By using the expression for the time-dependent coefficient  from Eq.~(\ref{SBE:10}), the equation of motion for the density matrix can be written as 
\begin{multline}\label{SBE:12}
    \dot{\rho}_{mn}^{\textbf{k}} = -i\left( \mathcal{E}_m^{\textbf{k}_{t}}c_{m,\textbf{k}} -\textbf{E}(t)\cdot\sum_{p}\textbf{d}_{mp}^{\textbf{k}_{t}} c_{p,\textbf{k}} \right)c{^*}_{n,\textbf{k}}(t) +  \\ i c_{m,\textbf{k}}(t) \left( \mathcal{E}_n^{\textbf{k}_{t}}c{^*}_{n,\textbf{k}} + \textbf{E}(t)\cdot\sum_{p}\textbf{d}_{pn}^{\textbf{k}_{t}} c{^*}_{p,\textbf{k}} \right).
\end{multline}
The above equation can be further written as
\begin{equation}\label{SBE:13}
	 \dot{\rho}_{mn}^{\textbf{k}} = -i\mathcal{E}_{mn}^{\textbf{k}_{t}}\rho^{\textbf{k}}_{mn}+i\textbf{E}(t)\cdot \left[\sum_q\left(\textbf{d}_{mq}^{\textbf{k}_{t}}\rho_{qn}^{\textbf{k}}-\textbf{d}_{qn}^{\textbf{k}_{t}}\rho_{mq}^{\textbf{k}}\right)\right],
\end{equation}
where  $\mathcal{E}^{\textbf{k}_{t}}_{mn}$ is the band gap energy between $m$ and $n$ energy bands at  $\textbf{k}_{t}$. 

Let us introduce a phenomenological term to account  for the interband decoherence between electrons and holes with a constant dephasing time $\textrm{T}_2$ as
\begin{equation}\label{SBE:14}
\dot{\rho}_{mn}^{\textbf{k}}  = - \left[ i \mathcal{E}_{mn}^{\textbf{k}_{t}}+\frac{(1-\delta_{mn})}{\textrm{T}_2} \right]\rho_{mn}^{\textbf{k}} + 
i \textbf{E}(t)\cdot\left[\sum_q \left(\textbf{d}_{mq}^{\textbf{k}_{t}}\rho_{qn}^{\textbf{k}} - \textbf{d}_{qn}^{\textbf{k}_{t}}\rho_{mq}^{\textbf{k}}  \right) \right].
\end{equation}
The above equation is dubbed  as the Semiconductor-Bloch equation (SBE), which is used to describe the laser-driven coherent dynamics of electrons in solids~\citep{haug2009quantum,houston1940acceleration,krieger1986time}. 
The range of the dephasing time $\textrm{T}_2$ varies from a few femtoseconds to tens of femtoseconds 
~\citep{taucer2017nonperturbative,liu2018driving,heide2022probing}.
There is another term to account for the relaxation of the electronic population from a conduction  band, $\textrm{T}_1$, which is typically on the order of a few hundred femtoseconds, i.e., $\textrm{T}_{1}   \gg  \textrm{T}_{2}$~\citep{hwang2008single}. 
As we are going to explore the electron dynamics induced by a laser pulse with a duration shorter than $\textrm{T}_1$, which allows us to disregard $\textrm{T}_1$ term in our study.

The total current at any $\mathbf{k}$-point in the Brillouin zone is defined as
\begin{equation}\label{SBE:15}
\begin{split}
\textbf{J}^{\textbf{k}}(t) &= \sum_{m,n \in \{c,v\}} \rho_{mn}^{\textbf{k}}(t)  \textbf{p}_{nm}(\textbf{k}_t,t), \\
&= \sum_{m\neq n} \rho_{mn}^{\textbf{k}}(t)~\textbf{p}_{nm}(\textbf{k}_t,t) + \sum_{m=n} \rho_{m,n}^{\textbf{k}}(t)~\textbf{p}_{nm}(\textbf{k}_t,t),\\
&= \textbf{J}_{inter}^{\textbf{k}}(t) + \textbf{J}_{intra}^{\textbf{k}}(t). 
\end{split}
\end{equation}
Here, \textbf{p}$_{nm}^{\textbf{k}} = \left\langle n, \textbf{k} \left| \nabla_{\textbf{k}}  \mathcal{H}_{\textbf{k}} \right| m, \textbf{k} \right\rangle$ is  the momentum matrix element.  
$\textbf{J}_{inter}^{\textbf{k}}(t)$ and $\textbf{J}_{intra}^{\textbf{k}}(t)$ are the  interband and intraband contributions to the total current, respectively. 
The total current can be calculated by integrating \textbf{J}(\textbf{k}, $t$) over the entire Brillouin zone.
The Fourier-transform of the time-derivative of the current is employed to obtain the high-harmonic spectrum as
\begin{equation}\label{SBE:16}
\mathcal{I}(\omega) = \left|\mathcal{FT}\left(\frac{d}{dt} \textbf{J}(t) \right) \right|^2 .
\end{equation}
 Here, $\mathcal{FT}$ stands for the Fourier transform.
 
 \begin{figure}
\includegraphics[width=\linewidth]{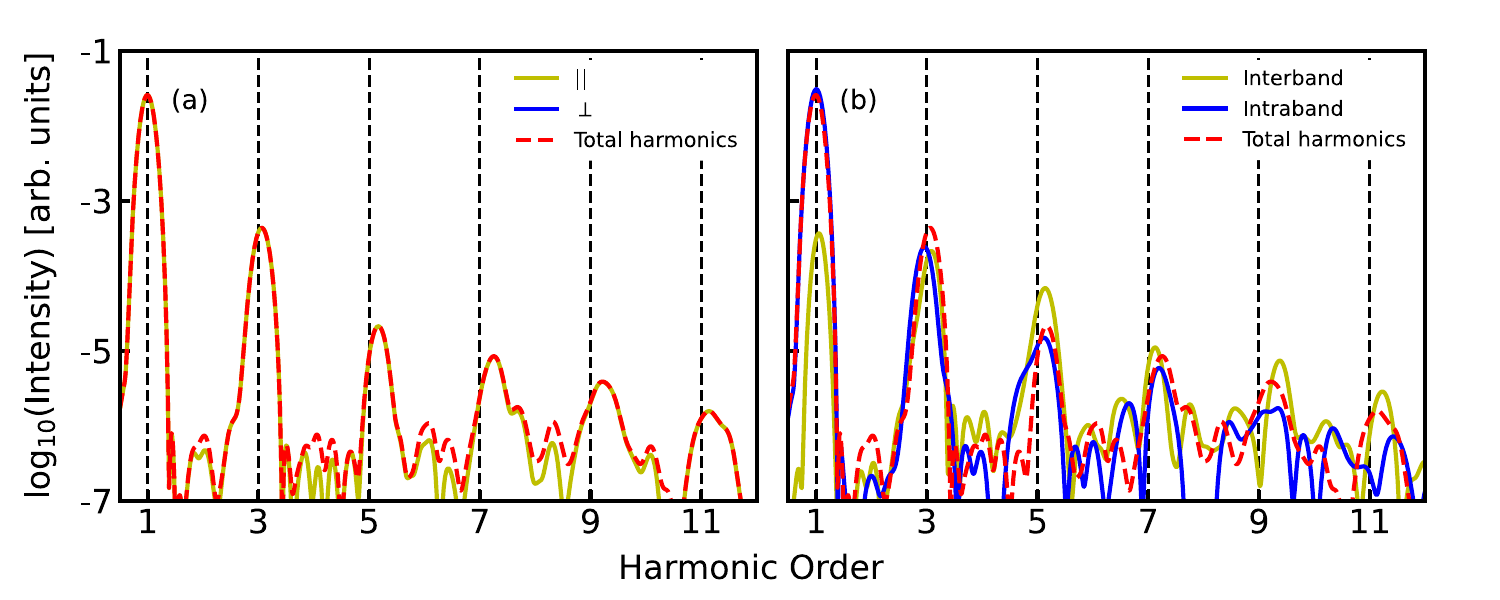}
\caption{High-harmonic spectra of monolayer graphene.
(a) Red (blue) color corresponds to the polarization of emitted radiation parallel (perpendicular) to the polarization of the harmonic generating probe pulse. (b) Intra- and interband resolved high-harmonics.}
\label{fig2.3}
\end{figure}

Let us briefly discuss the advantages of using the Houston basis state to describe laser-driven electron dynamics 
in solids. 
\begin{itemize}
   \item Intra- and inter-band contributions can be naturally separated during the electron dynamics. 
    \item Fewer energy bands are needed to obtain the converged results and to describe the electron dynamics.
\end{itemize}

In the following, we will validate the SBE formalism by simulating the high-harmonic spectrum for 
monolayer graphene and its comparison with available experimental results. 
Figure~\ref{fig2.3} shows the high-order harmonic spectrum of graphene irradiated with a linearly polarized pulse along $\mathsf{X}$-axis with 3.2 $\mu$m wavelength. 
The laser pulse is eight-cycle long with a sin-squared envelope and has 0.1 TWcm$^{-2}$ 
intensity, which is  below the damage threshold of graphene~\citep{currie2011quantifying}. 
Owing to the inversion symmetry of graphene, odd-order harmonics are generated~\citep{boyd2020nonlinear}.
Moreover, graphene exhibits reflection symmetry along an axis parallel  to the laser pulse, resulting in harmonics that are entirely parallel to the laser pulse as evident from Fig.~\ref{fig2.3}(a).
Our results shown in Fig.~\ref{fig2.3} are consistent with the earlier reports~\citep{al2014high,yoshikawa2017high,murakami2018high,mrudul2021high}. 
Furthermore, our findings  are in accordance with  the observed ellipticity-dependence of the harmonics in  experiments~\citep{yoshikawa2017high,taucer2017nonperturbative}. 
Thus, the employed method based on SBE  formalism is robust and reliable in capturing  
important aspects of the electron dynamics in solids. 

Intra- and intra-band resolved harmonics are presented in Fig.~\ref{fig2.3}(b). 
A strong coupling between interband and intraband driven dynamics in graphene 
has been reported~\citep{al2014high,taucer2017nonperturbative,liu2018driving}. 
Contrary to semiconductors with broad bandgaps~\citep{wu2015high}, the interband and intraband transitions occur at the same energy regimes for graphene~\citep{stroucken2011optical}.
The intraband contribution dominates the lower-order harmonics, whereas the interband contribution dominates the higher-order harmonics, which can be attributed to a greater joint density of states~\citep{lanin2017mapping,vampa2014theoretical,wu2015high}.

\section{Lattice Dynamics}\label{section:2.3}
So far, we have discussed laser-driven electron dynamics in solids with their equilibrium lattice structure.
Solids with fixed atomic positions are not a practical situation. There are a number of reasons, which cause atoms to vibrate from their equilibrium positions. 
Thus, it is essential to know how the electron dynamics are affected by the atomic vibrations in a solid. 

Lattice vibrations, also known as phonons, are responsible  for numerous physical properties of solids, including thermal conductivity, specific heat,  temperature-dependence of optical spectra and electron-phonon 
related phenomena, such as the resistivity of metals, and superconductivity to name a few~\citep{baroni2001phonons}.
Therefore, comprehending  lattice dynamics in solids has numerous applications in various fields, including materials science, solid-state physics, and nanotechnology as it 
plays a key role in designing materials with specific thermal and mechanical properties; 
and in the development of semiconductor devices, such as transistors and solar cells.
Thus, the knowledge of phonons in solids is regarded as one of the most persuasive pieces of evidence that our present quantum description of solids is accurate.

The theory of lattice dynamics is one of the most well-established branches of contemporary  condensed-matter physics.   
There are very few physical properties of solids that could have been accomplished without a firm basis of 
the theory of lattice dynamics. 
The fundamental theory of lattice dynamics goes back to the 1930s, and Born and Huang's
 treatise is regarded as the authoritative source in this topic~\citep{born1954dynamical}. 
These early formulations focused primarily on defining the generic qualities of the dynamical matrices, such as their symmetry and/or analytical properties, without ever addressing their links with the electronic properties that really determine them. 
In the following, we will develop  the Born model  to simulate coherent lattice dynamics in graphene. 

\subsection{Phonons in Monolayer Graphene}\label{section:2.3.1}
Let us start our discussion about lattice dynamics by writing  the Hamiltonian for a vibrating solid as
 $\mathcal{H}_{\textrm{L}} = \mathcal{T}_{\textrm{L}} + \mathcal{V}_{\textrm{L}}$ with 
 $\mathcal{T}_{\textrm{L}}~(\mathcal{V}_{\textrm{L}})$ as the kinetic (potential) energy of the lattice in a solid. 
The kinetic energy is defined as $\mathcal{T}_{\textrm{L}} = \sum_{npi} M_{n} \dot{u}^{2}_{i}(n,p)/2$ 
with $M_{n}$ as the mass of the $n^{\textrm{th}}$ atom, $u_{i}(n,p)$  as the small displacement of the 
$n^{\textrm{th}}$ atom in the $p^{\textrm{th}}$ cell along $i^{\textrm{th}}$ direction from its equilibrium position. 
We will assume that the atoms are displaced slightly away from their equilibrium positions to define the 
potential energy. The  potential energy 
within this assumption can be written using Taylor series expansion as
\begin{equation}\label{BM:2}
\mathcal{V}_{\textrm{L}} = \mathcal{V}_{0} +   \sum_{npi} \left( \frac{\partial \mathcal{V}_{L}}{\partial u_{i} (n,p)}   \right)_{0} u_{i} (n,p)  + \dfrac{1}{2}\sum_{npi} \sum_{mqj} \left( \frac{\partial^2 \mathcal{V}_{L}}{\partial u_{i} (n,p)  \partial u_{j} (m,q)}   \right)_{0} u_{i} (n,p) u_{j} (m,q) +  \ldots 
\end{equation}
Here, $\mathcal{V}_{0}$ is the potential energy corresponding to the equilibrium lattice position  and
``0'' stands for the equilibrium lattice configuration.  
The second term on the right-hand side will be zero for the stable lattice
configurations. 
The term $\left( \frac{\partial^2 \mathcal{V}_{L}}{\partial u_{i} (n,p)  \partial u_{j} (m,q)}   \right)_{0} $ stands for 
the force acting on the $n^{\textrm{th}}$ atom in the $p^{\textrm{th}}$ cell along $i^{\textrm{th}}$ direction due to $m^{\textrm{th}}$ atom in the $q^{\textrm{th}}$ cell in the $j^{\textrm{th}}$ direction. 
The adiabatic approximation is used in writing such expressions, 
which states that the electrons immediately assume a configuration corresponding to the displaced atoms, whereas the change in the electronic  energy in the deformed solid contributes to an effective internuclear potential $\mathcal{V}_{\textrm{L}}$ ~\citep{dobbs1957theory}. 
In the following, we will employ harmonic approximation for vibrating atoms 
and higher-order terms beyond second-order in the expansion will be ignored.

The equation of motion of the $n^{\textrm{th}}$ atom in the $p^{\textrm{th}}$ cell can be written as \begin{equation}\label{BM:5}
   M_{n} \ddot{u}^{2}_{i}(n,p) = \sum_{mqj} f_{ij}\left(n,p;m,q \right) u_{j} (m,q),
\end{equation}
where $f_{ij}\left(n,p;m,q \right)  =   \left( \frac{\partial^2 \mathcal{V}_{L}}{\partial u_{i} (n,p)  \partial u_{j} (m,q)}   \right)_{0}$. 
A linear superposition of the traveling plane waves of wave vector $\textbf{k}$ and frequency $\omega$
can be considered as the solution of the above equation as 
\begin{equation}\label{BM:6}
 u_{i} (n,p) = \sqrt{M_{n}} \ u_{ni} \exp\left[-i \{ \omega t - \textbf{k} \cdot \textbf{R}_{n}(p) \} \right].
\end{equation}
Here,  prefactor $\sqrt{M_{n}}$ is used to simplify the mathematics  and
$\textbf{R}_{n}$ is the lattice vector corresponding to the $n^{\textrm{th}}$ atom. 
Using the above ansatz of the displacement, the equation of motion of the lattice dynamics can be simplified  as  
\begin{equation}\label{BM:7}
  \omega^{2} u_{ni} = \sum_{mqj} f_{ij}\left(n,p;m,q \right) u_{mj} \exp\left[-i\textbf{k} \cdot \{ \textbf{R}_{n}(p) - \textbf{R}_{m}(q) \}\right].
\end{equation}
The force acting on an atom in a solid is given as
$F_{i}  =  \textrm{D}_{ij}(\textbf{k}) u_{j}$ with $\textrm{D}_{ij}$ is the dynamical matrix given as
\begin{equation}  \label{BM:9}
\text{D}_{ij}(\textbf{k}) =  \sum_{\mathbf{R}_{l}} f_{ij}(\textbf{R}_l) e^{-i \mathbf{k.R}_{l}},
\end{equation}
where  $\textbf{R}_l$ is the relative lattice vector and $\textbf{u}  = ( u_{x}^{1}, u_{y}^{2}, u_{x}^{1},  u_{y}^{2})^{\textrm{T}}$. The equation of motion in compact matrix form can be recast  in the form of  an eigenvalue problem as
$\omega^{2} \textbf{u} = \textbf{D}(\textbf{k}) \textbf{u}$. 
The eigenvalues of this equation provide phonon dispersion  
and  the eigenvectors yield information about the active phonon mode.

Let us simulate the phonon dispersion curve for the monolayer graphene. 
For this purpose, we need to define the dynamical matrix for  graphene consisting of  A and B types of carbon atoms. 
The dynamical matrix within nearest-neighbor interaction is defined as
\begin{equation}\label{BM:11}
\textbf{D}(\textbf{k}) = \begin{bmatrix}
    D_{xx}^{\textrm{AA}}  & D_{xy}^{\textrm{AA}}  & D_{xx}^{\textrm{AB}}    & D_{xy}^{\textrm{AB}}   \\
   D_{yx}^{\textrm{AA}}  & D_{yy}^{\textrm{AA}}   &   D_{yx}^{\textrm{AB}}    &  D_{yy}^{\textrm{AB}} \\
    D_{xx}^{\textrm{BA}}  & D_{xy}^{\textrm{BA}}   & D_{xx}^{\textrm{BB}}  & D_{xy}^{\textrm{BB}} \\
    D_{yx}^{\textrm{BA}}  & D_{yy}^{\textrm{BA}} & D_{yx}^{\textrm{BB}}   & D_{yy}^{\textrm{BB}} \\
\end{bmatrix}
= \begin{bmatrix}
    A_0  & B_0   & C     & D   \\
    B_0  & A_1   & D     & B1  \\
    C^*  & D^*   & A_0   & B_0 \\
    D^*  & B_1^* & B_0   & A_1 \\
\end{bmatrix}.  
\end{equation}
Here,  AA and AB represent the interaction between the same and different types of atoms within 
the primitive unit cell,  respectively.  The elements of $\textbf{D}(\textbf{k})$ are calculated using Eq.~(\ref{BM:9})  and can be expressed  as

$A_0$ = $\dfrac{3}{2} \alpha_s$ + 3 $\alpha_{\phi} \left[ 1 - \cos\left( \dfrac{\sqrt{3}}{2} \text{k}_x a \right) \cos\left( \dfrac{1}{2} \text{k}_y a \right) \right]$,

$A_1$ = $\dfrac{3}{2} \alpha_s$ +  $\alpha_{\phi} \left[ 3 -  \cos\left( \dfrac{\sqrt{3}}{2} \text{k}_x a \right) - 2 \cos\left( \text{k}_y a \right) \right] \cos\left( \dfrac{1}{2}\text{k}_y a \right) $,

$B_0$ =    $ -\ \sqrt{3} \alpha_{\phi}   \ \sin\left( \dfrac{\sqrt{3}}{2} \text{k}_x a \right) \sin\left( \dfrac{1}{2} \text{k}_y a \right) $,

$B_1$ =  $ -\ \dfrac{3}{2} \  \alpha_s \  \exp\left( \dfrac{i \text{k}_x a}{2 \sqrt{3}} \right)  \cos\left( \dfrac{1}{2} \text{k}_y a \right) $,

$C $ $ $   =  $ -\ \alpha_s \left[ \exp\left( \dfrac{ - i \text{k}_x a } { \sqrt{3}}  \right) + \dfrac{1}{2} \exp\left( \dfrac{  i \text{k}_x a } { 2 \sqrt{3}}  \right)  \cos\left( \dfrac{1}{2} \text{k}_y a \right)  \right] $,~~\textrm{and}

$D $ $ $   =  $  -\ i \dfrac{\sqrt{3}}{2}   \ \alpha_s  \ \exp\left( \dfrac{  i \text{k}_x a } { 2 \sqrt{3}}  \right)  \sin\left( \dfrac{1}{2} \text{k}_y a \right)  $.

\begin{figure}
\includegraphics[width=\linewidth]{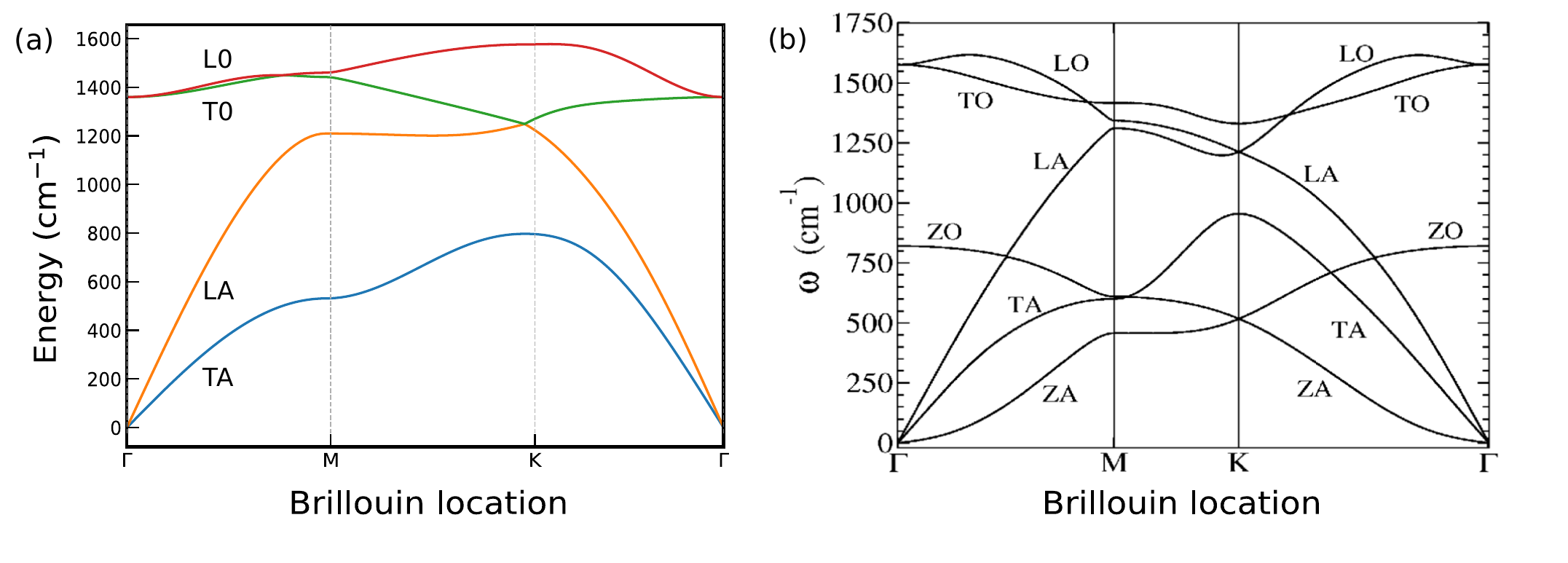}
\caption{(a) Phonon  dispersion along high-symmetry points of graphene calculated from the dynamical matrix using fitted constants: $\alpha_s$ = 445 N/m and $\alpha_{\phi}$ = 102 N/m.    
(b) Phonon  dispersion of graphene obtained using \textit{ab initio} simulation~\citep{marquina2013ab}.}
\label{fig2.4}
\end{figure}

In the following, we will focus on in-plane phonon modes  shown in  Fig.~\ref{fig2.4}(a) as 
graphene is a two-dimensional material.
The in-plane phonon dispersion is calculated using the dynamical matrix formalism for  fitted constants 
$\alpha_s$ = 445 N/m and $\alpha_{\phi}$ = 102 N/m. 
For comparison purposes, we have also presented the phonon dispersion obtained
from \textit{ab initio} simulation as shown in  Fig.~\ref{fig2.4}(b)~\citep{marquina2013ab}. 
The overall nature of the in-plane phonon dispersion  is in well agreement with the \textit{ab initio} simulations at the $\mathsf{\Gamma}$ point as evident from  Fig.~\ref{fig2.4}.

The eigenvectors obtained from Eq.~(\ref{BM:11}) show 
the active phonon modes at the particular \textbf{k} point.  
The in-plane acoustic and optical phonon modes at $\mathsf{\Gamma}$ point are degenerate as shown in 
Fig.~\ref{fig2.5}.  
 The acoustic phonon modes do not lead to any change in the structure of graphene. 
Also, the acoustic phonons have zero frequency at $\mathsf{\Gamma}$ 
and thus can not be excited by any optical means at  $\mathsf{\Gamma}$. 
Therefore, we are going to focus on the two degenerate in-plane optical phonon modes. 
In the following, we will treat phonon dynamics classically and assume that the atoms perform harmonic oscillations for short displacements from their equilibrium positions. The displacement vector for a particular phonon mode can be expressed as 
\begin{equation}\label{eq:ampl}
\textbf{q}(t)=   \textrm{q}_{0}~\Re \left\{ \exp({i \omega _{\textrm{ph}}t}) \right\}~\hat{\textbf{e}}.
\end{equation}
Here, $\textrm{q}_{0}$  is the maximum displacement of an atom from its equilibrium position, $\omega_{\textrm{ph}}$  is the energy of the  
phonon mode and $\hat{\textbf{e}}$ is the normalized eigenvector for a particular phonon mode. 
The eigenvectors corresponding to the in-plane longitudinal optical ($\textsf{iLO}$) and in-plane transverse optical ($\textsf{iTO}$) phonon modes can be extracted from 
Figs.~\ref{fig2.5}(c) and \ref{fig2.5}(d) as $\hat{\textbf{e}}_{\textsf{iLO}}$ = $[1,0,-1,0]/\sqrt{2}$ and $\hat{\textbf{e}}_{\textsf{iTO}}$ = $[0,1,0,-1]/\sqrt{2}$, respectively;  
in which the first (last) two elements are components of A (B) atom. 

\begin{figure}
\begin{center}
  \includegraphics[width=0.6\linewidth]{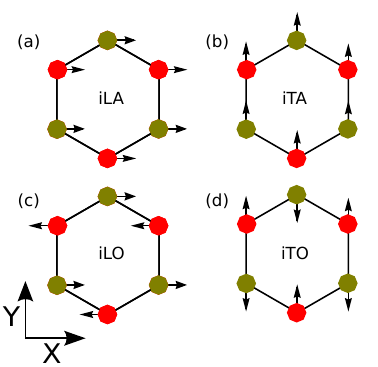}
\end{center}
\caption{Sketches of atomic vibrations associated with the degenerate E$_{2g}$ phonon  modes in real-space for (a),(b) acoustic and (c),(d)  optical Phonon modes at the $\mathsf{\Gamma}$ point. 
Here, modes are labeled as (a) in-plane longitudinal acoustic ($\textsf{iLA}$), (b) in-plane transverse acoustic ($\textsf{iTA}$), (c) in-plane longitudinal optical ($\textsf{iLO}$),  and (d) in-plane transverse optical ($\textsf{iTO}$) phonon mode, respectively.}
\label{fig2.5}
\end{figure}

Alternatively, the valence-force-field model can be used to describe 
the optical phonon at $\mathsf{\Gamma}$ point~\citep{ando2006anomaly,suzuura2002phonons}.
Let $\textbf{u}(\textbf{R})$ = $(u_{x},u_{y})$ be the relative displacement of two sub-lattice atoms A and B as
\begin{equation}\label{BM:12}
    \textbf{u}(\textbf{R}) = \dfrac{1}{\sqrt{2}}\left[\textbf{u}_{\textrm{A}}(\textbf{R}) - \textbf{u}_{\textrm{B}}(\textbf{R}) \right].
\end{equation}
We will limit our discussion to the long-wavelength limit under which $\textbf{R}$ goes to $\bm{r}$.
Hence, we have   
\begin{equation}\label{BM:13}
    \textbf{u}(\bm{r}) = \sum_{\textbf{k},\mu} \sqrt{\dfrac{1}{2NM\omega_{\textrm{ph}}}} \left( p_{\textbf{k}\mu}+ p^{\dagger}_{-\textbf{k}\mu}  \right) \bm{e}_{\mu}(\textbf{k}) e^{i\textbf{k}\cdot\bm{r}}, 
\end{equation}
where $N$ is the number of unit cells, $M$ is the mass of the carbon atom, $\mu$ denotes the modes 
and $p_{\textbf{k}\mu}$ ($p^{\dagger}_{-\textbf{k}\mu}$) is the annihilation (creation) operator for the phonon mode with  momentum $\textbf{k}$ and  mode $\mu$.
The eigenvector $ \bm{e}_{\mu}$ can be defined as 
$$ e_{\textsf{L}}(\textbf{k}) = i[\cos \{\phi(\textbf{k})\}, \sin \{\phi(\textbf{k})\}],~~\textrm{and}~~
e_{\textsf{T}}(\textbf{k}) = i[-\sin \{\phi(\textbf{k})\}, \cos \{\phi(\textbf{k})\}].$$
Here, ``$\textsf{T}~(\textsf{L})$'' stands for in-plane transverse (longitudinal) optical phonon mode. 
The corresponding phonon Hamiltonian within harmonic approximation 
in  second quantization can be written as 
\begin{equation}\label{BM:14}
 \mathcal{H}_{\textrm{L}} =  \sum_{\textbf{k},\mu} \omega_{\textrm{ph}} \left( p^{\dagger}_{\textbf{k}\mu} p_{\textbf{k}\mu} + \dfrac{1}{2}  \right).
\end{equation}

\subsection{Time-Dependent Tight-Binding Approach}\label{section:2.3.2}
After the coherent excitation of  the in-plane optical phonon mode, the lattice dynamics cause temporal variations in the relative distance between carbon atoms ($\textbf{d}_i$). 
As a result of this change, the hopping term in Eq.~(\ref{tb:4}) becomes time-dependent.  
To account for the time variation in the hopping term, we model the hopping term as an exponentially decaying function of the relative displacement between nearest-neighbor atoms as $\gamma(t)$ = $\gamma_0~e^{-(|\textbf{d}_{i}(t)|-a)/\delta}$, in which $\delta$ is the  width of the decay function chosen to be 0.184$a_0$~\citep{moon2013optical}. 
In this case, the corresponding time-dependent Hamiltonian within the tight-binding approximation can be written as~\citep{mohanty2019lazy, rodriguez2021direct}
\begin{equation}\label{tdtbh:1}
\mathcal{H}_{t} =  - \sum_{i\in nn} \gamma(t) e^{i\textbf{k}\cdot \textbf{d}_i(t)} \hat{a}_{\textbf{k}}^{\dagger} \hat{b}_{\textbf{k}} + \textrm{H.c.}
\end{equation}
At each time-step during the HHG probe  pulse and lattice vibration, time-dependent tight-binding Hamiltonian 
$\mathcal{H}_{t} $ is diagonalized and updated eigenstates and eigenenergies are obtained.
The coherent phonon dynamics lead to an additional time dependence on the matrix elements.
The dipole matrix elements are calculated at each time step using updated eigenstates.
The interaction among laser, electrons, and coherently excited phonon mode in graphene is modeled by solving the following equations within the density-matrix framework.  
By updating the modified Hamiltonian as a result of the coherent lattice dynamics,   SBEs in the co-moving frame 
$|n,\textbf{k}_t \rangle$, is extended from Eq.~(\ref{SBE:14}), and  equations of motion read as
\begin{subequations}\label{tdtbh:2}
\begin{align}
     \dot{\rho}_{vv}^{\textbf{k}} &= i\textbf{E}(t)\cdot \textbf{d}_{vc}^{\textbf{k}_{t}}\rho_{cv}^{\textbf{k}} + \textrm{c.c.}\\
     \dot{\rho}_{cv}^{\textbf{k}} &= \left[-i \mathcal{E}_{cv}^{\textbf{k}(t)}\ + \frac{1}{\textrm{T}_2}\right]\rho_{cv}^{\textbf{k}} + i\textbf{E}(t)\cdot\textbf{d}_{cv}^{\textbf{k}_{t}} \left[\rho_{vv}^{\textbf{k}}-\rho_{cc}^{\textbf{k}}\right]. \end{align}
\label{SBE}
\end{subequations} 
Note that $\rho_{cc}^{\textbf{k}}(t) = 1 - \rho_{vv}^{\textbf{k}}(t)$, and $\rho^{\textbf{k}}_{vc}(t) = \rho^{\textbf{k}*}_{cv}(t)$.

The dipole-matrix elements and the eigenenergies are smoothly updated in the SBEs at each consecutive time step as long as the maximum displacement of the atoms is small, and the time step is too small compared to the phonon oscillation period. 
The coupled differential equations described in Eq.~(\ref{SBE}) are solved 
using the fourth-order Runge-Kutta method with a time-step of 0.01 fs. 
We sampled the Brillouin zone with 251$\times$251 grid.  
The current at any \textbf{k} point in the Brillouin zone is calculated as defined in Eq.(~\ref{SBE:15}).
The high-harmonic spectrum is simulated from the current as defined in Eq.(~\ref{SBE:16}). 

The time-frequency map of the harmonic spectra is also explored to understand the interplay of coherent lattice dynamics, subcycle electron dynamics, and transiently evolving lattice structure. 
The time-frequency map is obtained by performing  Gabor transformation as 
\begin{equation}\label{tdtbh:3}
    \textrm{G}(\omega,t) = \int dt^{\prime}~\dfrac{\exp\left[ - (t-t^{\prime})^{2}/ 2\zeta^{2} \right]}{\sqrt{2\pi}\zeta} ~\textbf{J}(t^{\prime})\exp(i\omega t^{\prime}).
\end{equation}
Here, we have considered $\zeta= 1/(3\omega)$ to get an adequate balance between the time and energy resolutions.
\cleardoublepage
\chapter{High-Harmonic Spectroscopy of Coherent Lattice Dynamics}\label{chap:Chapter3}

Strong-field driven HHG  is a  nonlinear frequency up-conversion process, which emits radiation at integer multiples of the incident laser frequency~\citep{ferray1988multiple}.  
Taking advantage of major technical advances in mid-infrared sources, the pioneering experiments  have extended HHG from gases to solids~\citep{ghimire2011observation}, stimulating intense research into probing electron dynamics in solids on the natural timescale. 
The availability of mid-infrared light sources also enables coherent excitation of a desired  
phonon mode by tuning the polarization and frequency of the laser pulse~\citep{mankowsky2016non}.  
Yet, the analysis of the effect of coherent lattice dynamics  
on high-harmonic generation in solids appears 
lacking. 
This situation stands in stark contrast to molecular gases, 
where  high-harmonic spectroscopy  has been 
extensively employed to probe nuclear motion 
in various molecules~\citep{patchkovskii2009nuclear, wagner2006monitoring, le2012theory, baker2006probing, lein2005attosecond, worner2011conical}. 

This chapter aims to fill this gap and highlight some of the capabilities offered by high-harmonic spectroscopy in time-resolving 
the interplay of femtosecond lattice and attosecond electronic motions. 
Various spectroscopic methods have been developed to excite and probe phonons, see e.g. ~\citep{dhar1994time, debnath2021coherent, graf2007spatially, virga2019coherent, koivistoinen2017time, rana2021four, brown2019direct, flannigan2018electrons, gierz2015phonon, moulet2017soft, geneaux2020attosecond}, but their temporal resolution is limited by the length of the pulses used. 
Large coherent bandwidth of high-harmonic signals offers sub-laser-cycle temporal resolution and 
the possibility to time-resolve the impact of lattice distortions on the faster electronic response.

One difficulty in tracking lattice vibrations via highly nonlinear optical response stems from their 
small amplitude. If the corresponding changes in both the band structure and  couplings are similarly small, the high-harmonic response hardly changes. Yet, large distortions are not needed if the excited phonon mode 
dynamically changes the symmetry of the unit cell.
This chapter shows how coherent lattice dynamics and the associated changes in the lattice symmetry are encoded  
in the electronic response and the harmonic signal, and how the sub-cycle temporal resolution inherent in the harmonic signal can be used to track the interplay of electronic and lattice dynamics. 

As discussed in Chapter~\ref{chp:Chapter2}, 
we analyze monolayer graphene with $\textbf{D}_{6h}$ point group symmetry as shown in Fig.~\ref{fig2.1}(a).  Monolayer graphene exhibits six phonon branches: three optical and three acoustic. Here we focus on the former.
Out of the three optical phonon modes, one  is out-of-plane,  and the two others are in-plane modes.     
We will consider only the in-plane modes. 
The lattice vibrations corresponding to the in-plane Longitudinal Optical ($\textsf{iLO}$), and the in-plane Transverse Optical ($\textsf{iTO}$) E$_{2g}$ modes are shown in Figs.~\ref{fig2.5}(c) and \ref{fig2.5}(d), respectively.
The two modes are  degenerate at the $\mathsf{\Gamma}$-point, with the phonon frequency equal to 194 meV~\citep{kim2013coherent}.
Both phonon modes are Raman active with 
oscillation period $\sim$ 21 femtoseconds  
and can be excited with a resonant pulse pair or
impulsively by a short pulse with bandwidth covering 
194 meV.
Moreover, it is possible to selectively excite either  $\textsf{iLO}$ or $\textsf{iTO}$ coherent phonon mode by tuning the polarization of the pump pulse either along  the $\mathsf{\Gamma-K}$ or $\mathsf{\Gamma-M}$ direction, respectively.

The coherent lattice dynamics is treated classically as discussed in section~\ref{section:2.3.1}. 
If the atoms are displaced from their equilibrium position beyond a certain distance, 
it can result in substantial alterations in the material's structure and electronic properties. 
Our objective is to prevent such modifications while considering the practical limitations of the experimental setup. Specifically, we have focused on the excitation of the phonon modes with amplitudes that are relatively small and lie within a range that is experimentally feasible~\citep{gierz2015phonon}. 
We have assumed that atoms exhibit harmonic oscillations for short displacements with respect to  
their equilibrium positions as discussed in  section~\ref{section:2.3.1}. 

Let us assume that the graphene lattice is in equilibrium without any coherent lattice dynamics
at time $t$ = 0. 
A specific phonon mode is coherently excited at  a subsequent time $t = t_{1}$, which induces time-dependent displacement of the atoms, causing the hopping term in the Hamiltonian to  become time-dependent. 
This results in modifying electronic band structure  
for various amplitudes of the in-plane  coherent phonon mode as shown in Fig.~\ref{fig:3.2}(b).
Our results on the modifications in the band structure are qualitatively the same as the one reported for the in-plane lattice vibrations in bilayer graphene as evident from Fig.~\ref{fig:3.2}(a) [see supplementary information in \cite{gierz2015phonon}]. 
Additionally, our findings are qualitatively the same for the different amplitude of lattice vibrations, which lead to an opening of the 
band gap  at $\textsf{K}$-point. 
A close agreement between previous work on bilayer graphene and present results on monolayer graphene
validates our approach for coherent lattice dynamics. 
A similar approach for the lattice dynamics within tight-binding Hamiltonian has been considered  where electron-phonon interaction Hamiltonian in second quantization is used~\citep{ando2006anomaly}.

\begin{figure}[h!]
\centering
 \includegraphics[width=\linewidth]{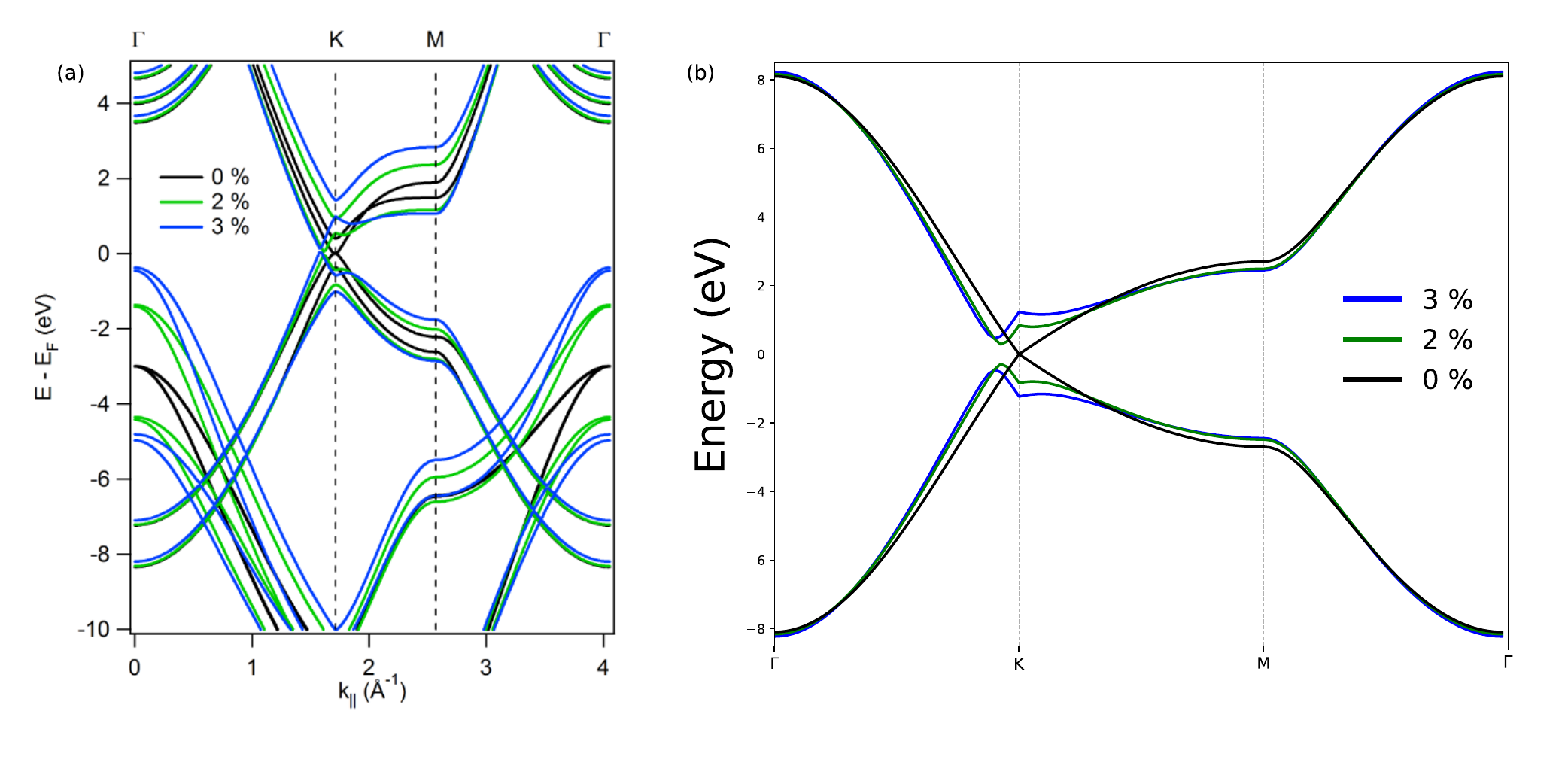}
 \caption{(a) Change in the electronic band structure of the 
 bilayer graphene for different amplitudes of the  in-plane coherent
 E$_{1u}$  phonon mode [adapted from supplementary information in \cite{gierz2015phonon}].  
 (b) Similar changes in the band structure of the monolayer graphene for different amplitudes of the  in-plane coherent E$_{2g}$ phonon mode. Results for the monolayer graphene  are obtained from our theoretical model  presented  in 
 Chapter~\ref{chp:Chapter2}.}
 \label{fig:3.2}
\end{figure}

Coherent lattice dynamics should in general introduce 
periodic modulations of the system parameters and thus of its high-harmonic response. In the frequency domain, such modulations 
add sidebands to the main peaks in the harmonic spectrum. 
We shall see that their position and polarization encode the information about the frequency and the symmetry of the excited phonon mode, respectively.
We adopt a convention to define zigzag and armchair directions of graphene in 
this  and the next chapters. The zigzag direction aligns with the $\mathsf{X}$-axis, which corresponds to the $\mathsf{\Gamma}-\mathsf{K}$ direction. On the other hand, the armchair direction aligns with the $\mathsf{Y}$-axis, corresponding to the $\mathsf{\Gamma}-\mathsf{M}$ direction [see Fig.~\ref{fig2.1}].

\section{Results}\label{section:2}

In this chapter, higher-order harmonics are generated from monolayer graphene, with or without coherent lattice dynamics, using a linearly polarized pulse with a  wavelength of 2.0 $\mu$m and peak intensity of $ 10^{11} $ W/cm$^2$. 
The pulse is 100 fs long and has a sin-squared envelope.  The laser parameters used in this chapter are below the damage threshold of graphene~\citep{currie2011quantifying}. 
Similar laser parameters have been used to investigate  electron dynamics in graphene via intense laser pulse
~\citep{heide2018coherent, higuchi2017light, yoshikawa2017high}. 
Both in-plane E$_{2g}$ phonon modes are considered here and a dephasing time $\mathrm{T}_2 = $  10 fs is used. 
Results presented in this chapter correspond to the above-mentioned parameters unless stated otherwise. 

High-harmonic spectra for monolayer graphene, with and without coherent lattice dynamics, are presented in Fig.~\ref{HHGlattice}. The spectrum corresponding to the graphene, without lattice dynamics, is shown by the gray shaded area as a reference.  
Owing to the inversion symmetry of the graphene, the reference spectrum in gray color exhibits only odd harmonics, 
which is consistent with earlier reports; e.g., Refs.~\citep{mrudul2021high,yoshikawa2017high,al2014high}.

We assume that coherent lattice dynamics are excited prior to a harmonic-generating  probe pulse.
When one of the E$_{2g}$ phonon modes in graphene is coherently excited, the harmonic spectra display sidebands along with the main odd harmonic peaks as reflected from Fig.~\ref{HHGlattice}. The energy difference between the adjacent sidebands matches the phonon energy ($\omega_{\textrm{ph}}$). 
The sideband intensity is sensitive to the phonon amplitude but is clearly visible already for amplitudes above 0.01 of the lattice constant.

\begin{figure}[h!]
\includegraphics[width=\linewidth]{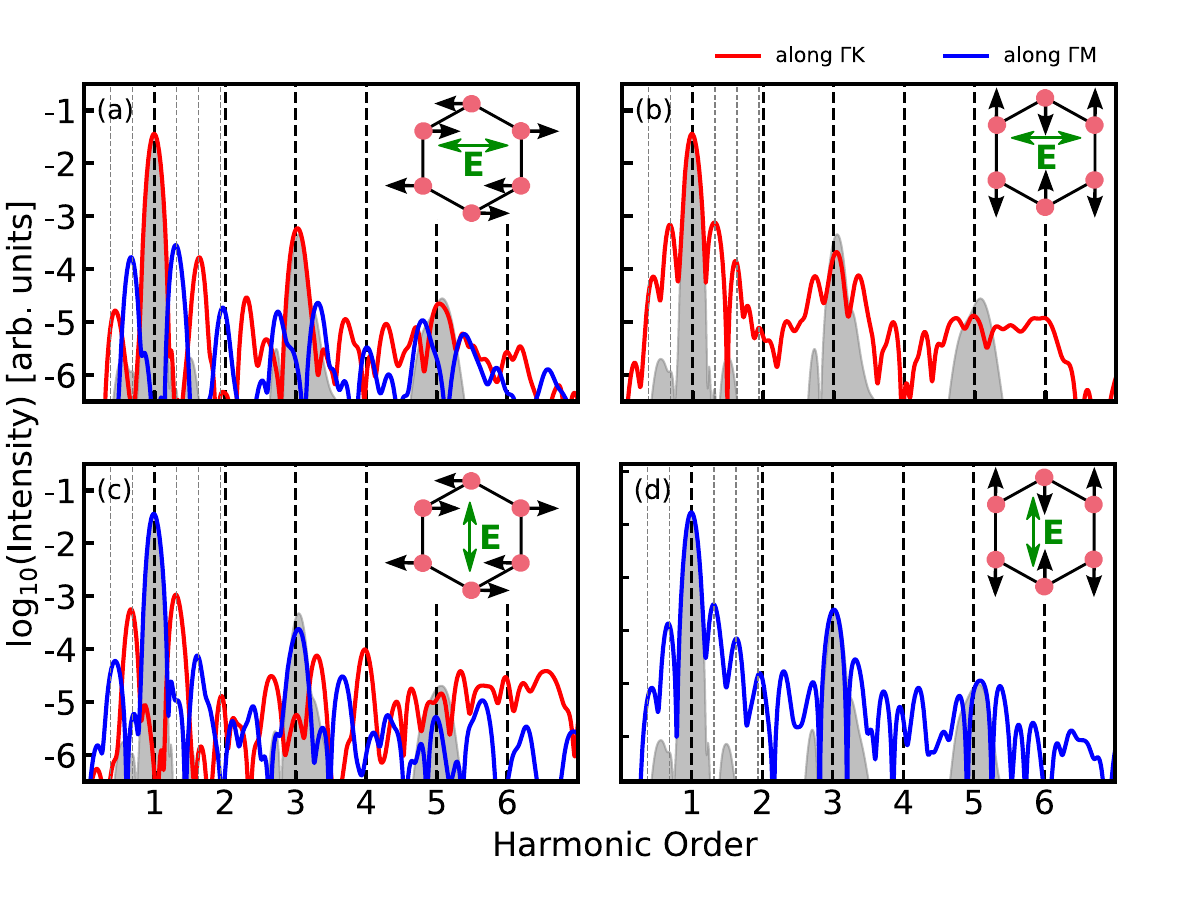}
\caption{High-harmonic spectra of monolayer graphene with and without coherent lattice dynamics. (a) and (c) High-harmonic spectra corresponding to the coherent $\textsf{iLO}$  phonon mode and the probe harmonic pulse is polarized along $\mathsf{\Gamma-K}$ and $\mathsf{\Gamma-M}$ directions, respectively. 
(b) and (d) Same as (a) and (c) except $\textsf{iTO}$  phonon mode is coherently excited. 
In all the cases, sidebands corresponding to the first harmonic are marked at frequencies ($\omega_{0} \pm  m \omega_{\textrm{ph}}$), where $\omega_0$  is the frequency of the harmonic generating probe pulse, and  $\omega_{\textrm{ph}}$ is the phonon frequency. 
The harmonics with the gray shaded area are the reference spectra and represent the spectra of graphene without phonon excitation. 
The unit cell of the graphene with the corresponding phonon eigenvector and polarization of the harmonic generating probe pulse are shown in the respective insets.  
The red (blue) color corresponds to the polarization of emitted radiation parallel (perpendicular) to the polarization of the harmonic generating probe pulse. The spectra correspond to a maximum 0.03a$_0$ displacement of atoms from their equilibrium positions during coherent lattice dynamics.}
\label{HHGlattice}
\end{figure}

As E$_{2g}$ phonon modes preserve the inversion center, only odd harmonics are generated. 
When the coherent $\textsf{iLO}$ phonon mode and the probe harmonic pulse (along  $\mathsf{\Gamma-K}$) are in the same direction, the even-order sidebands are polarized along $\mathsf{\Gamma-K}$ (red color), whereas the odd-order sidebands are polarized perpendicular to $\mathsf{\Gamma-K}$ (blue color), i.e., along $\mathsf{\Gamma-M}$ direction [see Fig.~\ref{HHGlattice}(a)]. 
When the polarization of the probe pulse changes from $\mathsf{\Gamma-K}$ to $\mathsf{\Gamma - M}$ direction, the polarization of the sidebands  remains the same with respect to the laser polarization. 
In this case, the even-order sidebands are polarized along $\mathsf{\Gamma-M}$  (blue color), whereas odd-order sidebands are polarized along $\mathsf{\Gamma-K}$ (red color) [see Fig.~\ref{HHGlattice}(c)].
In both cases, the main harmonic peaks are always polarized along the direction of the probe pulse. 

\begin{figure}[h!]
\includegraphics[width=\linewidth]{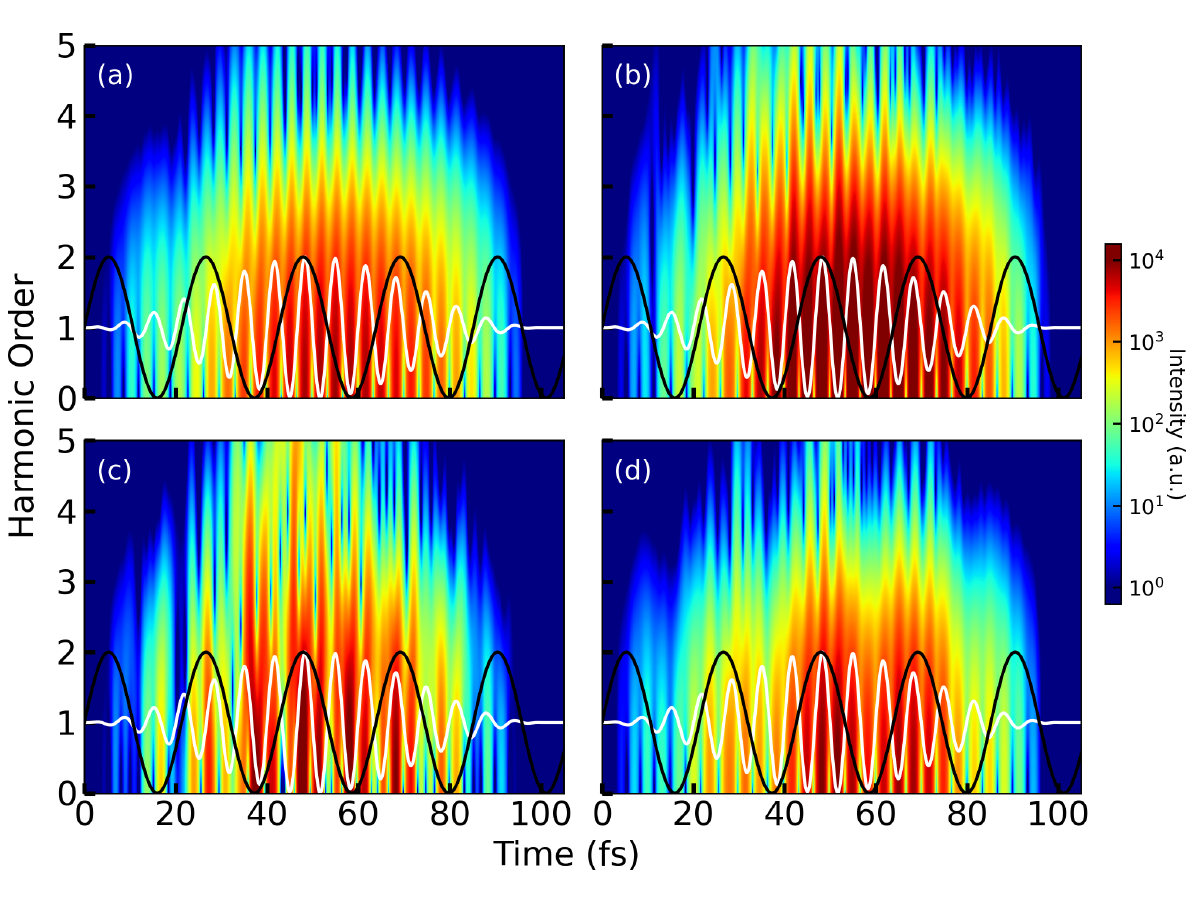}
\caption{Time-frequency representation of the high-harmonic generation.  
Time-frequency map of the current corresponds to (a) graphene without phonon excitation,   
(b) parallel ($\mathsf{X}$) and (c) perpendicular ($\mathsf{Y}$)  components in graphene  with the  $\textsf{iLO}$ coherent phonon mode.  
(d) Time-frequency map of the current in graphene  with the  $\textsf{iTO}$ coherent phonon mode. 
The electric field of the probe pulse is shown by a white curve, which is along the $\mathsf{X}$ direction. 
The black curve represents the periodic displacement of atoms from the equilibrium position with phonon frequency $\omega_{\textrm{ph}}$.}
\label{gabor}
\end{figure}

The situation is simpler in the case of coherent $\textsf{iTO}$ phonon mode excitation. 
Both the main harmonic peaks and the sidebands are polarized along the direction of the probe pulse [see Figs.~\ref{HHGlattice}(b) and \ref{HHGlattice}(d)]. 
Thus, we see that the polarization of the sidebands yields information about the symmetries of the excited phonon modes.  

Let us analyze the time-frequency map of the harmonic spectra to know how high-harmonic spectroscopy is sensitive to an interplay of the coherent lattice and sub-cycle electron dynamics. 
The time-frequency maps are obtained by performing  the Gabor transformation and are presented in Fig.~\ref{gabor}. 
Two bursts of the harmonics per cycle, at the extrema of the electric field, are evident from the time-frequency map of graphene without lattice dynamics [Fig.~\ref{gabor}(a)].  
In this case, the Gabor profile does not vary periodically with $\omega_{\textrm{ph}}$. When $\textsf{iLO}$ phonon mode is coherently excited, 
currents in both parallel and perpendicular directions, with respect to the  probe pulse, are generated [Figs.~\ref{HHGlattice}(a) and \ref{HHGlattice}(c)]. 
The Gabor profile of the parallel ($\mathsf{X}$ component) and the perpendicular ($\mathsf{Y}$ component) are presented in Figs.~\ref{gabor}(b) and \ref{gabor}(c), respectively. 
The profile of the perpendicular component exhibits modulation with a time period equal to 21.3 fs, which matches with  $\omega_{\textrm{ph}}$. 
However, the  parallel profile is not showing any periodic modulation of frequency  $\omega_{\textrm{ph}}$ (see black curve). 
Moreover,  the harmonic bursts are not exactly at the extremum  of the probe electric field but are slightly shifted, which 
could be attributed to the electron-phonon interaction. In the case of the coherent $\textsf{iTO}$ phonon mode, current 
along the probe pulse's direction  is generated [Figs.~\ref{HHGlattice}(b) and \ref{HHGlattice}(d)]. 
Thus, the resultant  Gabor profile of the current display periodic modulation with frequency $\omega_{\textrm{ph}}$. 
The reason behind the periodic modulation of the Gabor profile can be attributed to the time-periodic Hamiltonian as given in Eq.~(\ref{tdtbh:1}).  

One phonon period corresponds to the three optical cycles of the probe pulse and two harmonic bursts per laser cycle are evident from the Gabor profile. 
It means that there are six harmonic bursts within one phonon period: three at the maxima of the cycle and three at the minima. 
This explains the generation of three frequency up ($\omega_{0} + m \omega_{\textrm{ph}}$) and three frequency up sidebands ($\omega_{0} - m \omega_{\textrm{ph}}$), and is consistent with our observations made in  Fig.~\ref{HHGlattice}.

 \begin{figure}
\includegraphics[width=\linewidth]{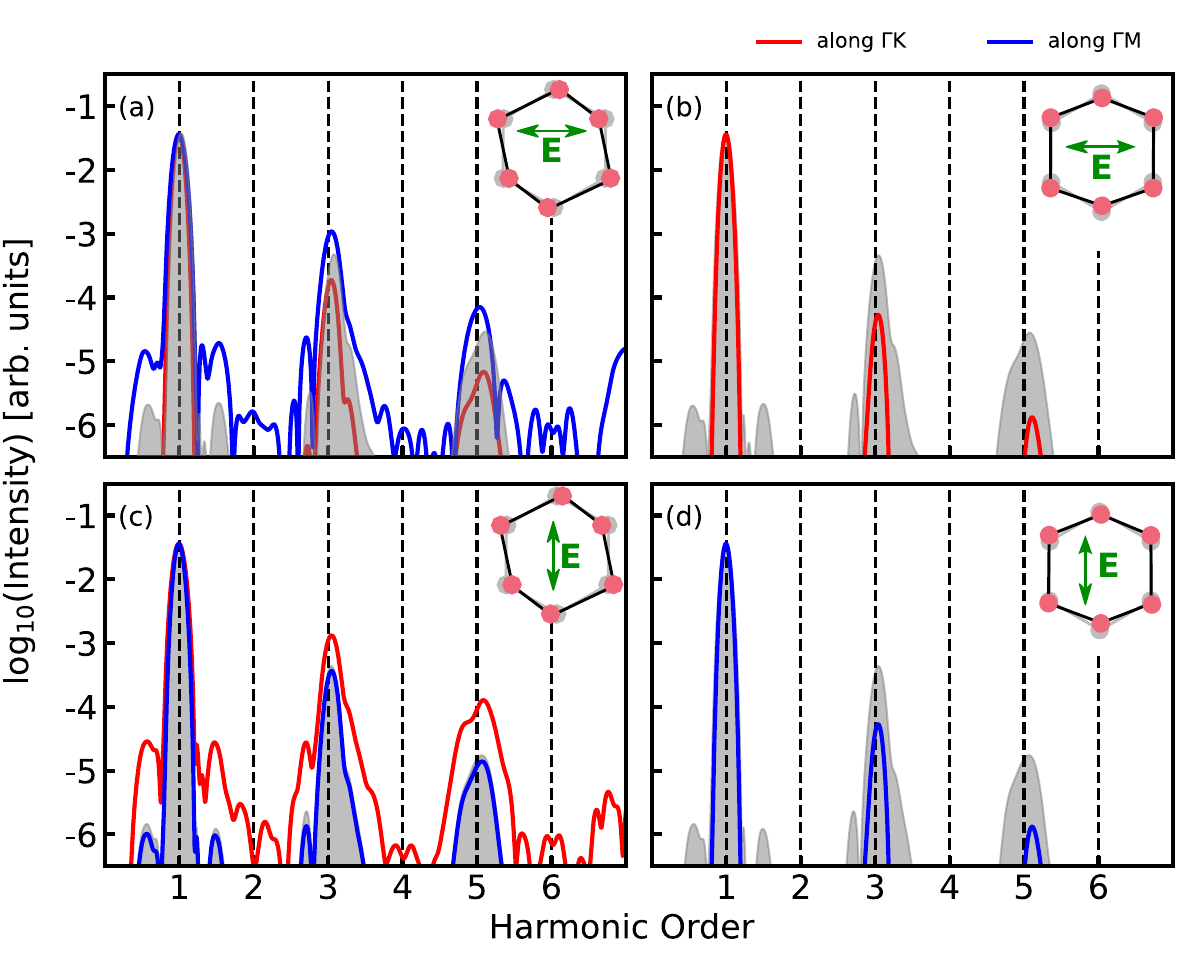}
\caption{High-harmonic spectra of  monolayer deformed graphene. 
(a) and (c) When the atoms in graphene are maximally displaced, from their equilibrium position,  along  $\textsf{iLO}$  phonon mode. (b) and (d) Similar to (a) and (b) but atoms are displaced along $\textsf{iTO}$   phonon mode. The harmonic spectrum of undeformed graphene is shown in the grey-shaded area for reference. The unit cell of the deformed graphene lattice and the polarization of the harmonic generating probe pulse are shown in the respective insets. The red (blue) color corresponds to the polarization of emitted radiation parallel (perpendicular) to the polarization of the harmonic generating probe pulse.}
\label{HHGdeformed}
\end{figure}

\subsection{Transient Breaking of the Symmetry Planes}\label{section:3.2}

We now investigate how the dynamical changes in symmetries  differ from similar static variations in the  high-harmonic spectra. 
Consider the static case  with the  maximum displacement of atoms, along a particular phonon mode direction, 3$\%$ of the lattice parameter from their equilibrium positions. 
Figure~\ref{HHGdeformed} compares high-harmonic spectra for the statically-deformed and undeformed graphene (grey color). 
The probe polarization is  along  $\mathsf{\Gamma-K}$ and $\mathsf{\Gamma-M}$ directions in the top and bottom panels of Fig.~\ref{HHGdeformed}, respectively.  
 
When the graphene is deformed along the $\textsf{iLO}$ phonon mode, odd harmonics are generated along parallel and perpendicular directions with respect to the laser polarization as shown in Figs.~\ref{HHGdeformed}(a) and \ref{HHGdeformed}(c), respectively.  
However, only odd harmonics,  parallel to the laser polarization, are  generated  when graphene  is deformed  in accordance with the $\textsf{iTO}$ mode [see Figs.~\ref{HHGdeformed}(b) and \ref{HHGdeformed}(d)]. 

The emergence of parallel and perpendicular components in the first case, and the parallel component in the second case can be explained as follows: the monolayer graphene has $\sigma_{x}$ and $\sigma_{y}$ symmetry planes, in addition to the inversion center. 
When the polarization of the probe laser is along the high symmetry direction ($\mathsf{\Gamma-K}$ or $\mathsf{\Gamma-M}$), there is no perpendicular component of the current.
However, if the probe pulse is polarized along any direction other than the high-symmetry directions, symmetry constraints allow the generation of  harmonics perpendicular to the direction of the laser polarization. 
It is straightforward  to see that the distortion due to the $\textsf{iLO}$ phonon mode breaks the symmetries of the reflection planes in graphene. 
The absence of the reflection symmetry planes along \textsf{X} and \textsf{Y} directions guarantees the generation of harmonics in both   $\mathsf{\Gamma-K}$ and $\mathsf{\Gamma-M}$ directions  as shown in Figs.~\ref{HHGdeformed}(a) and \ref{HHGdeformed}(c). 
On the other hand, the $\textsf{iTO}$ phonon mode preserves both the symmetry planes and as a result harmonics along the laser polarization are only allowed [Figs.~\ref{HHGdeformed}(b) and \ref{HHGdeformed}(d)].

In short, the presence or absence of the perpendicular current is a result of the transient breaking of the symmetry planes, which can be correlated to the results in Fig.~\ref{HHGlattice}.  
To understand the mechanism behind the sideband generations and associated polarization properties during coherent lattice dynamics, we need to consider the changes in the symmetries dynamically  during the probe pulse. 

\subsection{Dynamical Symmetries}\label{section:3.3}
To understand the symmetry constraint on the polarization of sidebands, let us consider the dynamical symmetries (DSs) of the system, accounting for the coherent lattice dynamics and the probe pulse.
We  apply the Floquet formalism to a periodically driven system, represented by the Hamiltonian  described by Eq.~(\ref{tdtbh:1}), which satisfies 
${\mathcal{H}}_{\textbf{k}}$(t) = ${\mathcal{H}}_{\textbf{k}}(t + \tau_{\textrm{ph}})$, where $\tau_{\textrm{ph}}$ is the time-period corresponding to $\omega_{\textrm{ph}}$. 
The Hamiltonian obeys TDSE and its  solution is  obtained in the basis of the Floquet states as $|\psi_{n}^{\rm F}(t)\rangle = e^{-i\epsilon_{n}^{\rm F}t}|\phi_{n}^{\rm F}(t)\rangle$. 
Here, $\epsilon_{n}^{\rm F}$ is the quasi-energy corresponding to the $n^{\rm th}$ Floquet state 
and  $|\phi_{n}^{\rm F}(t)\rangle$ is the time-periodic part of the wave function, such that $|\phi_{n}^{\rm F}(t+\tau_{\textrm{ph}})\rangle = |\phi_{n}^{\rm F}(t)\rangle$. 
The DSs in a Floquet system are the combined spatiotemporal symmetries, which provide different kinds of  selection rules as discussed in Refs.~\citep{neufeld2019floquet,nagai2020dynamical}. 

In the presence of the probe pulse, the laser-graphene interaction within tight-binding approximation can be modeled with the Peierls substitution as 
${\mathcal{H}}_{\textbf{k}}(t) \rightarrow {\mathcal{H}}_{\textbf{k}+\mathcal{A}(t)}(t)$.
For the sake of simplicity, we employ  a perturbative approach to understand the polarization of the sidebands as the strength of the sidebands is much weaker in comparison to the main harmonic peaks. 
Let us expand ${\mathcal{H}}_{\textbf{k}+\mathcal{A}(t)}(t)$ in terms of $i\mathcal{A}(t)\cdot \textbf{d}_{i}(t)$ as 
\begin{equation}	
{\mathcal{H}}_{\textbf{k}+\mathcal{A}(t)}(t) \approx {\mathcal{H}}_{\textbf{k}}(t) + \mathcal{A}(t)\cdot \nabla_{\textbf{k}} {\mathcal{H}}_{\textbf{k}}(t). \label{eq:h_approx}
\end{equation}
The second term in the above equation can be treated as perturbation as ${\mathcal{H}}_{\textbf{k}}^\prime(t) = \mathcal{A}(t)\cdot\hat{\textbf{J}}(t)$ with 
$\hat{\textbf{J}} = \nabla_{\mathcal{A}(t)} {\mathcal{H}}_{\textbf{k}+\mathcal{A}(t)}$ being the current operator in the Bloch basis. 
Higher-order terms are neglected in Eq.~(\ref{eq:h_approx}).
 
By following Ref.~\cite{nagai2020dynamical} and assuming the electron initially is in the Floquet state $|\phi_{i}^{\rm F}\rangle$, we can  solve the time-dependent Schr\"odinger equation within first-order perturbation theory and the $\mu^{\rm th}$-component of the current can be written as 
\begin{equation}
	\begin{split}
	   J_\mu(t) &= \langle \phi_{i}^{\rm F}(t)| \hat{J}_\mu(t) | \phi_{i}^{\rm F}(t)\rangle \\
		&-\sum_{e\neq i} \int_{-\infty}^t  i dt' e^{-i\omega_{ei}(t-t')}\chi^{\rm F}_{\mu\nu}(t,t') A_\nu(t') 
		+ \textrm{c.c.} 
	\end{split}
\end{equation}
Here, $\chi^{\rm F}_{\mu\nu}(t,t') = \langle \phi_{i}^{\rm F}(t)| \hat{J}_\mu(t) | \phi_{e}^{\rm F}(t)\rangle \langle \phi_{e}^{\rm F}(t')| \hat{J}_{\nu}(t') | \phi_{i}^{\rm F}(t')\rangle$. 
From the above equation, it is apparent that the second term correlates to the generations of the sidebands via  the Raman process.  

The symmetry constraint for the $m^{\rm th}$-order sideband in the harmonics can be written as $\mathcal{X}^t\textbf{E}_{s,m}(t) [\mathcal{X}^t\textbf{E}(t)]^\dagger$ = $\textbf{E}_{s,m} \textbf{E}^\dagger(t)$, provided spatial symmetries of $\mathcal{X}^t$, and probe pulse are the same~\citep{nagai2020dynamical}. 
Here, $\textbf{E}_{s,m}(t)$ and $\textbf{E}(t)$ are, respectively, the electric fields associated with $m^{\rm th}$-order sideband and the probe laser, and $\mathcal{X}^t$ is the dynamical symmetry operation. 
The quantity $\textbf{E}_{s,m}(t)\textbf{E}(t)^\dagger$ is denoted by  $\mathcal{R}_m(t)$ and known as the Raman tensor ~\citep{nagai2020dynamical}. Thus the selection rules for the sidebands depend on the invariance of the Raman tensor under operation with the DSs of the Floquet system.

There are two DSs corresponding to the coherent $\textsf{iLO}$ phonon mode as shown in Fig.~\ref{DS}. 
We define $\tau_n$ as the time translation of $\tau_{\textrm{ph}}/n$, $\mathcal{C}_{n\mu}$ is the rotation of 2$\pi/n$ with respect to the $\mu$-axis, $\hat{\sigma}_\mu$ is the reflection with respect to the $\mu$-axis, and ${\mathcal{T}}$ is the time-reversal operator. 
The symmetry operations $\mathcal{D}_{1} = \hat{\sigma}_{x} \cdot \tau_{2}$ [see Fig.~\ref{DS}(a)] and $\mathcal{D}_{2} = \hat{\sigma_{x}}$ [see Fig.~\ref{DS}(b)] leave the system invariant.   

\begin{figure}
\includegraphics[width=\linewidth]{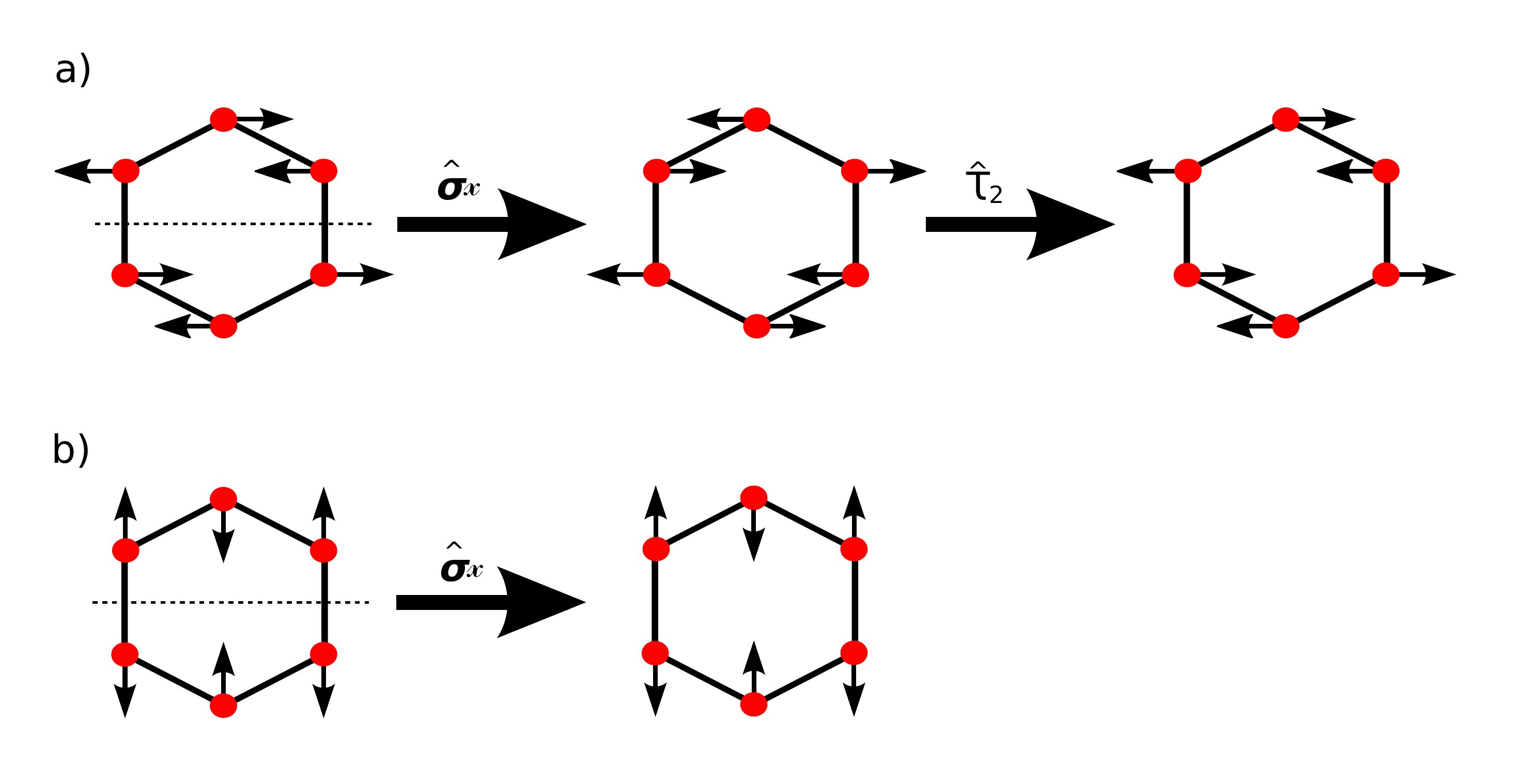}
\caption{Schematic representations of the dynamical symmetries of the Floquet Hamiltonian 
(a) ${\mathcal{D}}_{1}$ = $\hat{\sigma_{x}}\cdot \hat{\tau}_2$ 
(b) ${\mathcal{D}}_{2}$ = $\hat{\sigma_{x}}$. The arrows show the displacements of the atom for a particular phonon mode.}
\label{DS}
\end{figure}

The selection rules for the sidebands and their polarization directions are obtained from the DSs as shown in Fig.~\ref{DS} and require  a condition as ${\mathcal{D}}\mathcal{R}_m(t) = \mathcal{R}_m(t)$.
We assume that the temporal part of the  $m^{\rm th}$-order sideband is $e^{i(\omega_0 \pm  m\omega_{\textrm{ph}})t+\phi_0}$  and of the probe laser pulse is $e^{i\omega_0 t}$. 
In such a situation, the Raman tensor is explicitly written as
\begin{equation}
	\mathcal{R}_m(t) = e^{i(\pm m \omega_{\textrm{ph}}t + \phi_0)} \begin{bmatrix}
		E_{s,m_{x}}E^{*}_{x} & E_{s,m_{x}}E^{*}_{y} \\
		E_{s,m_{y}}E^{*}_{x} & E_{s,m_{y}}E^{*}_{y}
	\end{bmatrix}.
\end{equation}

When the probe laser is polarized along the \textsf{X}-axis, the invariance condition for the Raman tensor 
${\mathcal{D}_1}\mathcal{R}_m(t) = \mathcal{R}_m(t)$ reduces to 
\begin{equation}
	e^{i(\pm m\omega_{\textrm{ph}}t)} \begin{bmatrix}
		E_{s,m_{x}}  \\
		E_{s,m_{y}}
	\end{bmatrix}
	= e^{i[\pm m(\omega_{\textrm{ph}}t +\pi)]} \begin{bmatrix}
		E_{s,m_{x}}  \\
		-E_{s,m_{y}}
	\end{bmatrix}.
\label{selm1}
\end{equation}

The selection rule for the $m^{\rm th}$-order sideband is as follows: when $m$ is odd (even), the polarization of the sideband will be along the \textsf{Y} (\textsf{X}) direction. 
Our observations in Fig.~\ref{HHGlattice}(a) are consistent with Eq.~(\ref{selm1}).

When the $\textsf{iLO}$  phonon mode is excited and the probe pulse is along  the $\mathsf{\Gamma-M}$ direction,  $ \hat{\sigma}_y \cdot \tau_2$ and $\mathcal{C}_{2Z}$ are the DSs, which leave the Raman tensor invariant. 
It is straightforward to see that the  selection rules for the $m^{\rm th}$-order sideband are deduced as follows: 
when $m$  is odd (even), the polarization of the sidebands will be along the \textsf{X} (\textsf{Y}) direction. 
On the other hand, when  the $\textsf{iTO}$  phonon mode is excited and the probe pulse is along $\mathsf{\Gamma-K}$ ($\mathsf{\Gamma-M}$) direction, $\hat{\sigma}_x$ ($\hat{\sigma}_y$) is the DS, which yields the Raman tensor  invariant [see Fig.~\ref{DS}(b)]. 
This symmetry restricts the polarization of the sidebands to be along the direction of the probe pulse. 
Our results are consistent with the observation made in Fig. \ref{HHGlattice}.

\subsection{Role of System Properties}
So far, we have established that high-harmonic spectroscopy is responsive to coherent lattice dynamics. 
Moreover, we have also examined how the system's DSs provide valuable insights into the symmetry of the excited phonon mode. At this juncture,  
it is natural to wonder how different  properties of the system affect  
the high-harmonic spectrum. In the following, we will investigate various aspects of the system properties on HHG from vibrating graphene.
The high-harmonic spectra are normalized with respect to the first harmonic throughout this section unless stated otherwise. 

\begin{figure}[h!]
\includegraphics[width=\linewidth]{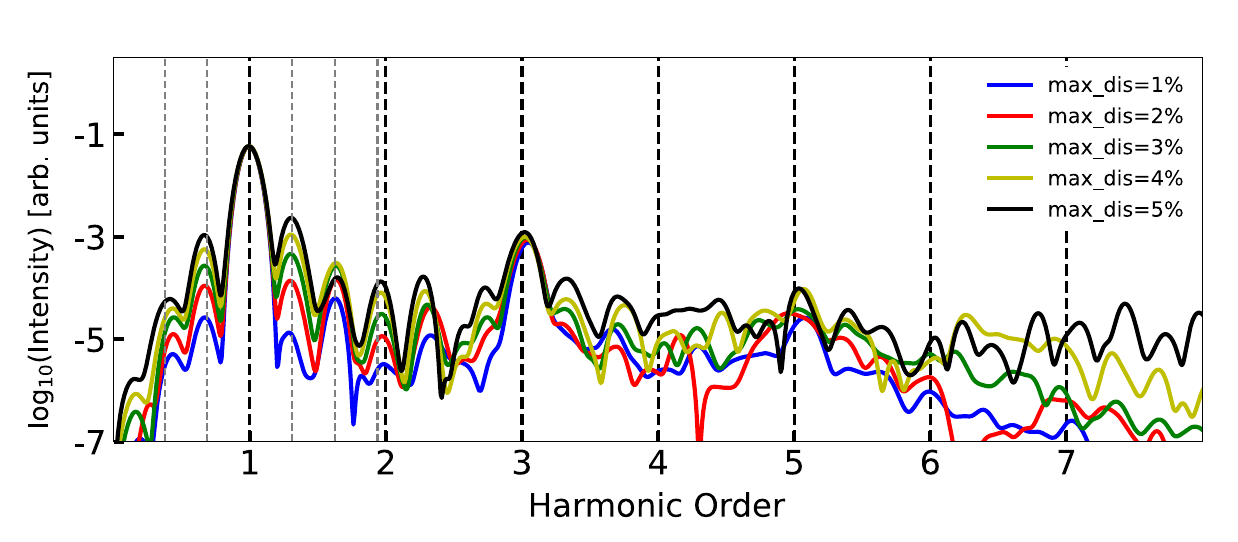}
\caption{Sensitivity of the high-harmonic spectra with respect to different amplitudes of the lattice vibrations. The spectra correspond to  graphene with the coherent $\textsf{iLO}$ phonon mode. 
The parameters of the probe laser pulse are same as in Fig.~\ref{fig:3.2}.}
\label{HHGlattice_ampl}
\end{figure}

Figure~\ref{HHGlattice_ampl} presents high-harmonic spectra corresponding to the coherently excited $\textsf{iLO}$ phonon mode for different amplitudes of the atomic displacement with respect to the equilibrium position. 
The intensity of the sidebands increases monotonically with  the displacement amplitude, whereas the intensity of the main harmonic peaks remains unaffected as evident from the figure.
An increase in the sideband intensity with the  displacement amplitude can be attributed to the 
 extended degree of symmetry breaking caused by the larger displacement of the atoms within graphene.  
The extended symmetry breaking leads to  an overall increase in the intensity of the sidebands.
Remarkably, the polarization properties of the sidebands remains unaffected despite the ``boost'' in  the symmetry breaking.

A similar trend is observed  for the  coherently excited $\textsf{iTO}$ phonon mode with the 
displacement amplitude ranging from 0.01a$_0$ to 0.05a$_0$ relative to the equilibrium positions (not shown here). 
It is important to emphasize that our findings are valid as long as the atomic vibrations with different amplitudes are within the harmonic approximation. 
Thus, it is pertinent to ask how the anharmonicity in the atomic vibrations  affects the harmonic spectra. 

\begin{figure}[h!]
\includegraphics[width=\linewidth]{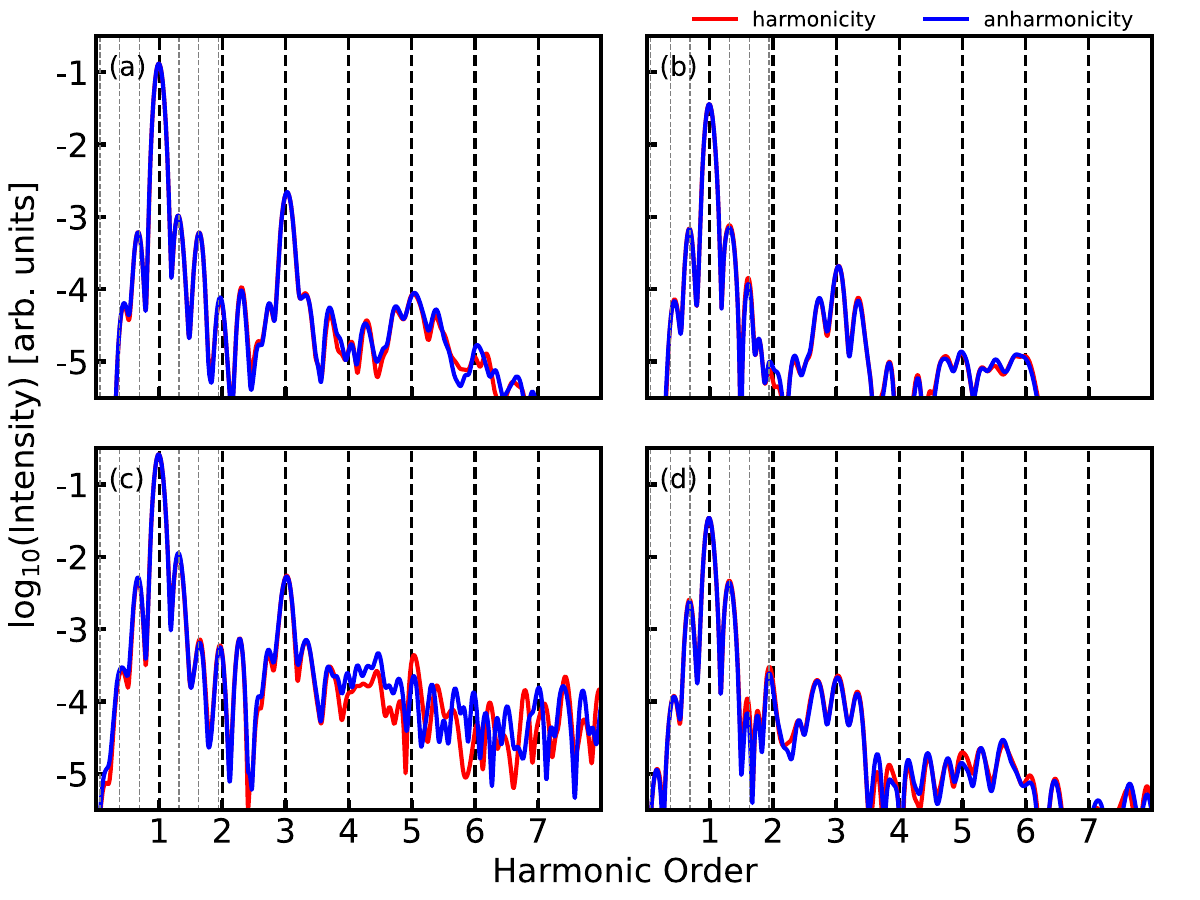}   
\caption{Effect of the anharmonicity in the high-harmonic spectroscopy of the coherent lattice dynamics in graphene. 
High-harmonic spectra corresponding to the coherent $\textsf{iLO}$  phonon mode with 
the maximum amplitude equal to (a) 0.03,  and (c) 0.05 of the lattice constant. 
(b) and (d) are the same as (a) and (c) except $\textsf{iTO}$ phonon mode is coherently excited.   
The  blue (red) color corresponds to the lattice dynamics with (without) an anharmonic effect. The probe harmonic pulse is polarized along $\mathsf{\Gamma-K}$ direction. 
 The parameters of the probe laser pulse are the same as in Fig.~\ref{fig:3.2}. }
\label{HHGlattice_anharmonicity}
\end{figure}

Until now, we have limited our discussion  of the lattice dynamics within harmonic approximation as the time-dependent  displacement vector for a particular phonon mode is linear in amplitude as shown in Eq.~(\ref{eq:ampl}). 
Let us extend our discussion beyond  the harmonic approximation  by adding a non-linear term in the displacement vector. The updated equation of motion for the lattice dynamics reads as  
\begin{equation}\label{eq:anharmonic}
\textbf{q}(t)=   \left[\textrm{q}_{0}~\sin(\omega _{\textrm{ph}}t) - \dfrac{\textrm{q}_{0}^{2}}{2} \left \{ \cos(2\omega _{\textrm{ph}}t) -1\right\} \right] \hat{\textbf{e}}.
\end{equation}
Here, $\textrm{q}_{0}^{2}$-term on the right-hand side  includes anharmonicity in our simulation. 
High-harmonic spectra corresponding to the coherent $\textsf{iLO}$ and $\textsf{iTO}$ phonon modes for two displacement amplitudes are shown in Fig.~\ref{HHGlattice_anharmonicity}. 
As evident from the figure that there is no meaningful effect of the anharmonicity on the generated  spectra for both phonon modes, even for an amplitude equal to 0.05 of the lattice constant.
Thus, high-harmonic spectroscopy 
is not appropriate to capture anharmonic effects in graphene within our theoretical framework.   

\begin{figure}[h!]
\includegraphics[width=\linewidth]{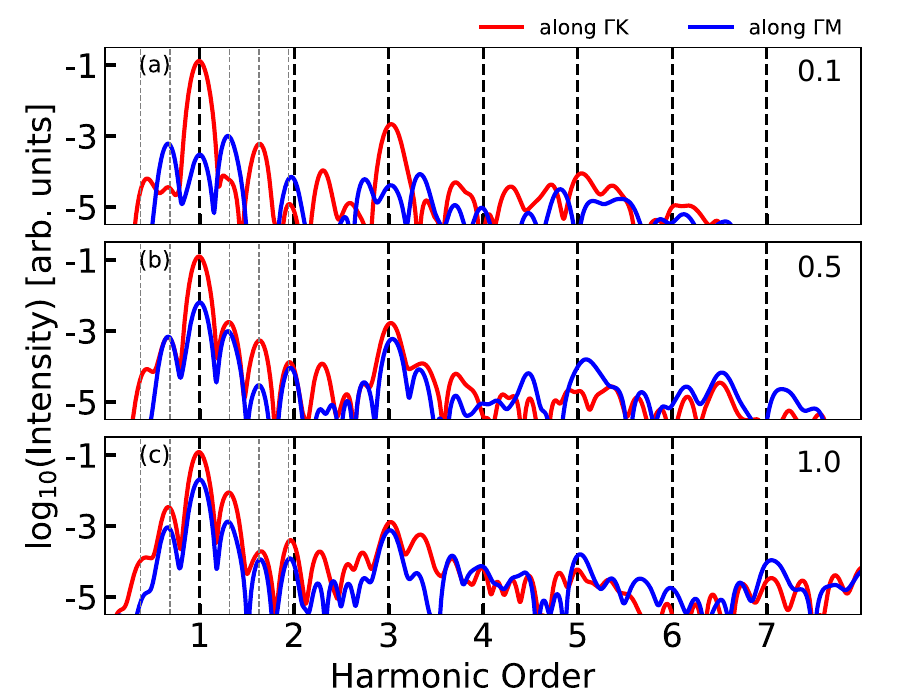}
\caption{High-harmonic spectra corresponding to graphene with the coherent $\textsf{iLO}$  phonon mode 
coupled with  $\alpha$ times $\textsf{iTO}$ phonon mode strength. 
The spectra shown in  (a), (b), and (c)  correspond to $\alpha = 0.1$, 0.5, and 1.0, respectively. 
The red (blue) color corresponds to the polarization of emitted radiation parallel (perpendicular) to the polarization of the harmonic generating probe pulse. The probe harmonic pulse is polarized along $\mathsf{\Gamma-K}$ direction. The parameters of the probe laser pulse are the same as in Fig.~\ref{fig:3.2}. }
\label{HHGlattice_coupling}
\end{figure}

During an excitation process of a coherent phonon mode, there is a possibility that other phonon modes can also get excited with different amplitude as both in-plane phonon modes in graphene are degenerate at $\mathsf{\Gamma}$-point. 
Thus, it is interesting to see how the coupling between the two modes affects the harmonic spectra. 
Let us introduce the coupling of the   $\textsf{iTO}$ phonon mode with $\textsf{iLO}$ phonon mode as 
 \begin{equation}\label{eq:coupled}
\textbf{q}_{\textsf{iLO}}(t)=   \textrm{q}_{0}~\Re \left\{ \exp({i \omega _{\textrm{ph}}t}) \right\}~\hat{\textbf{e}}_{\textsf{iLO}}, ~~~~\textrm{and}~~~~ \textbf{q}_{\textsf{iTO}}(t)= \alpha \textrm{q}_{0}~\Re \left\{ \exp({i \omega _{\textrm{ph}}t}) \right\}~\hat{\textbf{e}}_{\textsf{iTO}},
\end{equation}
where $\alpha$  is a real number, which determines the strength of the coupling between two phonon modes. 
We primarily focus on the coherently excited $\textsf{iLO}$ phonon mode and its coupling with the $\textsf{iTO}$ phonon mode.

Figure~\ref{HHGlattice_coupling} presents high-harmonic spectra for different coupling strengths ranging from 0.1 to 1.0, i.e., 
weak coupling to strong coupling regime.  
It is established that when the $\textsf{iLO}$ phonon is coherently excited, 
odd-order sidebands are polarized perpendicular to the probe pulse  ($\mathsf{\Gamma-M}$) direction, whereas 
the even-order sidebands are polarized parallel  to the probe pulse ($\mathsf{\Gamma-K}$) direction. 
However, additional symmetry restrictions apply to the polarization of the sidebands when the $\textsf{iTO}$ phonon mode is coupled with the $\textsf{iLO}$ mode. 
In this scenario, reflection-plane symmetry perpendicular to the probe pulse is absent, resulting in 
the generation of harmonics along $\mathsf{\Gamma-K}$ as well as $\mathsf{\Gamma-M}$ directions as evident from the figure. 
Moreover, the polarization properties of the sidebands also alter as the coupling strength  increases. 
This means that the polarization of the emitted harmonics are elliptical in nature, and the ellipticity increases as the coupling strength increase.  
Thus, the coupling between the phonon modes can be extracted by analyzing the polarization properties of the main harmonics as well as their sidebands. 

\begin{figure}[h!]
\includegraphics[width=\linewidth]{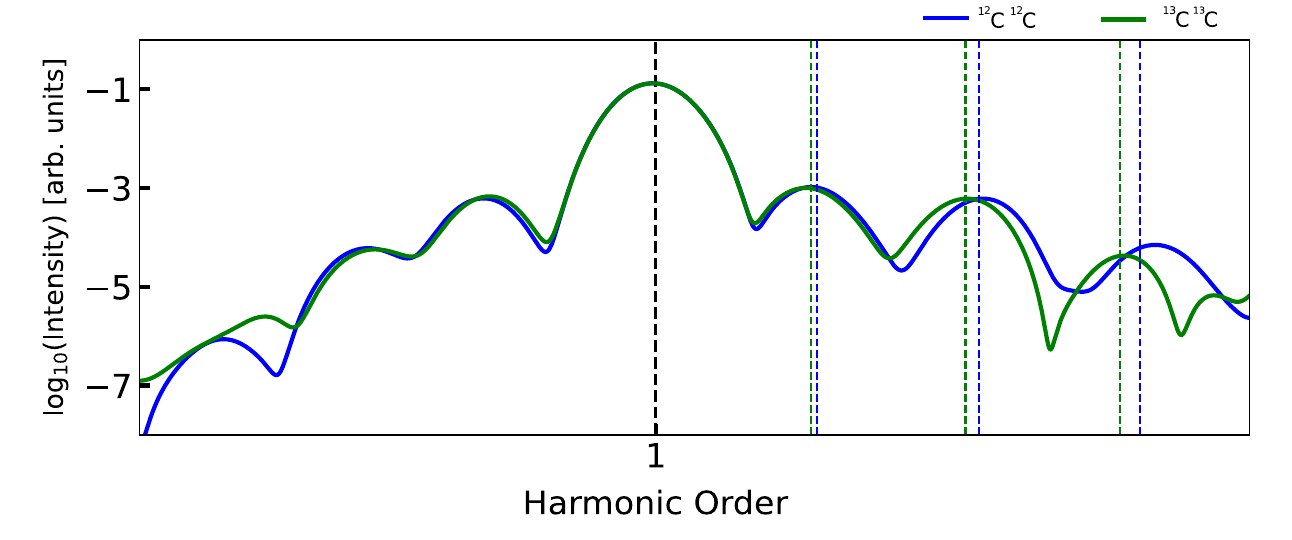}
\caption{Effect of the carbon isotope on high-harmonic spectroscopy of the coherent lattice dynamics corresponding to the 
in-plane $\textsf{iLO}$ phonon mode in graphene. The parameters of the probe pulse are same as in Fig.~\ref{fig:3.2}. }
\label{HHGlattice_isotope}
\end{figure}

After analyzing how different properties of the excited coherent phonon mode affect the generated harmonic spectra, it is 
important to know how HHG is sensitive to a change in the phonon frequency. 
For this purpose, we replace $^{12}$C by $^{13}$C in graphene and the corresponding change in the frequency 
can be estimated as 
\begin{equation}
\omega^{13_{\textrm{C}}}_{\textrm{ph}} = \omega^{12_{\textrm{C}}}_{\textrm{ph}} \sqrt{\dfrac{m_{12_{\textrm{C}}}}{n_{12}m_{12_{\textrm{C}}} + n_{13}m_{13_{\textrm{C}}}}},
\end{equation}
where $n_{12}$ and $n_{13}$ represent the atomic fraction of $^{12}$C and $^{13}$C carbon atoms, respectively. Similarly, $m_{12_{\textrm{C}}}$ and $m_{13_{\textrm{C}}}$ 
represent the atomic mass of $^{12}$C and $^{13}$C carbon atoms, respectively. 
We consider graphene entirely composed of $^{13}$C 
and the corresponding frequency of the in-plane phonon mode comes out as 186.4 meV.
In this case, it is assumed that the coherent $\textsf{iLO}$ phonon mode with  a frequency of 186.4 meV is excited by the pump pulse and the resulting high-harmonic spectra is presented in Fig.~\ref{HHGlattice_isotope}. 
The main harmonic peak is unaffected by the isotope substitution as expected. 
However, the sidebands exhibit sensitivity toward a change in the   isotope. 
The shift in the peak of a sideband increases as the order of the sideband increase, i.e., 
the third sideband is most affected by the isotope substitution.  
It is instructive to use a probe laser pulse with a longer wavelength to separate the peaks of  the sidebands associated with
graphene composed of $^{13}$C and $^{12}$C atoms,  
and to avoid the effect of the peak broadening to discern isotopes.   

\begin{figure}[h!]
\includegraphics[width=\linewidth]{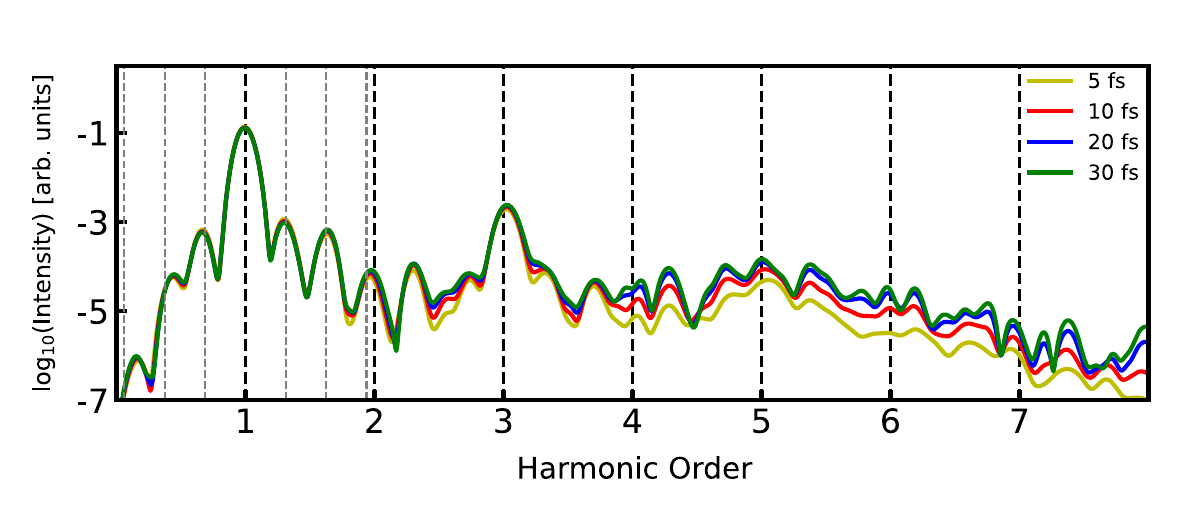}
\caption{Variations in the high-harmonic spectra corresponding to graphene with  the coherent $\textsf{iLO}$ phonon mode for the different dephasing time. The parameters of the probe laser pulse are same as in Fig.~\ref{fig:3.2}. }
\label{HHGlattice_dephasing}
\end{figure}

Before we start our discussion on how the HHG is responsive to different parameters of the probe laser pulse,  let us 
investigate how the emitted harmonics and sidebands are sensitive to the dephasing time -- a phenomenological term accounting for the decoherence between electrons and holes 
in the semiconductor-Bloch equations [see Eq.~\eqref{tdtbh:2}]. 
Figure~\ref{HHGlattice_dephasing} presents the harmonic spectra corresponding to the coherent $\textsf{iLO}$  phonon mode for different dephasing times. 
The low-order harmonics are insensitivity to the change in the dephasing time as the underlying mechanism for HHG 
from vibrating graphene is dominated by the  intraband process. 
However, the intensity of higher-order harmonics are boosted as the dephasing time increases. 
It is known that the population in the conduction band increases as  the dephasing time increases, which leads to an enhancement in the harmonic yield. 
Note that the increase in the harmonic yield does not alter the polarization properties of the emitted harmonics as well as the sidebands. 

\subsection{Impact of Laser Parameters}
In the previous section, we examined how different properties of the vibrating graphene, such as displacement amplitude, 
anharmonicity, coupling between phonon modes, isotope-induced frequency change, and dephasing time affect the harmonic spectra.  
In this section, we will investigate how various parameters of the probe pulse influence HHG from the vibrating graphene. 
Along this line,  high-harmonic spectra of graphene corresponding to the coherent $\textsf{iLO}$ phonon mode  for different duration of the probe pulses are shown in Fig.~\ref{HHGlattice_duration}. 
It is evident that  the generation of sidebands is insensitive to the pulse duration. 
The main harmonic peaks and the sidebands become sharper  as the pulse duration increases. 
Owing to the recombination of an electron with a hole,  a short burst of radiation is emitted during each half-cycle of the laser pulse. As the pulse duration becomes larger, multi-cycle accumulations of the radiation burst cause 
sharper peaks in the  spectra. 
Note that the polarization properties of the main harmonics and the sidebands remain unaffected as the pulse duration increases. 

\begin{figure}[h!]
\includegraphics[width=\linewidth]{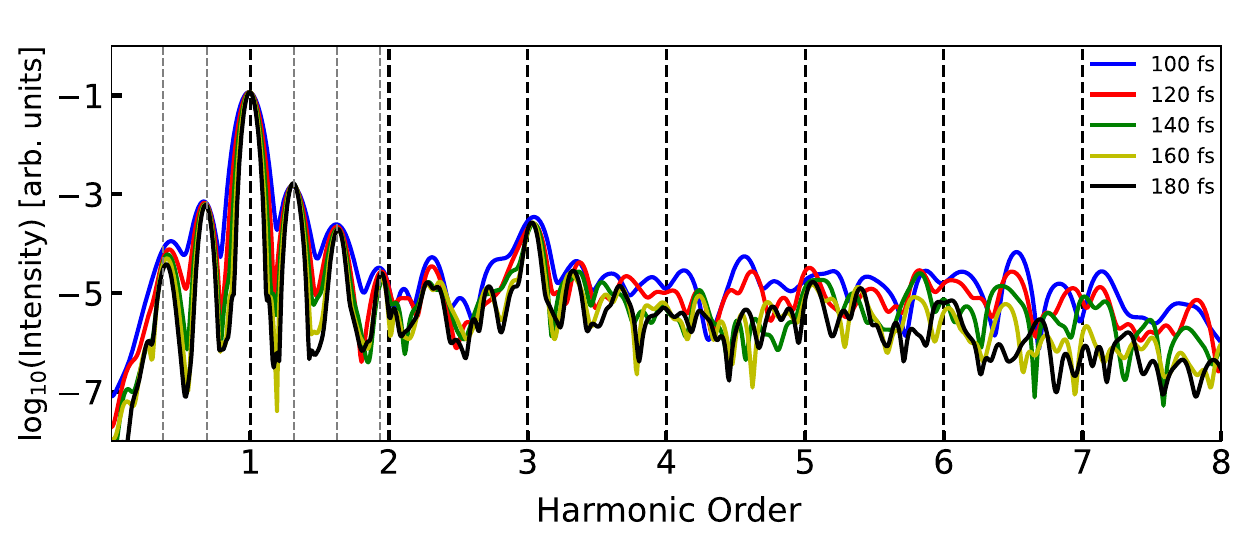}
\caption{High-harmonic spectra corresponding to the coherent $\textsf{iLO}$ phonon mode for the different probe pulse duration. The rest of the probe pulse parameters are same as in Fig.~\ref{fig:3.2}. }
\label{HHGlattice_duration}
\end{figure}

\begin{figure}[h!]
\includegraphics[width=\linewidth]{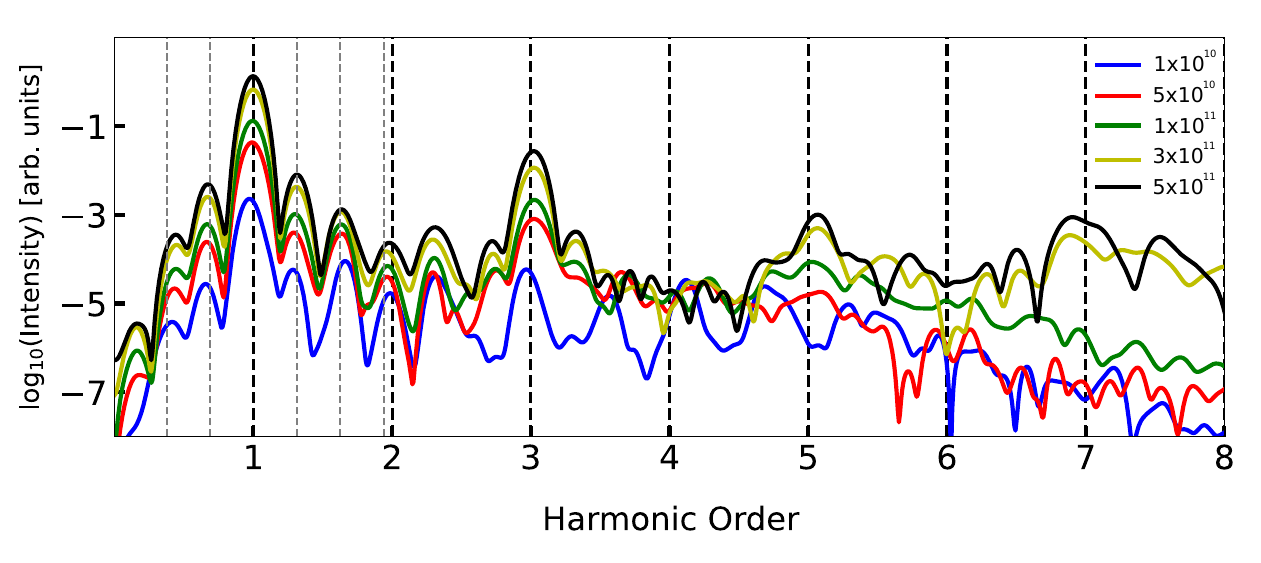}
\caption{High-harmonic spectra corresponding to the coherent $\textsf{iLO}$ phonon mode for the different intensity of the probe pulse. The rest of the probe pulse parameters are same as in Fig.~\ref{fig:3.2}.}
\label{HHGlattice_intensity}
\end{figure}

How the intensity of the probe pulse changes the harmonic generation from the graphene with coherent $\textsf{iLO}$ phonon mode is addressed in Fig.~\ref{HHGlattice_intensity}. 
The yield of the main harmonics and the sidebands increases as the intensity increases from 10$^{10}$ to 5$\times 10^{11}$ W/cm$^{2}$.
The drastic boost in the yield can be attributed to the significant increase  in the conduction band population as the excitation probability increases with  intensity. 
However, an interesting observation can be made when the harmonics are normalized with respect to the first harmonic corresponding to  10$^{10}$ W/cm$^{2}$ intensity. 
The intensity of the sidebands decreases as the intensity increases. 
At higher intensities, the effect of the electron dynamics becomes more dominant, suppressing the contributions from the coherent lattice dynamics. 
In other words, the intensity of the main harmonic peaks increases due to a higher population of electrons in the conduction band. 
However, the influence of the lattice dynamics does not increase at the same rate at higher intensities. The intensity of the probe pulse does not alter other properties of the spectra. 

\begin{figure}[h!]
\includegraphics[width= 0.9 \linewidth]{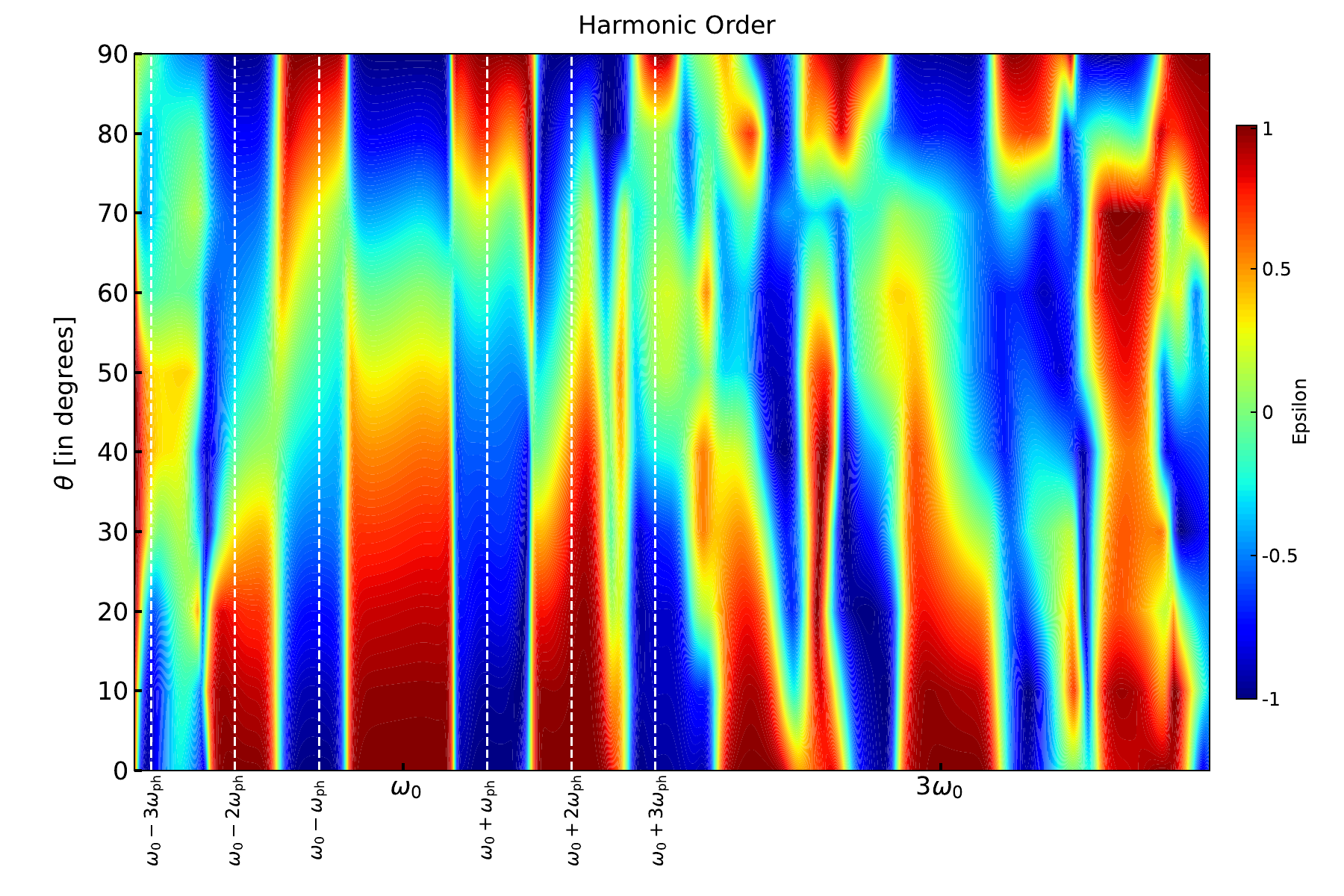}
\caption{Sensitivity of the harmonic generation for polarization direction of the probe pulse. 
Polarization of the main harmonics and the sidebands  along 
$\mathsf{\Gamma-K}$ ($\mathsf{\Gamma-M}$)  is  represented by Epsilon equal to 1 (-1) in the colorbar. 
The rest of the probe pulse parameters are same as in Fig.~\ref{fig:3.2}. 
The harmonic spectrum corresponds to the coherent $\textsf{iLO}$ phonon mode.  }
\label{HHGlattice_ilo}
\end{figure}

\begin{figure}[h!]
\includegraphics[width= 0.9 \linewidth]{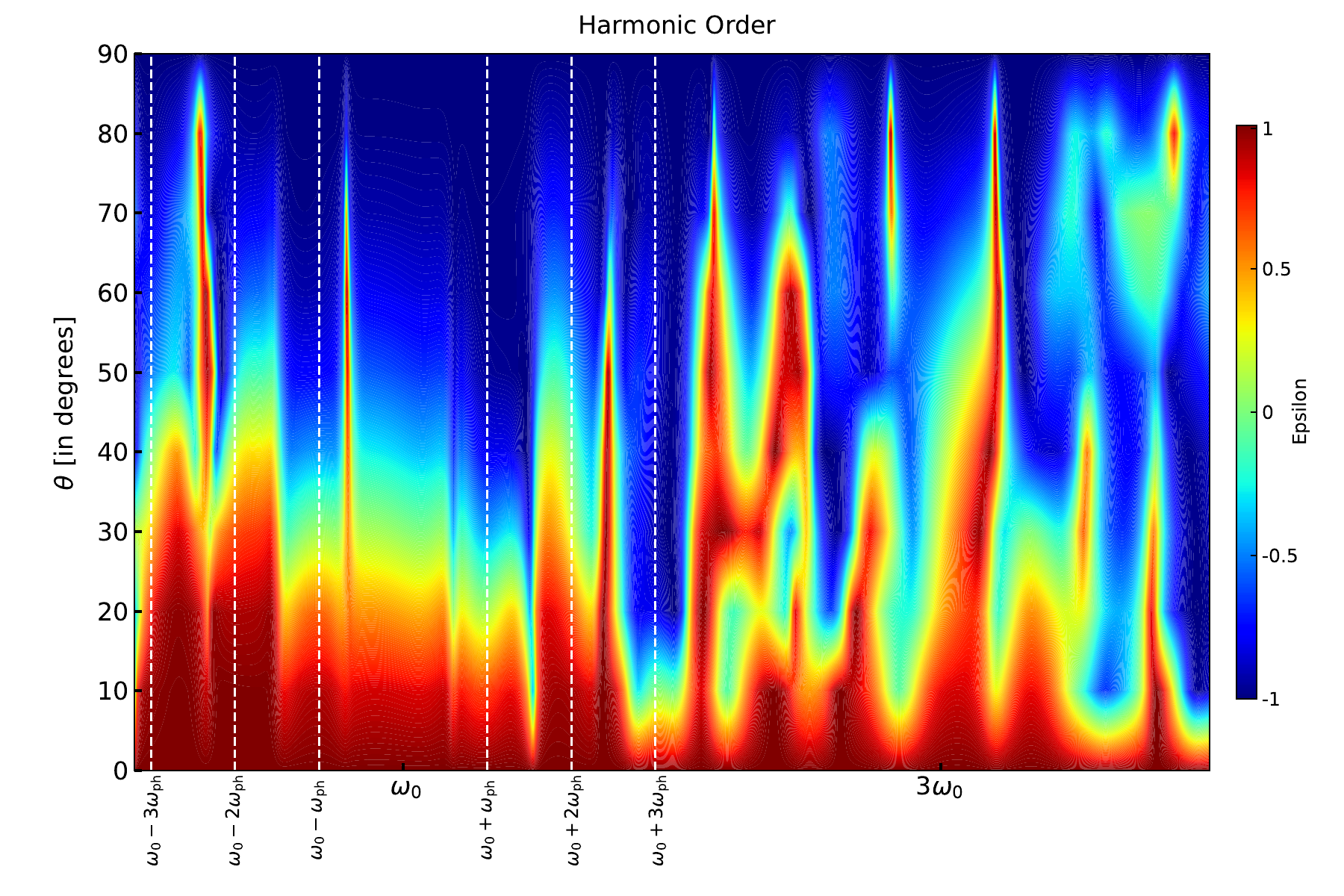}
\caption{Same as Fig.~\ref{HHGlattice_ilo}  for the coherent $\textsf{iTO}$ phonon mode.}
\label{HHGlattice_ito}
\end{figure}

After analyzing  the effects of the pulse duration and intensity of the probe pulse, it is interesting 
to see how variation in the polarization of the probe pulse impacts the harmonic spectra. 
For this purpose,  we tune the linear polarization of the probe pulse from 
$\mathsf{\Gamma-K}$ to $\mathsf{\Gamma-M}$ direction and  the corresponding high-harmonic spectra 
associated with the coherent $\textsf{iLO}$ and $\textsf{iTO}$ phonon modes are depicted in  Figs.~\ref{HHGlattice_ilo} and \ref{HHGlattice_ito}, respectively. 
As discussed earlier, the six-fold symmetry of graphene reduces  to the two-fold symmetry when  atoms vibrate along the $\textsf{iLO}$ phonon mode and there is no  reflection plane symmetry. 
Moreover, the even- (odd-) order sidebands are polarized parallel (perpendicular) to the polarization of the probe pulse, whereas the main harmonics are always polarized along the probe pulse. 
This means that the  harmonics and the sidebands are emitted along $\mathsf{\Gamma-K}$ and $\mathsf{\Gamma-M}$ directions when the polarization of the probe pulse is not aligned along $\mathsf{\Gamma-K}$ direction. 
Let's consider a situation  when the probe pulse is polarized  at  45$^{\circ}$  with respect to the $\mathsf{\Gamma-K}$ direction. This can be seen as two orthogonally polarized probe pulses along $\mathsf{\Gamma-K}$ and $\mathsf{\Gamma-M}$ directions with an amplitude equal to 1/$\sqrt{2}$ times of the original probe pulse along 45$^{\circ}$.
It means that the emitted harmonics and their sidebands have both components with the same magnitude  resulting in Epsilon equal to zero as evident from Fig.~\ref{HHGlattice_ilo}. 
A similar analysis is applicable to the harmonic spectra corresponding to the $\textsf{iTO}$ phonon mode  as shown in 
Fig.~\ref{HHGlattice_ito}. 
 
\section{Summary}\label{section:3}
To summarize, we have established that high-harmonic spectroscopy is responsive to the coherent lattice dynamics in solids. 
The  high-harmonic spectrum is modulated by the frequency of the excited phonon mode within the solid. Both  in-plane E$_{2g}$ Raman-active phonon modes of the monolayer graphene lead to the generation of higher-order sidebands, along with the main harmonic peaks. 
In the case of  $\textsf{iLO}$  phonon mode excitation, the even- and odd-order sidebands are polarized parallel and perpendicular to the polarization of the probe harmonic pulse, respectively.  
In the case of  the $\textsf{iTO}$  phonon mode, all sidebands are polarized along the probe harmonic pulse's polarization. 
The polarization of the sidebands are dictated by the dynamical symmetries of the combined system, which includes the phonon modes and probe laser pulse. 
Therefore, the polarization properties  are a sensitive probe of these dynamical symmetries.  
The presence of a high-harmonic signal perpendicular to the polarization of the probe pulse is a signature of  lattice excitation-driven symmetry breaking of the reflection plane. 
The present work is paving the way  for probing phonon-driven processes in solids and non-linear phononics with sub-cycle temporal resolution. 

\cleardoublepage
\chapter{Unveiling Phase and Chirality of Circular Phonons via High-Harmonic Spectroscopy}\label{Chapter4}
Vibrations of atoms within molecules and solids are fundamental processes that regulate matter's physical, optical, and chemical properties. 
When light triggers atomic vibrations, atoms exhibit periodic oscillations in a particular fashion, and these light-induced vibrations could potentially lead to modifications in various symmetries of solids. 
These modifications are dynamic in nature and result in several transient phenomena, such as light-induced superconductivity~\citep{hu2014optically, mitrano2016possible}, 
vibrationally-induced magnetism~\citep{rini2007control, nova2017effective}, 
and switching of electrical polarization~\citep{mankowsky2017ultrafast}, to name but a few. 
Thus, time-resolved mapping of the interplay of lattice vibration with electronic motion on the electronic timescale 
is essential to comprehend several ubiquitous phenomena in solids, such as structural phase transition~\citep{hase2015femtosecond}, thermal
and optical properties~\citep{katsuki2013all, Fultz2010, gambetta2006real}; 
and predicting new concepts in solids.

In the previous chapter, we have explored how the coherently excited either the $\textsf{iLO}$ or $\textsf{iTO}$ phonon mode of graphene leads  modulation in  the high-harmonic spectrum, generated by 
a linearly polarized pulse. 
However, in practical scenarios, there is a possibility that both the phonon modes get coherently  
excited with certain phase differences. 
Under this scenario, eigenvectors for coherent atomic vibrations given in  Eq.~(\ref{eq:ampl}) 
can be recasted as 
\begin{equation}\label{eq:circular}
\textbf{q}_{\textsf{iLO}}(t)=   \textrm{q}_{0}~\Re \left\{ \exp({i \omega _{\textrm{ph}}t}) \right\}~\hat{\textbf{e}}_{\textsf{iLO}}, ~~~~\textrm{and}~~~~ \textbf{q}_{\textsf{iTO}}(t)=   \textrm{q}_{0}~\Re \left\{ \exp({i \omega _{\textrm{ph}}t}+\phi) \right\}~\hat{\textbf{e}}_{\textsf{iTO}}.
\end{equation}
Here, $\phi$ is the phase difference between the two phonon modes.  
Specifically, a left-circular phonon (LCP) mode emerges  
when the phase difference between the $\textsf{iLO}$ and $\textsf{iTO}$ phonon modes is 90$^{\circ}$  
[see Fig.~\ref{fig:1}(c)]. On the other hand, a phase difference of -90$^{\circ}$ leads to the generation 
of a right-circular phonon (RCP) mode  as depicted in Fig.~\ref{fig:1}(d). 
Similarly, we will observe the formation of elliptical phonon modes for other values of $\phi$.

\begin{figure}
\includegraphics[width=  \linewidth]{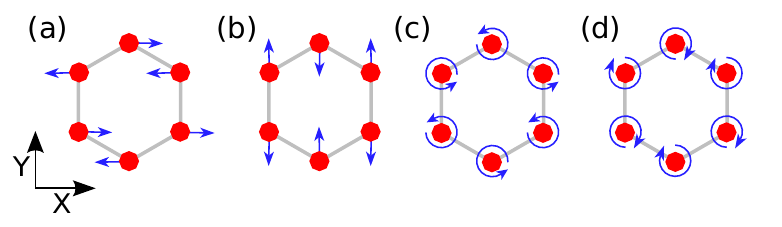}
\caption{Sketch of coherent atomic vibrations associated with (a) $\textsf{iLO}$ phonon, (b) $\textsf{iTO}$ phonon, (c) left-circular phonon (LCP), and  (d) right-circular phonon (RCP) modes. }
\label{fig:1}
\end{figure}

This chapter focuses on highlighting the capabilities of high-harmonic spectroscopy in the time-resolved mapping of coherent superposition of phonon modes with fixed phase differences. 
It aims to explore the interplay between coherent lattice vibration and electronic motion as well as the transient evolution of symmetries in solids during lattice dynamics. 
We will demonstrate that the phase difference between both phonon modes 
can be successfully extracted using high-harmonic spectroscopy. 
In addition, our study investigates the optical response in graphene when circular phonon modes are coherently excited. 
In the later part of this chapter, we will demonstrate  that the coherent lattice dynamics lead to the dynamical-symmetry alterations, which results in the generation of symmetry-forbidden harmonics.  
Moreover,  coherent lattice dynamics lead to the generation of higher-order sidebands along with the main harmonic peaks in the high-harmonic spectrum. 
The frequency and symmetry of the coherently excited phonon mode are imprinted  in the position and polarization of the sidebands, respectively. 
Note that probing transiently-evolving lattice-electronic dynamics and DSs 
of solids in the presence of light in a single experimental setup is challenging. 
This chapter addresses this crucial problem.  

Similar to previous chapter,  
we choose two-dimensional graphene with $\textbf{D}_{6\textrm{h}}$ symmetry in which
two in-plane Raman active E$_{2\textrm{g}}$ ($\mathsf{G}$)  phonon modes are considered. 
Note that both phonon modes are degenerate at $\mathsf{\Gamma}$ point with 
the frequency of 194 meV, which is equivalent to an oscillation period of $\sim$ 21 femtoseconds.  
Linearly-polarized pulse with a  2.0 $\mu$m wavelength and peak intensity of 1$ \times 10^{11} $ W/cm$^2$ is used to generate higher-order 
harmonics in graphene with and without coherent phonon dynamics. 
The pulse duration is 100 fs, which is  much longer than an oscillation period of in-plane phonon dynamics $\simeq$ 21 fs. Moreover, the lifetime of these phonon modes is around 1 ps~\citep{kang2010lifetimes}. 
The parameters for the circularly-polarized probe pulse are also the same unless stated otherwise.

\section{Results and Discussion}
\subsection{Extraction of the Phase}
\begin{figure}[h!]
\includegraphics[width=\linewidth]{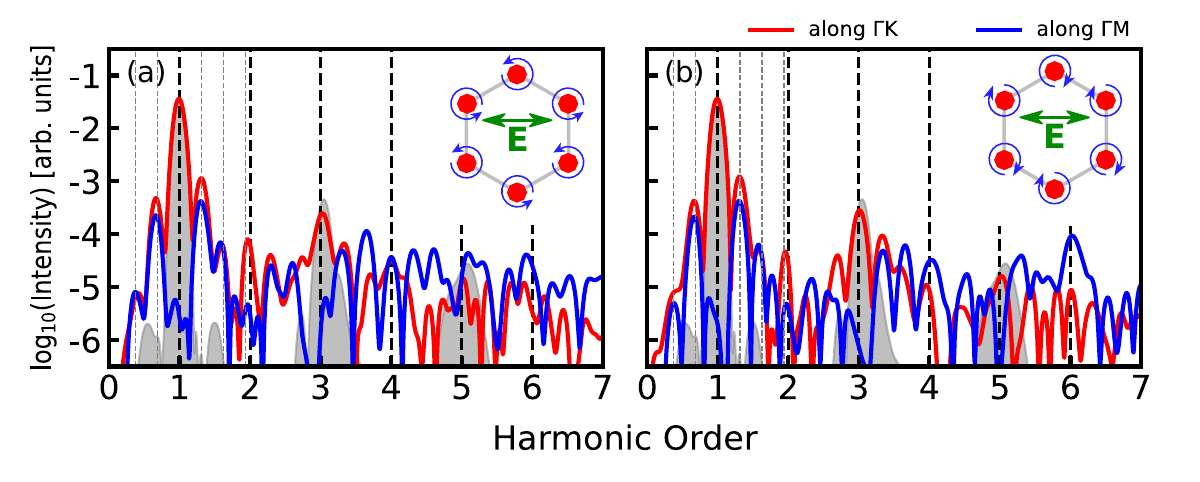}
\caption{High-harmonic spectra of monolayer graphene with and without coherent phonon dynamics.  
The spectra of the graphene with (a) left-handed phonon (LCP)  and (b) right-handed phonon (RCP) modes. 
In both spectra, sidebands corresponding to the prominent harmonic peaks are identified at frequencies 
($\omega_{0} \pm  n \omega_{\textrm{ph}}$) with $\omega_{\textrm{ph}}$ as the phonon frequency, $\omega_0$  as the frequency of the linearly polarized probe  pulse and 
$n$ as  an integer. 
Graphene's unit cell with the eigenvector of a particular coherent phonon mode and polarization of the harmonic generating probe pulse are shown in the respective insets.
Results are presented for $\textrm{T}_2 = $  10 fs and a maximum 0.03a$_0$ displacement of  atoms from their equilibrium positions during coherent phonon dynamics where a$_0$ is the lattice parameter of the equilibrium structure.}
\label{fig:2}
\end{figure}

Figures~\ref{fig:2}(a) and \ref{fig:2}(b) present  high-harmonic spectra for  
graphene with coherent LCP and RCP modes, respectively. 
In both cases, the spectra exhibit  sidebands' generation alongside the main harmonic peaks and 
subsequent sidebands are separated by the frequency of excited phonon mode.  
Coherent excitation of either LCP or RCP mode leads to the generation of the sidebands, 
perpendicular to the probe pulse's polarization  as evident from the spectra.  
Furthermore, all  sidebands exhibit elliptical polarization, indicating the influence of the coherently excited circular phonon modes. 
The elliptical nature of the sidebands can be explained using the dynamical symmetry of the 
the temporarily-evolving system as discussed in the previous chapter. 
Note that the main harmonic peaks are polarized parallel to  the  probe pulse's polarization. 
Harmonic spectra in gray color correspond to high-harmonic spectra without phonon dynamics throughout the chapter  stated otherwise.

\begin{figure}
\includegraphics[width= 0.9\linewidth]{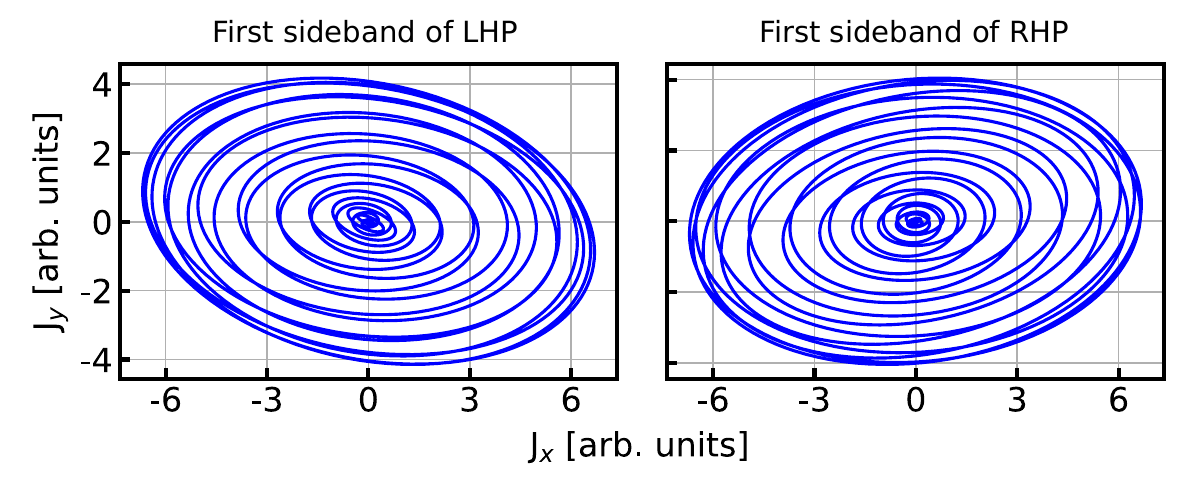}
\caption{Projection of the $x$ and $y$ components in time domain 
of the first sideband associated with the first main harmonic peak corresponding to
(a) LCP and (b) RCP modes. 
The phase difference between the $x$ and $y$ components of the first sideband is 95$^\circ$ (87$^\circ$) for the LCP (RCP) mode.  The current in the time domain corresponding to the sideband is 
extracted from the simulated harmonic spectra using a Gaussian function. }
\label{fig:3}
\end{figure}

\begin{figure}
\includegraphics[width= 0.9\linewidth]{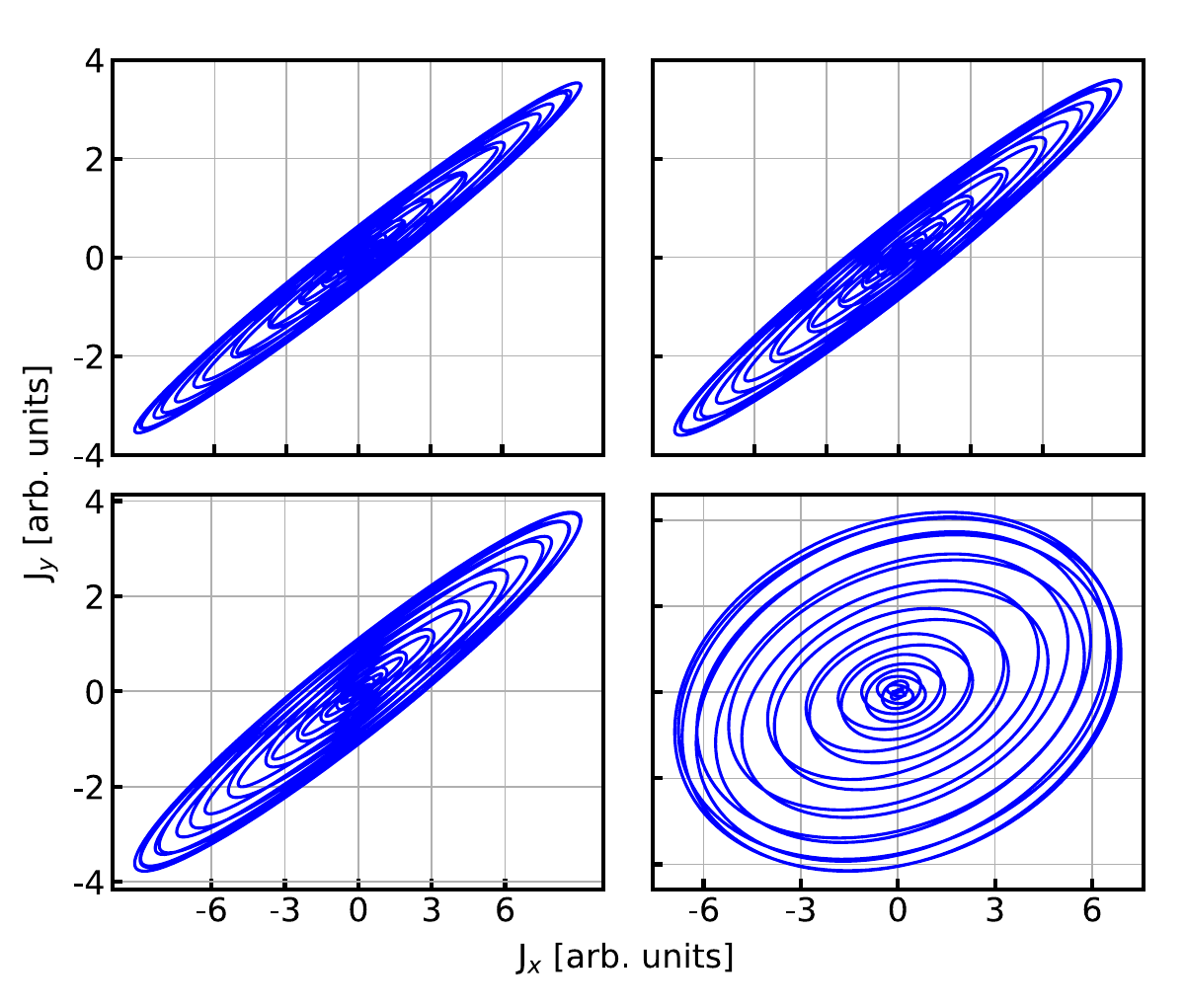}
\caption{Similar to Fig.~\ref{fig:3} with various  phase difference values between $\textsf{iLO}$ and $\textsf{iTO}$ phonon modes: (a) 10$^\circ$, (b) 15$^\circ$, (c) 20$^\circ$ and (d) 75$^\circ$.  
The extracted  phase differences from the simulated harmonic  spectra are 10$^\circ$, 14$^\circ$, 18$^\circ$, and 76$^\circ$.}
\label{fig:4}
\end{figure}

Time-domain projection of the $x$ and $y$ components of the first sideband corresponding 
to the first harmonics of graphene with LCP and RCP modes are shown in Figs.~\ref{fig:3}(a) and \ref{fig:3}(b), respectively.
The phase difference between the $x$ and $y$ components is determined as 
$\phi = \sin^{-1}(x_{1}/x_{2})$ with 
$x_{1}$ as  the point where the curve first intersects the $x$-axis and $x_{2}$ is the 
maximum projection point along the $x$-axis.
The extracted phase difference between the $x$ and $y$ components  associated with the 
first sideband of the first main harmonic is 95$^\circ$ (87$^\circ$) for the LCP (RCP) mode, which
is in excellent agreement with the theoretical value with a minor deviation of 5$^\circ$. 
To demonstrate the robustness of our proposed method, we examined the superposition of phonon modes with phase difference values of 10$^\circ$, 15$^\circ$, 20$^\circ$, and 75$^\circ$. 
The corresponding phase differences extracted  from the simulated harmonic spectra are  
10$^\circ$, 14$^\circ$, 18$^\circ$, and 76$^\circ$. 
The close agreement between the theoretical and extracted phase values establishes that high-harmonic spectroscopy is a potential probe  for determining  the phase differences between different phonon modes. 
The high-harmonic spectra corresponding to coherently excited  LCP and RCP modes 
appear similar in both qualitative and quantitative aspects. 
It should be noted that Fig.~\ref{fig:2} may give a misconception that the handedness of the phonon mode cannot be discerned from the high-harmonic spectra, which is not true as discussed below. 

\subsection{Chirality of the Circular Phonons}
It is known that chiral light, i.e., left- and right-handed polarized light, probes the chirality in the matter.
Following the same concept, we also employ a chiral laser pulse to probe the  chirality of the coherently excited phonon mode. 
High-harmonic spectra corresponding to LCP and RCP modes with the chiral probe pulse are shown in 
Figs.~\ref{fig:5}(a) and \ref{fig:5}(b), respectively. 
The appearance and polarization of the sidebands are distinct with the chirality of the phonon mode.
The first sideband to the right, and the second sideband to the left of the first main harmonic are absent when the excited LCP mode is probed by the right-handed circular pulse [see Fig. 5(a)].
On the other hand, probing of the RCP mode by a right-handed circular pulse forbids the generation  
of the second sideband to the right, and the first sideband to the left of the first  harmonic as visible in Fig. 5(b). Importantly, the sidebands in both cases are separated by the frequency of the phonon modes, and the polarization of the sidebands differs between the two cases.

\begin{figure}
\includegraphics[width=\linewidth]{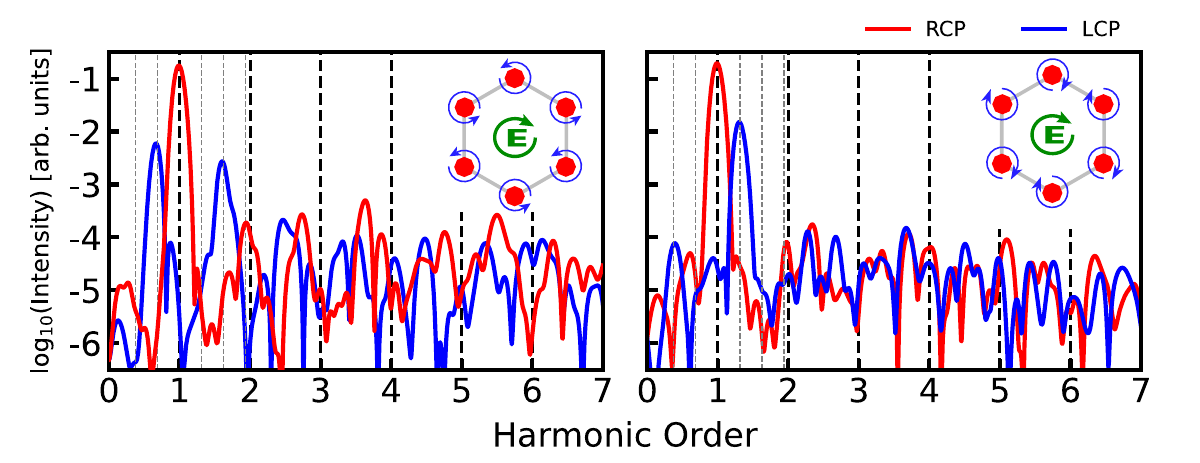}
\caption{High-harmonic spectra corresponding to (a) left-handed phonon (LCP)  and (b) right-handed phonon (RCP) modes.  The unit cell of the graphene with the eigenvector of a particular phonon mode and polarization of the harmonic generating right-handed circularly polarized probe pulse are shown in the respective insets.}
\label{fig:5}
\end{figure}

To elucidate the origin of the chiral sidebands, we employ  Floquet formalism to study the response of graphene with a coherently excited circular phonon mode, probed by a circularly polarized  pulse. 
The Floquet formalism allows us to determine the DSs of the system, which dictate the selection rules governing the occurrence of the sidebands as discussed in the previous chapter. The $m^{\rm th}$-order sideband satisfies a symmetry constraint given by $\mathcal{X}^t\textbf{E}_{s,m}(t) [\mathcal{X}^t\textbf{E}(t)]^\dagger = \textbf{E}_{s,m} \textbf{E}^\dagger(t)$, where $\mathcal{X}^t$ represents a DS, and $\textbf{E}_{s,m}(t)$ and $\textbf{E}(t)$ denote the electric fields associated with the 
$m^{\rm th}$-order sideband and the probe laser, respectively. The spatial symmetries of 
$\mathcal{X}^t$ and the probe pulse need to be the same for this symmetry condition to hold true~\citep{nagai2020dynamical}. In the subsequent analysis, we adopt the same approach as discussed in the previous chapter to investigate the properties of the sidebands by utilizing the DSs within the Floquet formalism. 

In the present case,  ${\mathcal{D}}_{n,\sigma_{m}}$ = $\hat{\tau}_{-n \sigma_{m}} \cdot \mathcal{R}_{n}$ is the DS,  which leaves the system with the circular phonon mode invariant. Here, $\mathcal{R}_{n}$ is the rotation by an angle of $2\pi/n$  with respect to $z$-axis,  $\hat{\tau}_{-n \sigma_{m}}$ is the time-translation operator and $\sigma_{m}$ is $\pm $ 1. 
Thus, the condition associated with the sidebands is determined as ${\mathcal{D}}_{n,\sigma_{m}}\mathcal{R}_m(t) = \mathcal{R}_m(t)$ with 
\begin{equation}
\mathcal{R}_m(t) = \mathbf{E}_{s,m}(t)\mathbf{E}(t)^\dagger
= \begin{bmatrix}
	E_{s,m_{x}}E^{*}_{x} & E_{s,m_{x}}E^{*}_{y} \\
	E_{s,m_{y}}E^{*}_{x} & E_{s,m_{y}}E^{*}_{y}
\end{bmatrix},
\label{eq1}
\end{equation}
which can be further simplified using the eigenvectors  as 
\begin{equation}
\mathcal{R}_m(t) = e^{(i m \omega_{\rm ph} t)}
\begin{bmatrix}
	1 & i \sigma_{p} \\
	- i \sigma_{s,m} &  \sigma_{p}\sigma_{s,m}
\end{bmatrix}. 
\label{eq2}
\end{equation}
Here,  $\sigma_{p}$ is the chirality of the probe pulse and $\sigma_{s,m}$ is the chirality of the 
$m^{\rm th}$-order sideband.  
The generalized selection rule governing the coherent excitation of the circular phonon mode with a circularly polarized pulse can be expressed as $\sigma_{s,m} = m \sigma_{\rm ph} + \sigma_{p} + 3N$ 
with  $\sigma_{\rm ph}$ as the chirality of the excited phonon and $N$ as an integer.

To illustrate a specific case, let us consider a coherently excited LCP mode probed by a right circularly polarized pulse as shown in Fig.~\ref{fig:5}(a). 
In this scenario, the selection rule simplifies to $\sigma_{s,m} = m - 1 + 3N$. 
By analyzing this expression for $N = 0$, we find that the second sideband with left-handed chirality, 
right to the first harmonic, is allowed. 
For $N = 1$, both first and third sidebands, left to the first harmonic, with left-  and right-handed chiralities 
become allowed. 
However,  only the third sideband with right-handed chirality,  right to the first harmonic,  is generated  when $N$ acquires  -1 value, and all other sidebands are forbidden based on the selection rules. 
Thus,  the properties of the simulated high-harmonic spectra  in Fig.~\ref{fig:5}(a) 
 corroborate with these derived  selection rules.
By using similar steps, selection rules for the coherently  excited RCP mode, probed by 
right-handed circular  pulse can be obtained. 
Present findings demonstrate that high-harmonic spectroscopy enables the characterization of both the phase difference and the chirality of the excited phonon mode and therefore provides valuable insights into the temporarily evolving system's  properties.

\subsection{Lattice-Driven Symmetry Alterations}

\begin{figure}[ht!]
\includegraphics[width=  \linewidth]{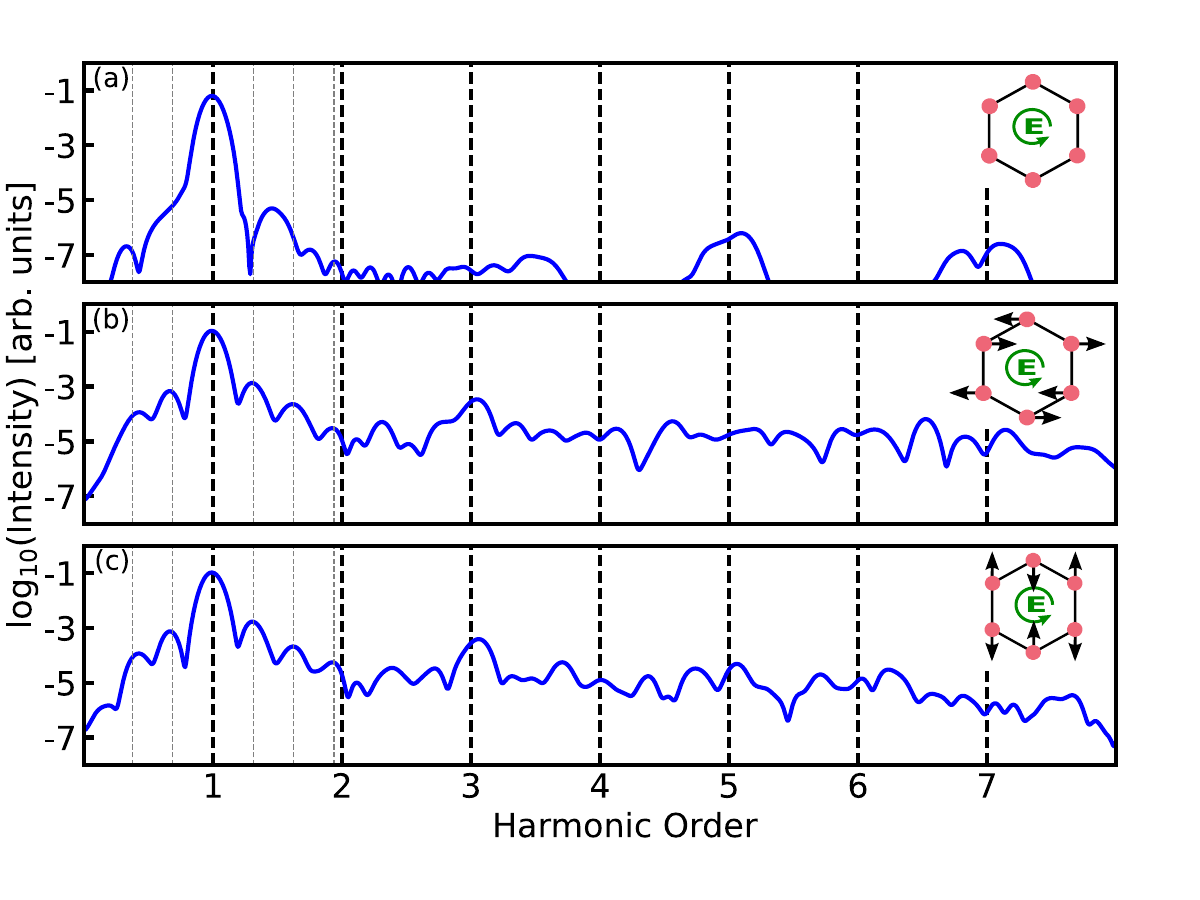}
\caption{High-harmonic spectra, generated by the left-handed circularly polarized laser pulse, of   graphene with and without coherent lattice dynamics.  (a)  The spectra of the graphene without  lattice 
dynamics. The spectra of graphene with the coherent  (b) $\textsf{iLO}$ and (c) $\textsf{iTO}$ phonon modes.  
The unit cell of the graphene with the eigenvector of a particular phonon mode and polarization of the 
harmonic generating probe pulse are shown in the respective insets. 
Results are presented for $\textrm{T}_2 = $  10 fs and a maximum 0.03a$_0$ displacement of  atoms from their equilibrium positions during coherent lattice dynamics where a$_0$ is the lattice parameter of the equilibrium structure.}
\label{spectra}
\end{figure}

High-harmonic spectrum corresponding to  graphene without lattice dynamics is presented in Fig.~\ref{spectra}(a).  
We have employed a left-handed circularly polarized laser pulse for HHG in graphene with and without phonon dynamics.
As dictated  by the symmetry constraints and selections rules, it is expected that circularly polarized  pulse
yields $(6m \pm 1)$-orders of harmonics from inversion-symmetric graphene with the six-fold symmetry~\citep{alon1998selection, chen2019circularly, neufeld2019floquet}. 
Here,  $m = 0, 1, 2, \ldots$ is a positive integer. 
In this case,  the third harmonic is symmetry forbidden. 
On the other hand, linearly polarized laser pulse leads to  $(2m + 1)$-orders of harmonics as shown earlier~\citep{mrudul2021high}.  
Our results shown in Fig.~\ref{spectra}(a)  are consistent with the selection rules and earlier report~\citep{alon1998selection, chen2019circularly}. 

After discussing HHG from graphene without lattice dynamics, let us investigate how the in-plane phonon modes 
affect the harmonic spectrum shown in  Fig.~\ref{spectra}(a). 
For this purpose, we coherently excite one of the  two degenerate   in-plane phonon modes,
and assume that the excitation is done prior to the probe harmonic pulse. Figure~\ref{spectra}(b) presents the harmonic spectrum  corresponding to 
coherently excited $\textsf{iLO}$ phonon mode.  
The spectrum in Fig.~\ref{spectra}(b) is drastically different from the one without lattice dynamics [see Fig.~\ref{spectra}(a)]. 
There are only odd harmonics in the spectrum as the $\textsf{iLO}$ phonon mode preserves the inversion symmetry in graphene. 
Moreover, the spectrum exhibits multiple sidebands along with the main odd harmonics as evident from Fig.~\ref{spectra}(b). 
Our findings remain  qualitatively  the same  for the amplitude of the lattice displacement ranging from 0.01a$_0$ to  0.05a$_0$ with respect to the equilibrium positions as evident from Fig.~\ref{spectra_variation}(a). As expected, the intensity of the sidebands increases as the amplitude of vibration increases.  
Moreover, dephasing time $\textrm{T}_2$ does not impact our findings significantly as the spectra are qualitatively  the same  for $\textrm{T}_2$ ranging from 5 to 30 fs [see  Fig.~\ref{spectra_variation}(b)].

The coherent excitation of the $\textsf{iTO}$ phonon mode also leads to  multiple sidebands along with the  odd harmonics as visible from Fig.~\ref{spectra}(c). 
Apparently, it seems that the spectra are insensitive to the symmetry of the excited phonon mode as  both $\textsf{iLO}$  and $\textsf{iTO}$ phonon modes yield similar harmonic spectra [see Figs.~\ref{spectra}(b) and ~\ref{spectra}(c)]. 
In the following, we will show that this is not the case, and the symmetry of the excited phonon mode is encoded in the polarization properties of the spectra. 

\begin{figure}
\includegraphics[width=  \linewidth]{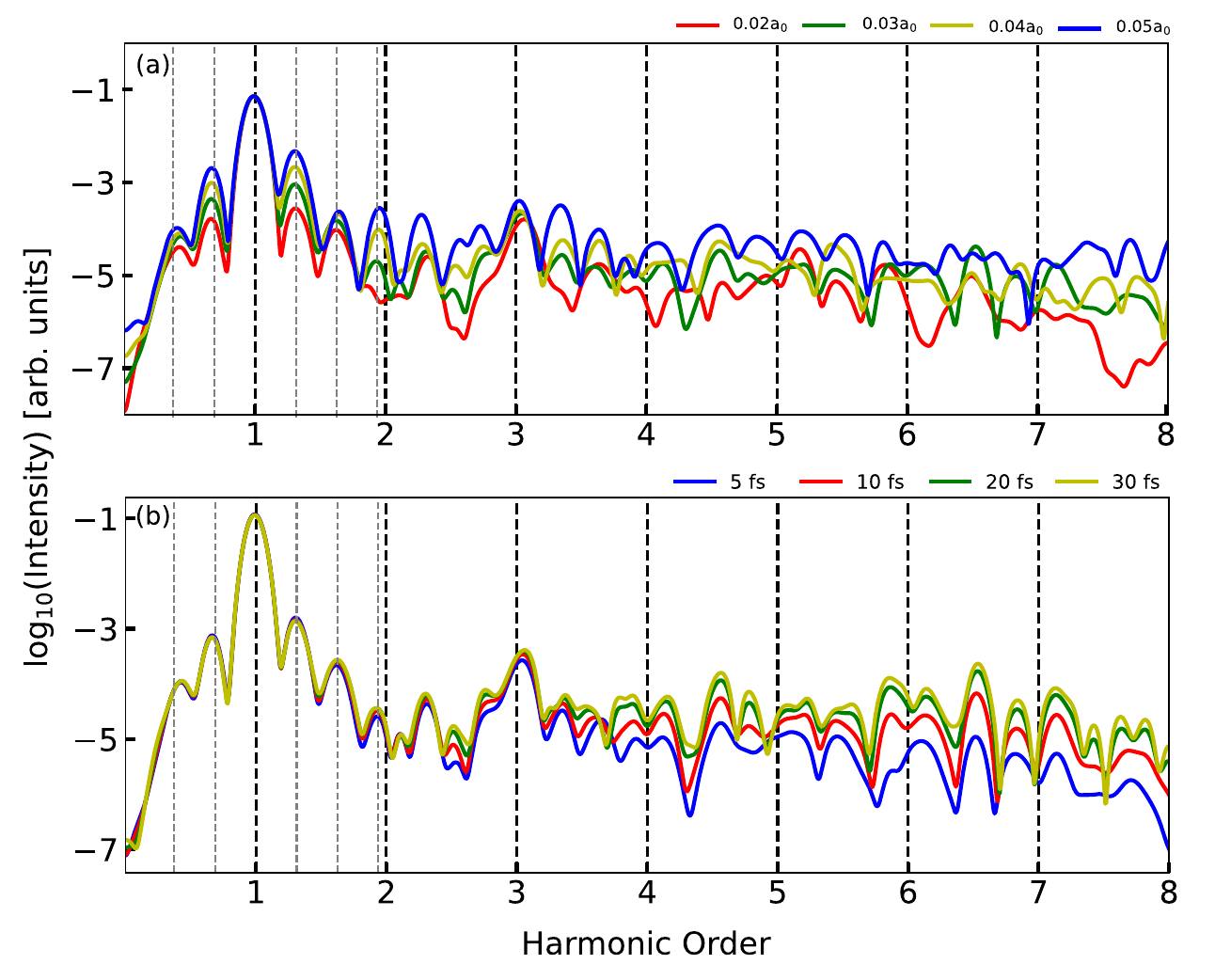}
\caption{High-harmonic spectra of graphene with $\textsf{iLO}$ phonon mode  for different (a) amplitudes of the atomic oscillation, and (b) dephasing time (T$_{2}$).}
\label{spectra_variation}
\end{figure}

Not only do coherent lattice dynamics lead to the generation of multiple sidebands  but also the forbidden harmonics become allowed. 
As stated earlier, the third harmonic is absent for  the circular laser-driven HHG from graphene without phonon [see Fig.~\ref{spectra}(a)].
However, dynamics of the coherent E$_{2\textrm{g}}$ phonon  mode reduces graphene's six-fold symmetry into two-fold dynamically, which 
allows the generation of $(2m \pm 1)$ harmonic orders.  The presence of  the third harmonic in both cases, graphene with $\textsf{iLO}$  or $\textsf{iTO}$ phonon mode, is a signature of the dynamical symmetry reduction as evident from Figs.~\ref{spectra}(b) and ~\ref{spectra}(c).  At a glance, it seems that the criteria for HHG are the same for linearly polarized laser pulse  [$(2m + 1)$ orders and third harmonic]  and the combination of the phonon-driven symmetry reduction  with circularly polarized laser pulse [$(2m \pm 1)$ orders and third harmonic].
To distinguish the two situations, let us analyze the  polarization properties of the emitted harmonics.

\begin{figure}
\includegraphics[width=\linewidth]{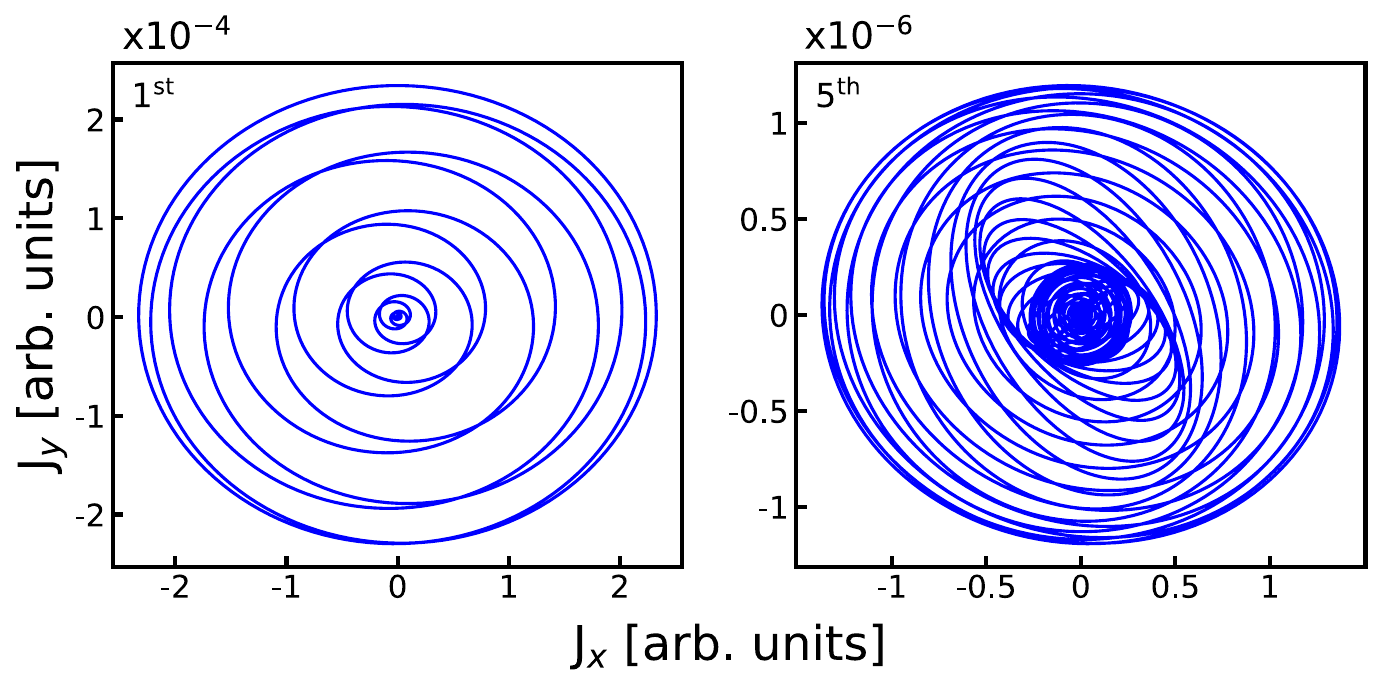}
\caption{ The projection of the $x$  and $y$ components, in the time domain,  of the first and  the fifth harmonics corresponding to the spectrum of  graphene without phonon dynamics as shown in Fig.~\ref{spectra}(a).}
\label{polarization_no_latt}
\end{figure}

Figure~\ref{polarization_no_latt} displays the projection of $x$ and $y$ components, in  the time domain, corresponding to the first and fifth harmonics of the spectrum in  Fig.~\ref{spectra}(a).  
It is known that the polarization of a given harmonic for a material with $l$-fold symmetry is
determined by $lm  + \sigma$, where $\sigma =  + (-)1$ represents the $m^{\textrm{th}}$ harmonic's  polarization, which is  the same (opposite) as the helicity of the driving laser pulse~\citep{alon1998selection}. 
It is straightforward to see that  $(lm - 1)^{\textrm{th}}$ and $(lm + 1)^{\textrm{th}}$ harmonics are circularly polarized with $\sigma = -1$ and $\sigma = 1$, respectively. In the present case,  the first and fifth harmonics are circularly  polarized  with opposite helicity and are consistent with  earlier findings~\citep{saito2017observation, chen2019circularly}.

\begin{figure}[ht!]
\includegraphics[width= \linewidth]{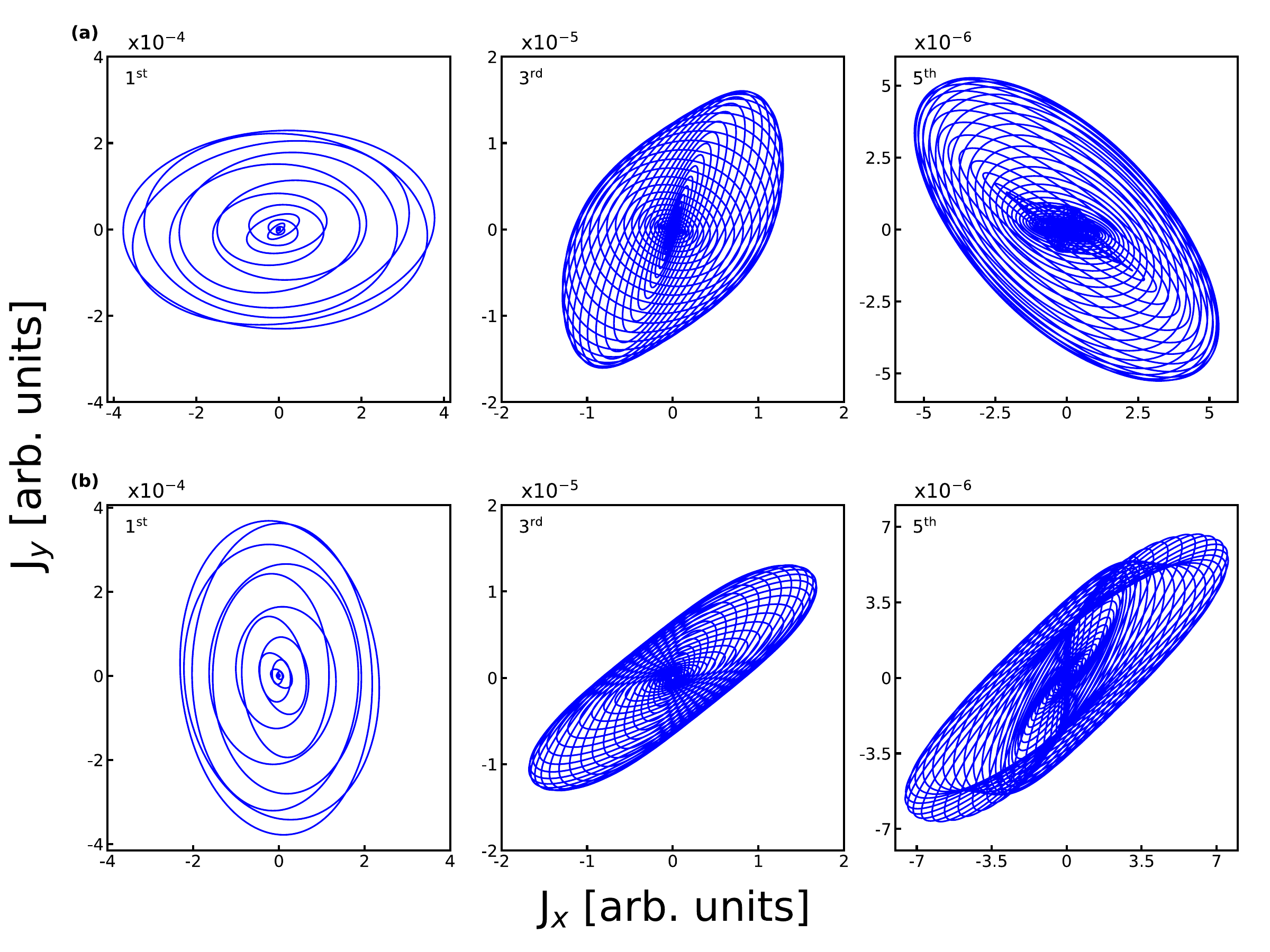}
\caption{Same as Fig.~\ref{polarization_no_latt} for the first, third and  
fifth harmonics  for graphene with (a) $\textsf{iLO}$ and (b) $\textsf{iTO}$ phonon modes. 
In the case of $\textsf{iLO}$ ($\textsf{iTO}$)  phonon mode, the ellipticities of the first, third, and fifth harmonics are 0.63 (0.62), 0.77 (0.79), and 0.93 (0.96), respectively. Also, the phase differences between $x$ and $y$  components  of the first, third and fifth harmonics are  90$^{\circ}$ (90$^{\circ}$), 40$^{\circ}$ (35$^{\circ}$), and 125$^{\circ}$ (45$^{\circ}$), respectively.}
\label{polarization_latt}
\end{figure}

As stated above, $(2m \pm 1)$ harmonic orders are allowed due to phonon-driven dynamical symmetry reduction from six-fold to two-fold, which leads to the generation of the third harmonic. Moreover, this dynamical symmetry reduction also alters the polarization properties of the emitted harmonics. The projected $x$ and $y$ components of the first, third, and fifth harmonics for graphene with $\textsf{iLO}$  and $\textsf{iTO}$ phonon modes are presented in Figs.~\ref{polarization_latt}(a) and  ~\ref{polarization_latt}(b), respectively. 
As evident from the figure, the ellipticity of the first harmonic reduces drastically from 1 for graphene without phonon to 0.63  for graphene with phonon. 
The change in the ellipticity can be understood as follow:  When $\textsf{iLO}$  phonon mode is excited, carbon atoms vibrate along the $\textsf{X}$ direction, which increases the velocity of the  electrons in the $\textsf{X}$ direction. It is known that the intraband current is proportional to the velocity,  and low-order harmonics in graphene are dominated by the intraband current~\citep{mrudul2021high, vampa2015semiclassical}.  Thus, the major axis of the ellipse is along the $\textsf{X}$ direction in the case of  the first harmonic, which reduces the ellipticity from 1 to 0.63. 

\begin{figure}[ht!]
\includegraphics[width= \linewidth]{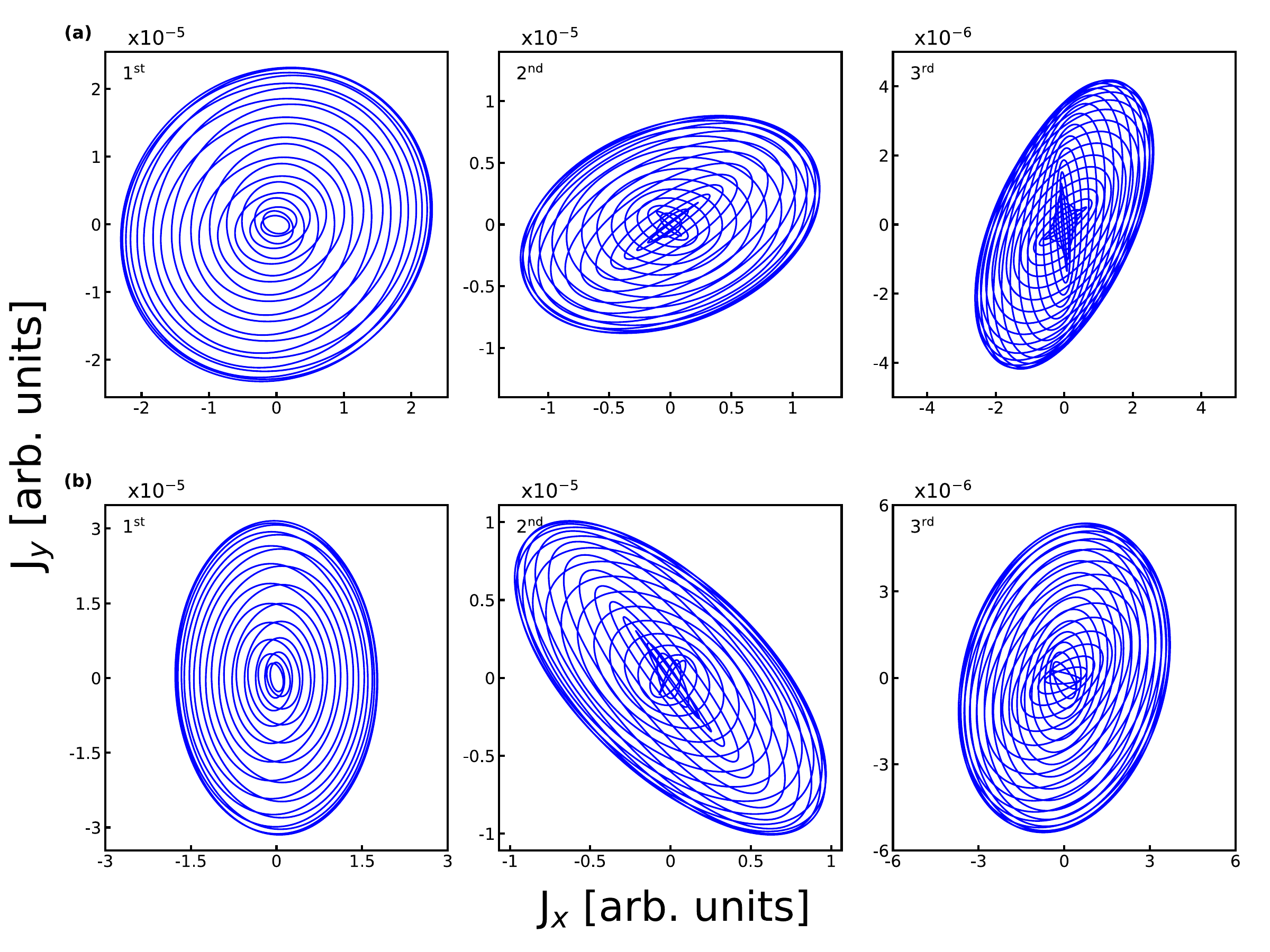}
\caption{Same as Fig.~\ref{polarization_no_latt} for the first, second, and third sidebands situated right to 
the first harmonic  for graphene with (a) $\textsf{iLO}$ phonon mode corresponding to Fig.~\ref{spectra}(b), and (b) $\textsf{iTO}$ phonon mode corresponding to Fig.~\ref{spectra}(c). The ellipticities of the first, second, and third sidebands corresponding to $\textsf{iLO}$  ($\textsf{iTO}$) phonon modes  are  0.98 (0.56), 0.72 (0.95), and 0.59 (0.67), respectively. 
The phase differences between $x$ and $y$  components  of the first, second, and third	sidebands corresponding to $\textsf{iLO}$ ($\textsf{iTO}$) phonon mode are 85$^{\circ}$ (90$^{\circ}$), 70$^{\circ}$ (135$^{\circ}$), and 60$^{\circ}$ (75$^{\circ}$), respectively.}
\label{polarization_side}
\end{figure}

Similarly,  the excitation of the $\textsf{iTO}$ mode leads  the vibrations of the atoms along  $\textsf{Y}$ direction. 
This provides an additional velocity component to electrons in the $\textsf{Y}$ direction, which translate to the major axis of the ellipse  along the $\textsf{Y}$ direction. The ellipticity of the fifth harmonic, corresponding to graphene without phonon to with $\textsf{iLO}$ ($\textsf{iTO}$) phonon mode, changes significantly, i.e., from 1 to 0.93 (0.96).  
Not only does the lattice dynamics modify the ellipticity of the harmonics  significantly  but also changes  the phase between the $x$ and $y$ components of the harmonics. In the case of $\textsf{iLO}$ ($\textsf{iTO}$) mode, the phase differences between the components for first and fifth harmonics are 90$^{\circ}$ (90$^{\circ}$) and  125$^{\circ}$ (45$^{\circ}$), respectively (see  Fig.~\ref{polarization_latt}).  The ellipticity and the phase difference of the third harmonic for graphene with $\textsf{iLO}$ ($\textsf{iTO}$) mode are 0.77 (0.79), and 40$^{\circ}$ (35$^{\circ}$), respectively. Thus, the changes in the ellipticity and phase indicate that the harmonics are sensitive to the symmetry of the excited phonon mode as reflected from Fig.~\ref{polarization_latt}.

After demonstrating how the information of the excited phonon mode and its symmetry are imprinted  in the  main harmonics and their polarization properties, 
let us analyze what information is encoded in the sidebands associated with prominent  harmonics.  
The  projection of $x$ and $y$ components of the first, second, and third sidebands associated with the first main harmonic corresponding to  graphene with $\textsf{iLO}$ phonon mode is shown in Fig.~\ref{polarization_side}(a). 
All three sidebands have nonzero $x$ and $y$ components as evident from the figure.
The same is true for the sidebands associated with  $\textsf{iTO}$ phonon excitation  as reflected  from Fig.~\ref{polarization_side}(b). 
Moreover, the ellipticities  of the first, second, and third sidebands of the first harmonic corresponding to graphene with $\textsf{iLO}$ ($\textsf{iTO}$)  phonon mode read as  0.98 (0.56), 0.72 (0.95), and 0.59 (0.67), respectively. 
Also,  the phases between the  $x$ and $y$ components of the first, second, and third sidebands of the $\textsf{iLO}$ ($\textsf{iTO}$) mode are 85$^{\circ}$ (90$^{\circ}$), 70$^{\circ}$ (135$^{\circ}$), and 60$^{\circ}$ (75$^{\circ}$), respectively. Thus, the analysis of Fig.~\ref{polarization_side} establishes that the polarization and the phase properties of the sidebands are different for different phonon modes. However, it is not obvious why the sidebands have nonzero $x$ and $y$ components, whereas a particular phonon mode ($\textsf{iLO}$ or $\textsf{iTO}$)  induces  atomic vibrations along  a particular direction ($\textsf{X}$ or $\textsf{Y}$).

To know the origin of the nonzero $x$ and $y$ components of the sidebands, we employ Floquet formalism  to graphene with  coherently excited phonon mode and  circularly polarized probe pulse. In the present case,  $\mathcal{D}_1 = \hat{\sigma}_y \cdot \mathcal{T}$ is the DS, which leaves the system with $\textsf{iLO}$ phonon mode invariant.  Here, 
$\hat{\sigma}_{y}$ is the reflection with respect to $y$-axis, and ${\mathcal{T}}$ is the time-reversal operator. 
Thus, the condition associated with the sidebands is determined as ${\mathcal{D}_1}\mathcal{R}_m(t) = \mathcal{R}_m(t)$. 
Let us substitute the expression of   $\mathbf{E}^\dagger(t) =  [\cos(\omega_{0} t) E^{\dagger}_{x}~~\sin(\omega_{0} t) E^{\dagger}_{y}]$.  The expression of the Raman tensor after substitution will read as
\begin{equation}
\mathcal{R}_{m}(t) = \sin[(m \omega_{\textrm{ph}}+\omega_{0})t]  
 \begin{bmatrix}
\cos(\omega_{0} t)	E_{s,m_{x}} \\
\sin(\omega_{0} t)	E_{s,m_{y}}	
\end{bmatrix}.
\label{eq3}
\end{equation}  
On operating $\mathcal{D}_1$ on $\mathcal{R}_{m}(t)$ leads the following expression of the invariant 
$\mathcal{R}_m(t)$ as
\begin{equation}
\sin[(m\omega_{\textrm{ph}}+\omega_{0})t] 
 \begin{bmatrix}
\cos(\omega_{0}t)	E_{s,m_{x}} \\
\sin(\omega_{0}t)	E_{s,m_{y}}
\end{bmatrix}
= \sin[-(n \omega_{\textrm{ph}}+\omega_{0})t]  
\begin{bmatrix}
-\cos(\omega_{0}t)	E_{s,m_{x}} \\
-\sin(\omega_{0}t)	E_{s,m_{y}}
\end{bmatrix}.
\label{selm2}
\end{equation}
Now it is straightforward  to notice that all the sidebands exhibit nonzero $x$ and $y$ components.  The ellipticity and the phase difference are estimated from the nonzero components and  our numerical results shown in Fig.~\ref{polarization_side} are consistent with the present analysis.

\section{Summary}
In conclusion, we have investigated the potential of high-harmonic spectroscopy to probe the impact of coherent lattice dynamics on electron dynamics in solids, with a focus on graphene. We have specifically explored the coherent excitation of in-plane Raman-active phonon modes, namely $\textsf{iLO}$ and $\textsf{iTO}$ modes.
Our findings reveal that high-harmonic spectroscopy offers valuable insights into the phase difference and 
``chirality'' of the excited phonon mode, allowing for a comprehensive understanding of the 
temporarily-evolving  system's properties. 
Importantly, after the excitation of circular phonon modes, graphene exhibits a piezo-optical effect, i.e., different responses to left-  and right-handed circularly polarized pulses.~\citep{tamaya2021piezo} Additionally, coherent phonons give rise to sidebands corresponding to prominent harmonic peaks 
and the positions of the sidebands are determined by the energy of the excited phonon modes. 
To further elucidate the properties of the sidebands, we apply the Floquet formalism to analyze the dynamical symmetries of the system. 
Furthermore, the polarization and ``chirality'' of the sidebands are dictated by the dynamical symmetries of the system, which consist of graphene with the excited phonon modes and the probe pulse. 
Therefore, the polarization properties of the sidebands serve as a sensitive probe of the system's dynamical symmetries. 

The six-fold symmetry of the graphene reduces to two-fold dynamically due to the coherent phonon excitation of a single phonon mode. This symmetry alteration allows for the generation of symmetry-forbidden harmonics of circularly polarized probe pulses. Overall, our findings demonstrate the potential of high-harmonic spectroscopy as a powerful tool for studying coherent phonon effects and their influence on electron dynamics in solids, paving the way for further advancements in the field.

\cleardoublepage
\chapter{Four-Dimensional Imaging of Lattice Dynamics using Inelastic Scattering}

Inelastic scattering of matter allows us to probe quasi-particles (QPs), such as phonons, magnons (quantized spin excitations), and polarons~\citep{squires1996introduction, Willis2009, schulke2007electron, egerton2011electron}. 
These QPs have finite energy and lifetime that carries information about the intra- and inter-QP interactions and their coupling strength, necessary to understand the materials' response to external stimuli. 
Thermal conductivity~\citep{lindsay2020thermal}, heat capacity~\citep{Dove1993, Fultz2010}, and phase transitions~\citep{Dove1993} are among many of the material properties that are often described by directly invoking various QPs. Optical, x-ray, neutron, and electron scattering methods -- for example, Raman scattering, inelastic x-ray scattering (IXS), inelastic neutron scattering (INS), and electron energy loss spectroscopy (EELS) -- have been routinely employed to measure these QPs~\citep{schulke2007electron, squires1996introduction, egerton2011electron}. 
Typically these measurements are performed in the momentum and energy domains (${\bf k}$-$\omega$), and lack information on temporal dynamics, i.e., ${\bf k}$-$t$ and ${\bf x}$-$t$ imaging -- time evolution of momentum or real-space coordinates, which ranges from femto- to several nano-seconds~\citep{gaffney2007imaging, sciaini2011femtosecond}.

To image ${\bf k}$-$t$ and ${\bf x}$-$t$ dynamics, a pump-probe setup having two ultrashort pulses, where the duration of the probe pulse must be shorter than the characteristic timescale of motion that is under probe, are required. 
Thanks to tremendous technological advancement, it has become possible to generate ultrashort x-ray and electron pulses,~\citep{hartmann2018attosecond, emma2, morimoto2018diffraction} and image the lattice dynamics in ${\bf k}$-$t$ and ${\bf x}$-$t$ domains. 
For example, the time-resolved x-ray and electron diffraction within  pump-probe configurations are used to image the lattice dynamics in the ${\bf k}$-$t$ domain~\citep{trigo2013fourier, clark2013ultrafast, elsaesser2014perspective, bredtmann2014x, fritz2007ultrafast, siwick2003atomic, wall2012atomistic, brown2019direct}, and the ultrafast electron microscopy has recently been demonstrated for  imaging in the ${\bf x}$-$t$ domain at an unprecedented spatiotemporal resolution~\citep{flannigan20124d, cremons2016femtosecond, fu2017imaging}. 
However, similar advances have not been taken for neutron sources to produce an ultrashort 
neutron pulse for imaging the lattice dynamics~\citep{pomerantz2014ultrashort}. 
At this juncture, it is not straightforward whether one can employ neutron sources within a pump-probe setup with sufficient atomic-scale spatiotemporal resolution to image lattice or spin dynamics ~\citep{pomerantz2014ultrashort}.

In this chapter, we  demonstrate that methods based on inelastic scattering are suitable to extract similar information as one could get from the time-resolved imaging of lattice dynamics in 
${\bf k}$-$t$ or ${\bf x}$-$t$ domains. 
Our approach is general and equally applicable to IXS, INS, and EELS. In general, all these inelastic scattering-based methods probe dynamical structure factor $\mathsf{S}(\mathbf{k}, \omega)$ in experiments. The inelastic scattering methods provide ${\bf k}$-$\omega$ resolved measurement of QPs and comprise a powerful way to investigate the correlated motion of atoms and electrons~\citep{schulke2007electron, squires1996introduction, egerton2011electron}. 
We should mention that in ideal conditions, irrespective of whether measurements are in ${\bf k}$-$\omega$, ${\bf k}$-$t$, or ${\bf x}$-$t$ domains, they provide similar information after coordinate transformation(s). 
However, in practice, one measurement domain may have an advantage over the other. For example, under static environmental conditions of temperature, pressure, or magnetic field, the four-dimensional (4D) ${\bf k}$-$\omega$ mapping of QPs is preferred because of its superior energy and momentum resolutions ($\sim$0.1\,meV and $\sim$0.5\,nm$^{-1}$)~\citep{Willis2009, Ehlers, ARCS, HERIX3}, from which one can readily extract the QP energy, group velocity, and linewidth. 
On the other hand, the ${\bf k}$-$t$ domain is useful for tracking the temporal evolution of atomic motions upon photoexcitation-induced structural phase transitions~\citep{wall2018ultrafast} or the measurement of long-wavelength phonon lifetime (of the order of tens of picoseconds, which is not easily accessible in the ${\bf k}$-$\omega$ domain). 
Moreover, mapping the acoustic phonon wavefronts or the nucleation of waves from defects and interfaces in nanostructures is better suited for the ${\bf x}$-$t$ domain~\citep{cremons2016femtosecond, cremons2017defect}. 

In the following, we show that $\mathsf{S}(\mathbf{k}, \omega)$ (obtained from experimental measurements or simulations)
encodes all the essential information to image the lattice dynamics in the ${\bf k}$-$t$ and ${\bf x}$-$t$ domains after coordinate transformation without the causality violation. 
In particular, as we illustrate, our approach is well-suited to image the first-order states [i.e., emission or absorption of a single phonon at ${\bf q}\simeq 0$ from inelastically scattering photons or disorder-activated continuum~\citep{carles1982new}] and second-order `squeezed' states~\citep{trigo2013fourier} in the ${\bf k}$-$t$ domain (squeezed states are generated in the entire reciprocal lattice immediately after pumping the sample with a visible or near-infrared pump pulse)~\citep{trigo2013fourier, henighan2016control}. 
The temporal evolution and decay of the measured intensity from the change in phonon occupation at a given ${\bf k}$ point due to electron-electron, electron-phonon, and phonon-phonon scattering channels~\citep{trigo2013fourier,stern2018mapping,murphy2019evolution} are not explicitly included within the current framework. 
However, we consider the finite lifetime of first- or second-order states by including the phonon linewidths. 
Moreover, our methodology allows for ${\bf x}$-$t$ imaging of the coherent phonon dynamics from a point-like nucleation site or an extended defect. 
Our approach to imaging dynamics in the ${\bf x}$-$t$ domain can be directly compared with the electron microscopy data, as we demonstrate later by an example. 

\section{Theoretical Framework}\label{5:2}

\subsection{Dynamical Structure Factor}\label{5:2.1}
The dynamical structure factor $\mathsf{S}(\mathbf{k}, \omega)$ is calculated using the following expression:
\begin{footnotesize}
	\begin{align} \label{eq:DDCS1}
		\mathsf{S}(\mathbf{k}, \omega) \propto & \sum_{s}\sum_{\mathbf{\tau}} \frac{1}{\omega_{s}}  \left| \sum_{d} \frac{{f_d({\bf k})}}{\sqrt{M_d}} \rm{exp}(-W_d)\rm{exp}(i\mathbf{k}\cdot\mathbf{d})(\mathbf{k}\cdot\mathbf{e}_{ds})\right|^2\times \langle n_s  + \frac{1}{2} \pm \frac{1}{2}\rangle \delta(\omega \mp \omega_s)\delta(\mathbf{k}-\mathbf{q} - \tau),
	\end{align}
\end{footnotesize}
where $f_d({\bf k})$ is the form factor for atom $d$ (can be replaced by the neutron scattering length $\overline{b_d}$ for inelastic neutron scattering), $\mathbf{k} = \mathbf{k}' - \mathbf{k}''$ is the wavevector or momentum transfer, and $\mathbf{k}''$ and $\mathbf{k}'$ are the final and incident wavevector of the scattered particle, respectively; $\textbf{q}$ is the phonon wavevector, $\omega$ is the energy transfer of x-ray, $\omega_s$ is the eigenvalue and 
$\mathbf{e}_{ds}$ is the eigenvector of the phonon corresponding to the branch index $s$, $d$ is the atom index in the unit cell, $\tau$ is the reciprocal lattice vector, $\exp(-2W_d)$ is corresponding to the Debye-Waller factor, and $n_s = \left[\exp\left(\frac{\hbar\omega_s}{k_{\rm B}T}\right)-1\right]^{-1}$ is the Bose-Einstein occupation factor ($k_{\rm B}$ is the Boltzmann's constant). The $+$ and $-$ signs in Eq.~\eqref{eq:DDCS1} correspond to phonon creation and phonon annihilation, respectively. 
The phonon eigenvalues and eigenvectors [simulation details are same as presented in Refs.~\cite{FBao_2016_1}, and~\cite{FBao_2016_2}] in Eq.~\eqref{eq:DDCS1} were obtained by solving dynamical matrix using Phonopy~\citep{Phonopy_2015}. 
Phonon dispersion curves, $\mathsf{S}(\mathbf{k}, \omega)$ and constant energy slices of $\mathsf{S}(\mathbf{k}, \omega)$ of silicon and germanium are shown in Figs.~\ref{silicon_dispersion} - \ref{Ge_sqw_HK0}.

\begin{figure}
\begin{center}
\includegraphics[width=0.7\linewidth]{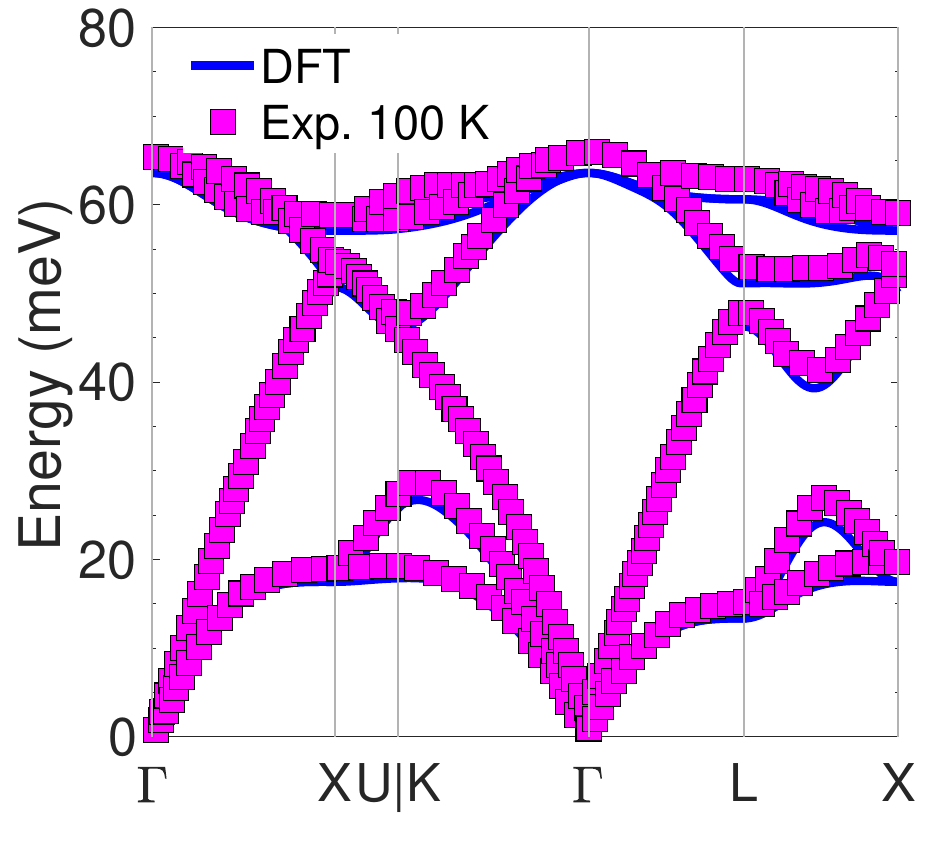}
\end{center}
\caption{Phonon dispersion of silicon along high-symmetry directions calculated using density functional theory (DFT) simulation (solid blue lines) and compared with the experimental data at 100\, K (magenta squares)~\citep{kim2018nuclear}.} \label{silicon_dispersion}
\end{figure}

\begin{figure}
	\begin{center}
		\includegraphics[width=0.8\linewidth]{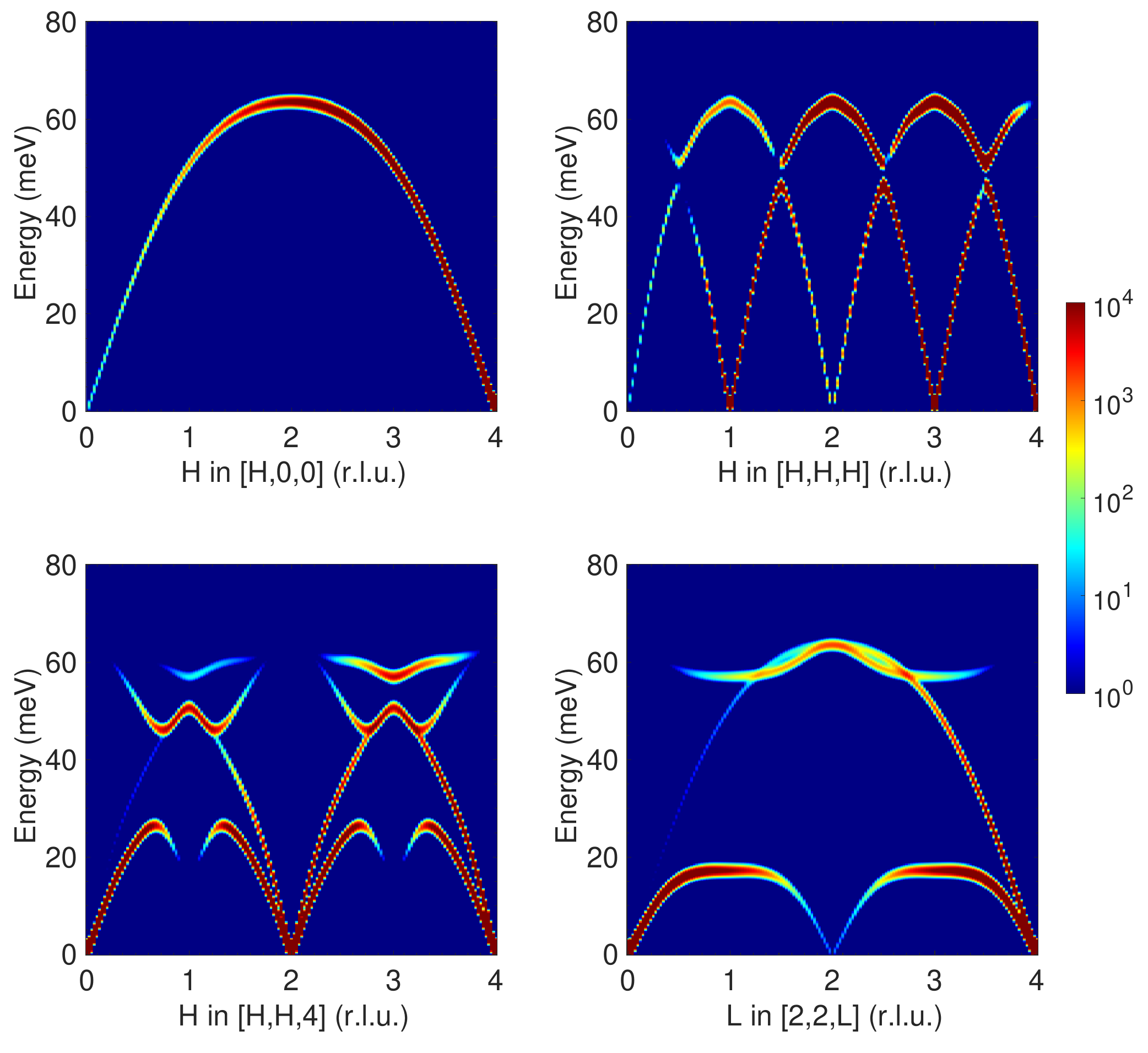}
	\end{center}
	\caption{Dynamical structure factor $\mathsf{S}(\mathbf{k}, \omega)$ of silicon along different high symmetry directions. All slices can be directly compared with single-crystal inelastic neutron/x-ray scattering measurements along the same reciprocal space directions.}
	\label{silicon_sqw}
\end{figure}

\begin{figure}
	\begin{center}
		\includegraphics[width=0.8\linewidth]{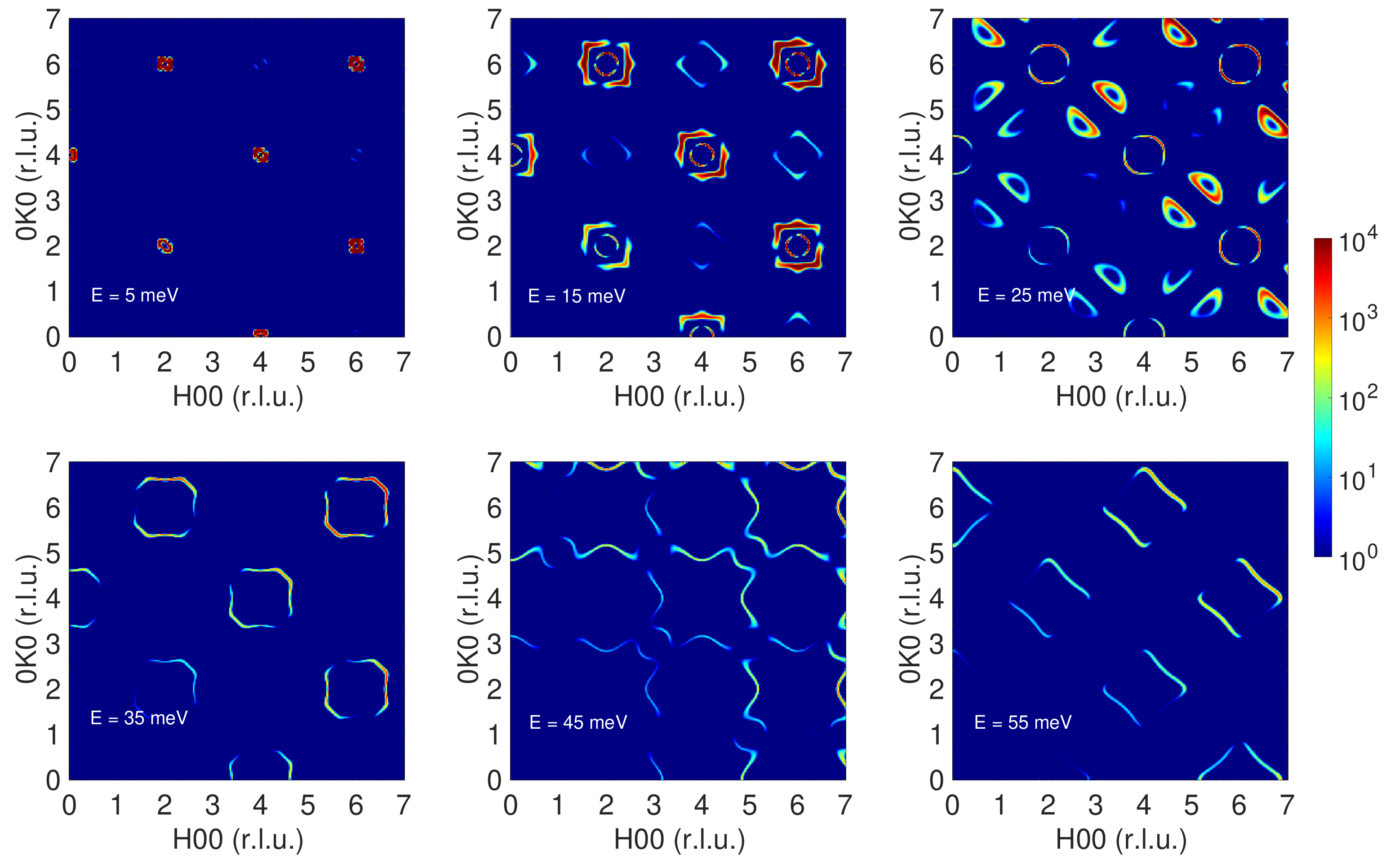}
	\end{center}
	\caption{Constant energy slices of dynamical structure factor $\mathsf{S}(\mathbf{k}, \omega)$ of silicon in the $(H,H,L)$ reciprocal plane. Energy is shown in each of the panels.}
	\label{silicon_sqw_HHL}
\end{figure}

\begin{figure}
	\begin{center}
		\includegraphics[width= \linewidth]{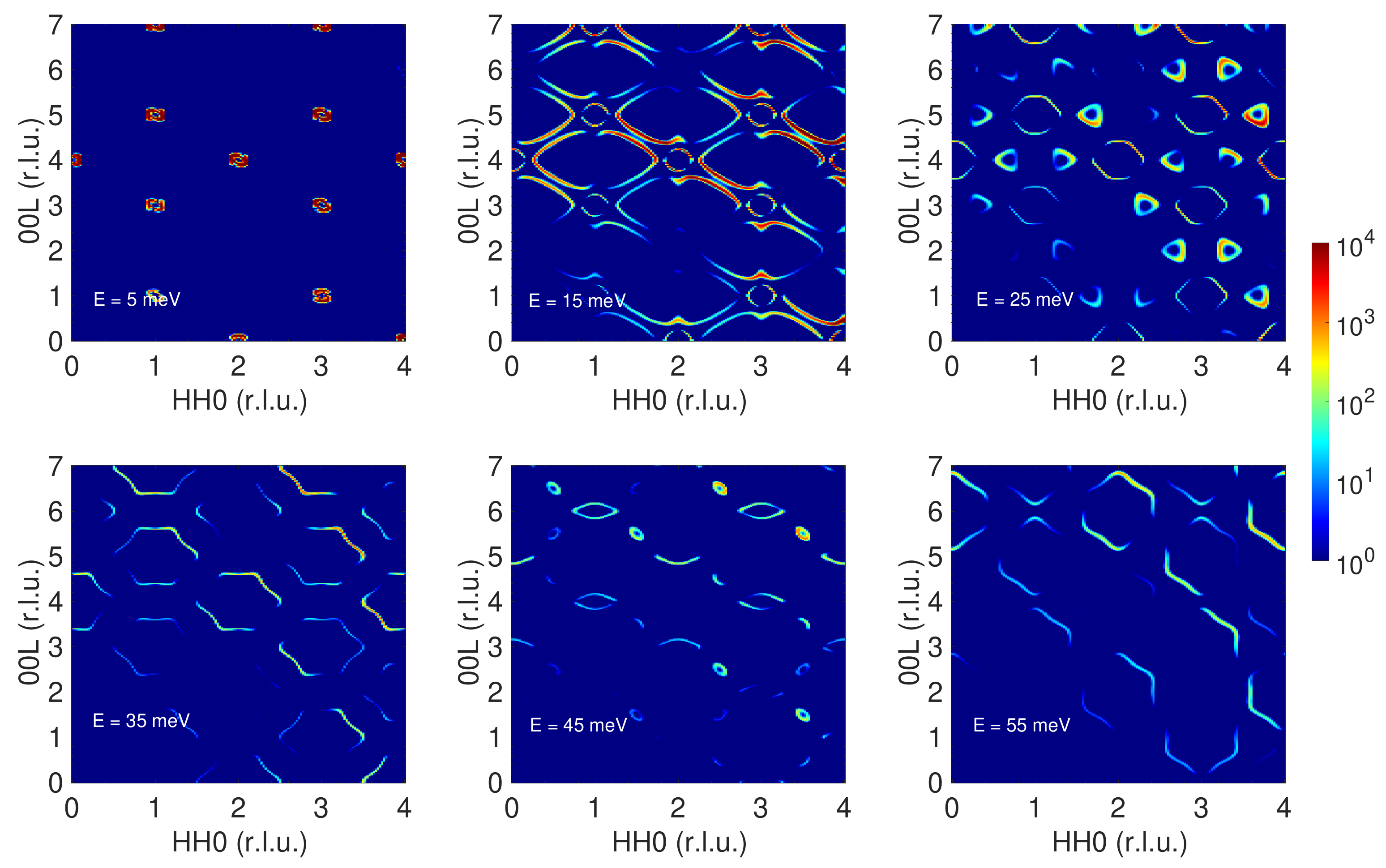}
	\end{center}
	\caption{Same as Fig.~\ref{silicon_sqw_HHL} for $(H,K,0)$ reciprocal plane.}
	\label{silicon_sqw_HK0}
\end{figure}

\begin{figure}
	\begin{center}
		\includegraphics[width= \linewidth]{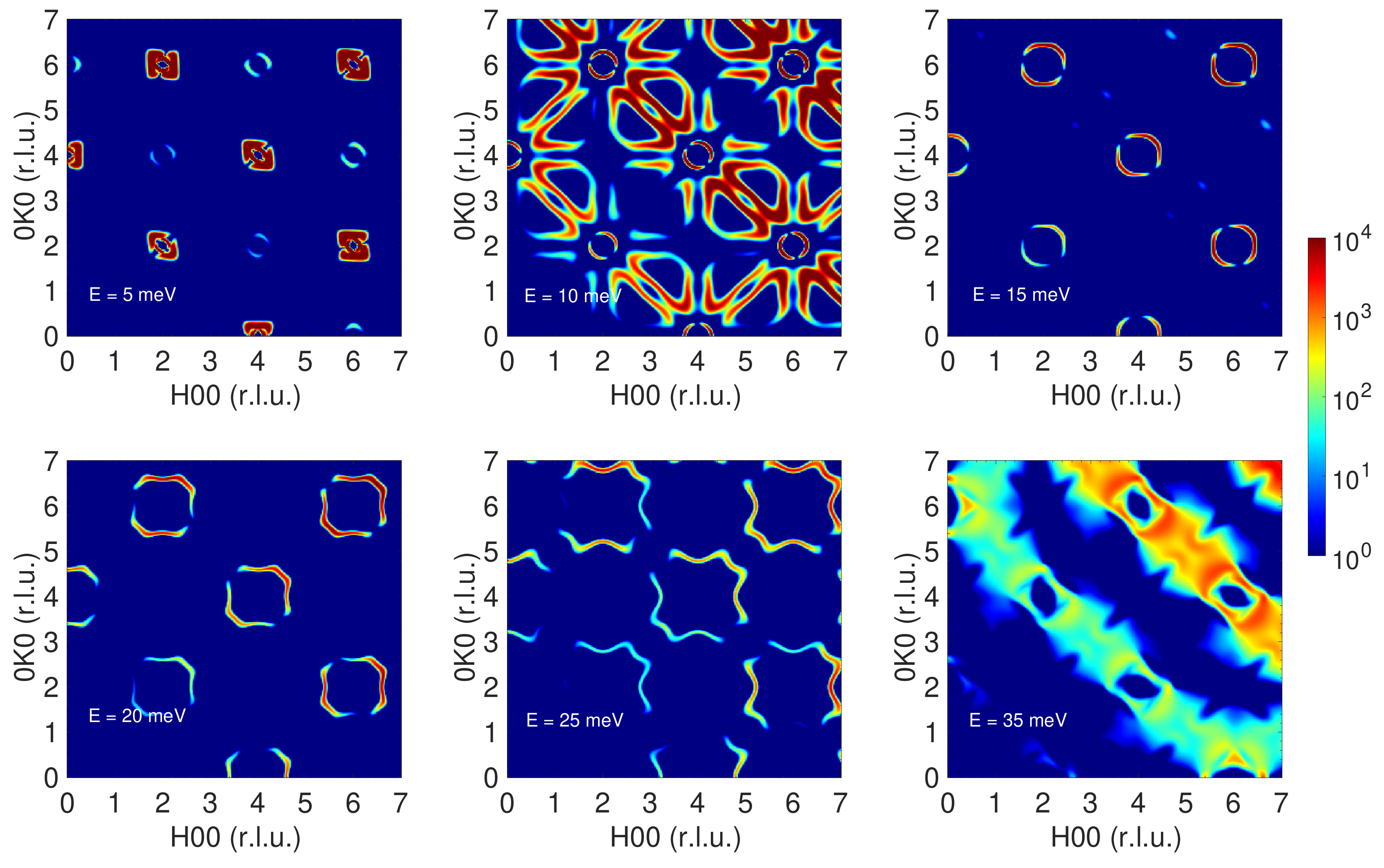}
	\end{center}
	\caption{ Same as Fig.~\ref{silicon_sqw_HHL} for germanium in the $(H,H,L)$ reciprocal plane. }
	\label{Ge_sqw_HHL}
\end{figure}

\begin{figure}
	\begin{center}
		\includegraphics[width= \linewidth]{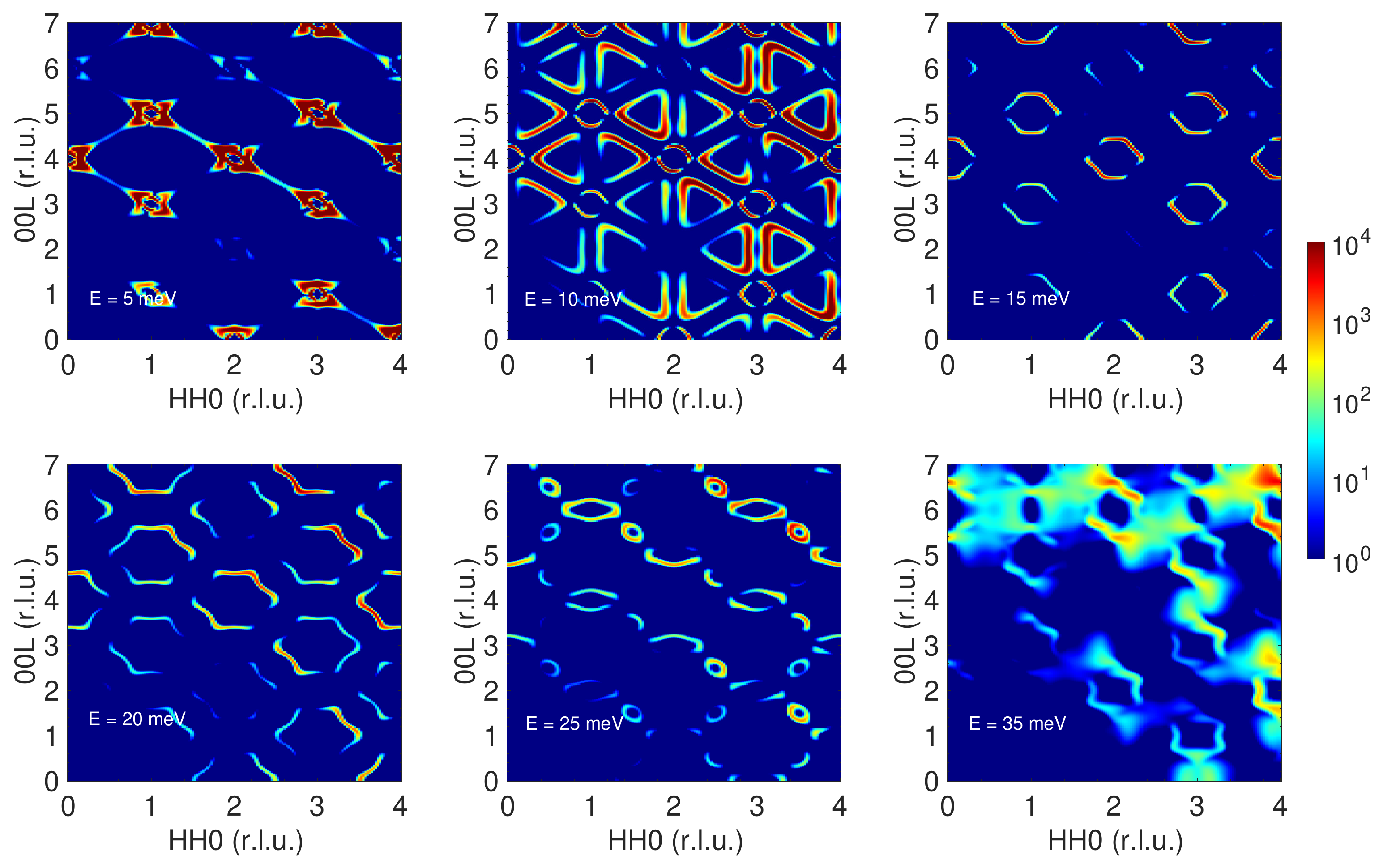}
	\end{center}
	\caption{Same as Fig.~\ref{silicon_sqw_HHL} for germanium in the $(H,K,0)$ reciprocal plane.}
	\label{Ge_sqw_HK0}
\end{figure}

\subsection{Equal-Time Correlation Function}\label{5:2.2} 
Let us write the expression of the dynamical structure factor for inelastic x-ray scattering as~\citep{schulke2007electron}
\begin{equation}\label{eqs1}
\mathsf{S}(\mathbf{k}, \omega) =  \sum_{j} \delta(E_{i}-E_{j}+\omega) \left[ \int d\mathbf{x} \int d\mathbf{x^{\prime}} \, 
\langle \Psi_{i} | \hat{n}(\mathbf{x^{\prime}} ) | \Psi_{j} \rangle \langle \Psi_{j} | \hat{n}(\mathbf{x}) | \Psi_{i} \rangle \, 
e^{i \mathbf{k} \cdot (\mathbf{x} - \mathbf{x^{\prime}})} \right], 
\end{equation}
where  $ | \Psi_{i} \rangle $ and $ | \Psi_{j} \rangle $ are eigenstates of the system under probe with 
$E_{i}$ and $E_{j}$ are corresponding eigenenergies, respectively, and $\hat{n}(\mathbf{x})$ is the density operator~\citep{dixit2014theory, dixit2012imaging}. After performing  the spatial Fourier transform, Eq.~\eqref{eqs1} reduces to  
\begin{equation}\label{eqs2}
S(\mathbf{k}, \omega) =  \sum_{j} \delta(E_{i}-E_{j}+\omega) \left[  
\langle \Psi_{i} | \hat{n}(-\mathbf{k}) | \Psi_{j} \rangle \langle \Psi_{j} | \hat{n}(\mathbf{k}) | \Psi_{i} \rangle \right].
\end{equation}
Let us express the delta function in terms of the Fourier transform as
\begin{eqnarray}\label{eqs3}
\mathsf{S}(\mathbf{k}, \omega) & = &  \frac{1}{2 \pi \hbar} \sum_{j} \int dt \, e^{-i \omega t} e^{i (E_{j}-E_{i}) t} \left[  
\langle \Psi_{i} | \hat{n}(-\mathbf{k}) | \Psi_{j} \rangle \langle \Psi_{j} | \hat{n}(\mathbf{k}) | \Psi_{i} \rangle \right] \nonumber \\ 
& = &  \frac{1}{2 \pi \hbar} \sum_{j} \int dt \, e^{-i \omega t} \left[  \langle \Psi_{i} | \hat{n}(-\mathbf{k}) | \Psi_{j} \rangle \langle \Psi_{j} | e^{i E_{j} t} \, \hat{n}(\mathbf{k}) \, e^{-i E_{i} t} | \Psi_{i} \rangle \right]  \nonumber \\  
& = &  \frac{1}{2 \pi \hbar} \sum_{j} \int dt \, e^{-i \omega t} \left[  \langle \Psi_{i} | \hat{n}(-\mathbf{k}) | \Psi_{j} \rangle \langle \Psi_{j} | e^{i \mathcal{H} t} \, \hat{n}(\mathbf{k}) \, e^{-i \mathcal{H} t} | \Psi_{i} \rangle \right] 
\end{eqnarray}
where $e^{-i \mathcal{H} t} | \Psi_{i} \rangle = e^{-i E_{i} t} | \Psi_{i} \rangle$ with $\mathcal{H}$ as a field-free Hamiltonian 
of the system. By using the Heisenberg representation of an operator $e^{i \mathcal{H} t} \, \hat{n}(\mathbf{k}) \, e^{-i \mathcal{H} t} = \hat{n}(\mathbf{k}, t)$, Eq.~\eqref{eqs3} simplifies to 
\begin{eqnarray}\label{eqs4}
\mathsf{S}(\mathbf{k}, \omega) & = &  \frac{1}{2 \pi \hbar} \sum_{j} \int dt \, e^{-i \omega t}  \left[  
\langle \Psi_{i} | \hat{n}(-\mathbf{k}, 0) | \Psi_{j} \rangle \langle \Psi_{j} | \hat{n}(\mathbf{k}, t) | \Psi_{i} \rangle \right]  \nonumber \\  
& = &  \frac{1}{2 \pi \hbar}  \int dt \, e^{-i \omega t}  \, 
\langle \Psi_{i} | \hat{n}(-\mathbf{k}, 0)\,  \hat{n}(\mathbf{k}, t) | \Psi_{i} \rangle.   
\end{eqnarray}
Equation~\eqref{eqs4} represents that the dynamical structure factor is a density-density correlation function, which
is related to the Van Hove correlation function~\citep{schulke2007electron, van1954correlations}.

For the energy-integrated inelastic x-ray scattering experiment, we can perform the integration over the scattered x-ray energy. This is equivalent to performing the integration over $\omega$ for a given fixed incident x-ray energy. Therefore, 
\begin{eqnarray}\label{eqs5}
\mathsf{S}(\mathbf{k})  & = &  \int d\omega \, \mathsf{S}(\mathbf{k}, \omega) \nonumber \\
& = &  \frac{1}{2 \pi \hbar}   \int d\omega \, \int dt \, e^{-i \omega t}  \, 
\langle \Psi_{i} | \hat{n}(-\mathbf{k}, 0)\,  \hat{n}(\mathbf{k}, t) | \Psi_{i} \rangle  \nonumber \\
& = & \langle \Psi_{i} | \hat{n}(-\mathbf{k}, 0)\,  \hat{n}(\mathbf{k}, 0) | \Psi_{i} \rangle  \nonumber \\
& = & \langle \hat{n}(-\mathbf{k}, 0)\,  \hat{n}(\mathbf{k}, 0) \rangle. 
\end{eqnarray}
After performing the energy integration, $\mathsf{S}(\mathbf{k})$ is the density-density correlation function at time zero. 
This can be also seen as the density-density correlation function at the equal time, i.e., the equal-time correlation function 
$\langle \hat{n}(-\mathbf{k}, t)\,  \hat{n}(\mathbf{k}, t) \rangle$~\citep{sinha2001theory}. 
Trigo \emph{et al.} had probed the equal-time correlation function using energy-integrated  
time-resolved diffuse x-ray scattering within pump-probe configuration~\citep{trigo2013fourier}.

\subsection{Connection between Response Function and Energy-Integrated Dynamical Structure Factor}\label{5:2.3} 
The dynamical structure factor is related to the imaginary part of the response function via Fluctuation-Dissipation theorem as~\citep{schulke2007electron} 
\begin{equation}\label{eqs6}
\Im[\chi (\mathbf{k}, \omega)] = -\pi [\mathsf{S}(\mathbf{k}, \omega) - \mathsf{S}(\mathbf{k}, -\omega)].
\end{equation}
Using $\mathsf{S}(\mathbf{k}, -\omega) = e^{-\beta \hbar \omega}\, \mathsf{S}(\mathbf{k}, \omega)$, above equation simplifies as 
\begin{equation}\label{eqs7}
\Im[\chi (\mathbf{k}, \omega)] = -\pi [1 - e^{-\beta \hbar \omega}] \mathsf{S}(\mathbf{k}, \omega), 
\end{equation}
where $\beta = (k_{\rm B}T)^{-1}$. The total response function is the sum of real and imaginary parts and can be written as $\chi (\mathbf{k}, \omega) = \Re[\chi (\mathbf{k}, \omega)]+ i\, \Im[\chi (\mathbf{k}, \omega)]$ 
where $\Re[\chi (\mathbf{k}, \omega)]$  is obtained  via 
Kramers-Kronig relation as~\citep{jackson2007classical} 
\begin{equation}\label{eqs11}
\Re[\chi(\mathbf{k},\omega)] = \frac{1}{\pi} \mathcal{P} \int_{-\infty} ^{\infty} d \omega^{\prime}~ 
\frac{\Im[\chi(\textbf{k},\omega ^{\prime})]}{(\omega^{\prime}-\omega)}.  
\end{equation} 
Here, $\mathcal{P}$ represents the principal value of the integral.  
Using Fourier relation, we can transform $\chi(\mathbf{k},\omega)$ from the $\mathbf{k}$-$\omega$ domain to the $\mathbf{k}$-$t$ domain as following
\begin{equation}\label{eqs8}
\chi (\mathbf{k}, t) = \int d\omega e^{i \omega t}\, \chi (\mathbf{k}, \omega) = \int d\omega e^{i \omega t}\, \left\{\Re[\chi(\mathbf{k},\omega)] + i\, \Im[\chi (\mathbf{k}, \omega)]\right\} = \textrm{I}_{2} + i\, \textrm{I}_{1}. 
\end{equation}
Let us consider $\textrm{I}_{1}  = \int d\omega e^{i \omega t}\, \Im[\chi (\mathbf{k}, \omega)]$, 
which can be simplified using Eq.~\eqref{eqs7} as
\begin{equation}\label{eqs9}
\textrm{I}_{1} =  -\pi \int d\omega e^{i \omega t}\, [1 - e^{-\beta \hbar \omega}] \mathsf{S}(\mathbf{k}, \omega). 
\end{equation}
In the following, we will ignore the pre-factor, $ -\pi [1 - e^{-\beta \hbar \omega}]$, as it will not affect our interpretation. 
Now using Eq.~\eqref{eqs4}, $\textrm{I}_{1}$ can be written as
\begin{eqnarray}\label{eqs10}
\textrm{I}_{1} & = &  \frac{1}{2 \pi \hbar} \int d\omega \int dt^{\prime} e^{i \omega t}\,  e^{-i \omega t^{\prime}}\,
\langle  \hat{n}(-\mathbf{k}, 0)\,  \hat{n}(\mathbf{k}, t^{\prime}) \rangle \nonumber \\
& = &  \frac{1}{2 \pi \hbar}  \int d\omega \int dt^{\prime} e^{i \omega (t-t^{\prime})}\,  \langle  \hat{n}(-\mathbf{k}, 0)\,  \hat{n}(\mathbf{k}, t^{\prime}) \rangle \nonumber \\
& = &  \int dt^{\prime} \delta(t-t^{\prime})\,  \langle  \hat{n}(-\mathbf{k}, 0)\,  \hat{n}(\mathbf{k}, t^{\prime}) \rangle \nonumber \\
& = &  \langle  \hat{n}(-\mathbf{k}, 0)\,  \hat{n}(\mathbf{k}, t) \rangle. 
\end{eqnarray}
Next we consider $\textrm{I}_{2}  = \int d\omega e^{i \omega t}\, \Re[\chi (\mathbf{k}, \omega)]$, 
which is simplified using Eq.~\eqref{eqs11} as
\begin{equation}\label{eqs12}
\textrm{I}_{2}  =  \frac{1}{ \pi} \int d\omega e^{i \omega t}\,  \mathcal{P} \int_{-\infty} ^{\infty} d\omega^{\prime}\,  
\frac{\Im[\chi(\mathbf{k}, \omega^{\prime})]}{(\omega^{\prime} - \omega)}. 
\end{equation}
By following the similar procedure as done for $\textrm{I}_{1}$, Eq.~\eqref{eqs12} is written as 
\begin{eqnarray}\label{eqs13}
\textrm{I}_{2} & = &   \frac{1}{\pi} \frac{1}{2 \pi \hbar} \int d\omega\,  \mathcal{P}\int_{-\infty} ^{\infty} d\omega^{\prime} \int dt^{\prime} 
\frac{1}{(\omega^{\prime} - \omega)}  e^{i(\omega t - \omega^{\prime} t^{\prime})} 
\langle  \hat{n}(-\mathbf{k}, 0)\,  \hat{n}(\mathbf{k}, t^{\prime}) \rangle \nonumber \\ 
& = &   \frac{1}{\pi} \frac{1}{2 \pi \hbar} (i \pi ) \int d\omega \int dt^{\prime} 
  e^{i \omega (t -  t^{\prime})} 
\langle  \hat{n}(-\mathbf{k}, 0)\,  \hat{n}(\mathbf{k}, t^{\prime}) \rangle.
\end{eqnarray}
To simplify Eq.~\eqref{eqs13}, we make use of the following result
\begin{equation}\label{eqs14}
 \mathcal{P}\, \int_{-\infty} ^{\infty} d\omega^{\prime} \frac{1}{(\omega^{\prime} - \omega)} e^{- i  \omega^{\prime} t^{\prime}}  = 
i \pi  e^{- i  \omega t^{\prime}},
\end{equation} 
and follow the similar steps as done for Eq.~\eqref{eqs9} to obtain
\begin{equation}\label{eqs15}
\textrm{I}_{2}   =  i   \langle  \hat{n}(-\mathbf{k}, 0)\,  \hat{n}(\mathbf{k}, t) \rangle. 
\end{equation}
Combining the results of Eqs.~\eqref{eqs10} and~\eqref{eqs15}, we can write
\begin{equation}\label{eqs16}
  \chi (\mathbf{k}, t)  =   \textrm{I}_{2} + i\, \textrm{I}_{1} = 2i\, \langle  \hat{n}(-\mathbf{k}, 0)\,  \hat{n}(\mathbf{k}, t) \rangle. 
\end{equation}
On comparing Eq.~(\ref{eqs4}) with the 
above equation, it is evident that $\chi (\mathbf{k}, t)$ is a complex quantity 
as $\langle  \hat{n}(-\mathbf{k}, 0)\,  \hat{n}(\mathbf{k}, t) \rangle$ is a Fourier transform of $\mathsf{S}(\mathbf{k}, \omega)$. 
Using the Heisenberg representation of the operator, as used in Eq.~(\ref{eqs3}),  Eq.~(\ref{eqs16}) is re-written as
\begin{equation}\label{eqs17}
  \chi (\mathbf{k}, t)  =  2i\, \sum_{j}\, \langle \Psi_{i} |  \hat{n}(-\mathbf{k}, 0) | \Psi_{j} \rangle \langle \Psi_{j} | \hat{n}(\mathbf{k}, 0) | \Psi_{i} \rangle\, e^{i(E_{j}-E_{i})t}. 
\end{equation}
Under the conditions in which $E_{j}-E_{i} \sim \omega $ is of the order of meV and the time duration of the dynamics is of the order of picoseconds, Eq.~\eqref{eqs17} simplifies to  
\begin{equation}\label{eqs18}
  \chi (\mathbf{k}, t)  =  2i\, e^{i \omega t}\, \sum_{j}\, \langle \Psi_{i} |  \hat{n}(-\mathbf{k}, 0) | \Psi_{j} \rangle \langle \Psi_{j} | \hat{n}(\mathbf{k}, 0) | \Psi_{i} \rangle\, = 2i\, e^{i \omega t}\, \langle  \hat{n}(-\mathbf{k}, 0) \hat{n}(\mathbf{k}, 0)  \rangle. 
\end{equation}
Therefore, $ \chi (\mathbf{k}, t)$ is proportional to the equal-time density-density correlation function, which is the  energy-integrated dynamical structure factor as shown in Eq.~(\ref{eqs5}). 
$ \chi (\mathbf{k}, t)$ represents a selected phonon branch in Eq.~\eqref{eqs18} as it depends on a particular value of $\omega$. 
Eq.~\eqref{eqs8}  is simplified to Eq.~\eqref{eqs18} to show the correspondence between the response function and the energy-integrated dynamical structure factor.

\subsection{Computational Details}\label{5:3}

Silicon is used in  this chapter to demonstrate our idea of 4D imaging. 
The $\mathsf{S}(\mathbf{k}, \omega)$ is simulated in the $({H}, {H}, {L})$ reciprocal plane following the same procedure as in our previous studies~\citep{FBao_2016_1,FBao_2016_2}. 
The range of energy transfer lies from  0 to 80\,meV with a step size  of 0.25\,meV (i.e., energy resolution), whereas the momentum transfer range varies from $(0,0,0)$ to $(4,4,7)$ reciprocal lattice units (r.l.u.)  with the step size of 0.025\,r.l.u. (see Fig.~\ref{silicon_sqw_HHL}). Here, $a$ = 0.543\,nm is used as the lattice parameter of silicon. 
After calculating $\mathsf{S}(\mathbf{k}, \omega)$ and  using the fluctuation-dissipation theorem, the imaginary part of the response function $\chi(\mathbf{k}, \omega)$ is obtained as, $\Im[\chi(\mathbf{k}, \omega)] = -\pi[\mathsf{S}(\mathbf{k}, \omega) - \mathsf{S}(\mathbf{k}, -\omega)]$. 
$\mathsf{S}(\mathbf{k}, \omega)$ and $\mathsf{S}(\mathbf{k}, -\omega)$ are related to each other by $\mathsf{S}(\mathbf{k}, -\omega) = \textrm{exp}(-\beta \hbar \omega)~\mathsf{S}(\mathbf{k}, \omega)$~\citep{schulke2007electron}. 

Fourier transform is performed to obtain $\chi(\mathbf{k}, t)$ from $\chi(\mathbf{k},\omega)$, which is complex in nature as discussed in section~\ref{5:2.3}. Equation~\eqref{eqs11} ensures that $\chi(\mathbf{k}, t)$ = 0 for $t < 0$ and enforces the causality. It should be emphasized that direct Fourier transform of $\mathsf{S}(\mathbf{k}, \omega)$ to $I(\mathbf{k}, t)$, i.e., $I(\mathbf{k}, t) = \hbar\int \mathsf{S}(\mathbf{k},\omega)\exp(i\omega t)\,d\omega$,~\citep{squires1996introduction} is not the time-evolution measured in time-resolved experiments, as $I(\mathbf{k}, t)$ violates causality. 
Note that there is always a limitation in the energy resolution while measuring $\mathsf{S}(\mathbf{k}, \omega)$ in an experiment. The discrete nature of the binned $\omega$ during the experiment causes the periodic nature of $\chi(\mathbf{k}, t)$. The time resolution of 
$\chi(\mathbf{k}, t)$ is estimated by Fourier law as $\Delta t = 2\pi\hbar/(80$\,meV) = 51.5\,fs and the time duration of the lattice dynamics is related to the range of the sampled values of $\omega$. Here, the time duration of the lattice dynamics in silicon is 8.2\,ps. Note that the $\Delta t$ and time duration are proportional to the energy transfer and energy binning (or resolution) and will vary from one experimental setup to another. 
For example, Abbamonte and co-workers have imaged density disturbances in water  with attosecond time resolution ($\Delta t = 41.3$\,attoseconds) with incident x-ray energy and resolution of 100\,eV and 0.3\,eV, respectively~\citep{abbamonte2004imaging}.

\section{Results and Discussion}\label{5:4}
\subsection{Response Function in Momentum Domain}\label{5:4.1}

Figure~\ref{fig1} represents snapshots of the real and imaginary parts of the normalized $\chi(\mathbf{k}, t)$ in the top and bottom rows, respectively. The dynamics can be seen as induced by the point-like source in ${\bf x}$, akin to the nucleation site in electron microscopy imaging. From this point-like source term, silicon absorbs the energy at $t = 0$, and phonon modes are generated in the entire reciprocal space (localized source in ${\bf x}$ is delocalized in ${\bf k}$). The snapshots are shown in the $(H,H,L)$ reciprocal plane for different time instances. We note that a (weak) optical pump pulse will also lead to the same $\chi(\mathbf{k}, t)$ snapshots, as the pump pulse will generate first- and second-order states in the entire reciprocal space. The first-order disorder-activated continuum is generated in the absence of perfect crystalline order~\citep{carles1982new}. In contrast, the second-order squeezed states are generated by the coupling of a photon (momentum ${\bf q}\simeq0$) with the two phonon modes of near-equal and opposite momenta, i.e., at $\mathbf{k}$ and $\mathbf{-k}$ due to the conservation of momentum~\citep{henighan2016control}. Thus in the present chapter, the simulated snapshots can be considered to arise from either a point-like source in ${\bf x}$ or the generation of first- and second-order states. As evident from Fig.~\ref{fig1}, these generated phonons propagate through the system and decay at different times for different $\mathbf{k}$ points according to their lifetime. It is possible to extract the lifetime of these phonon modes at different $\mathbf{k}$ values from the real and imaginary parts of the normalized $\chi(\mathbf{k}, t)$ as we discuss below.

 \begin{figure}
	\begin{center}
		\includegraphics[width=\linewidth]{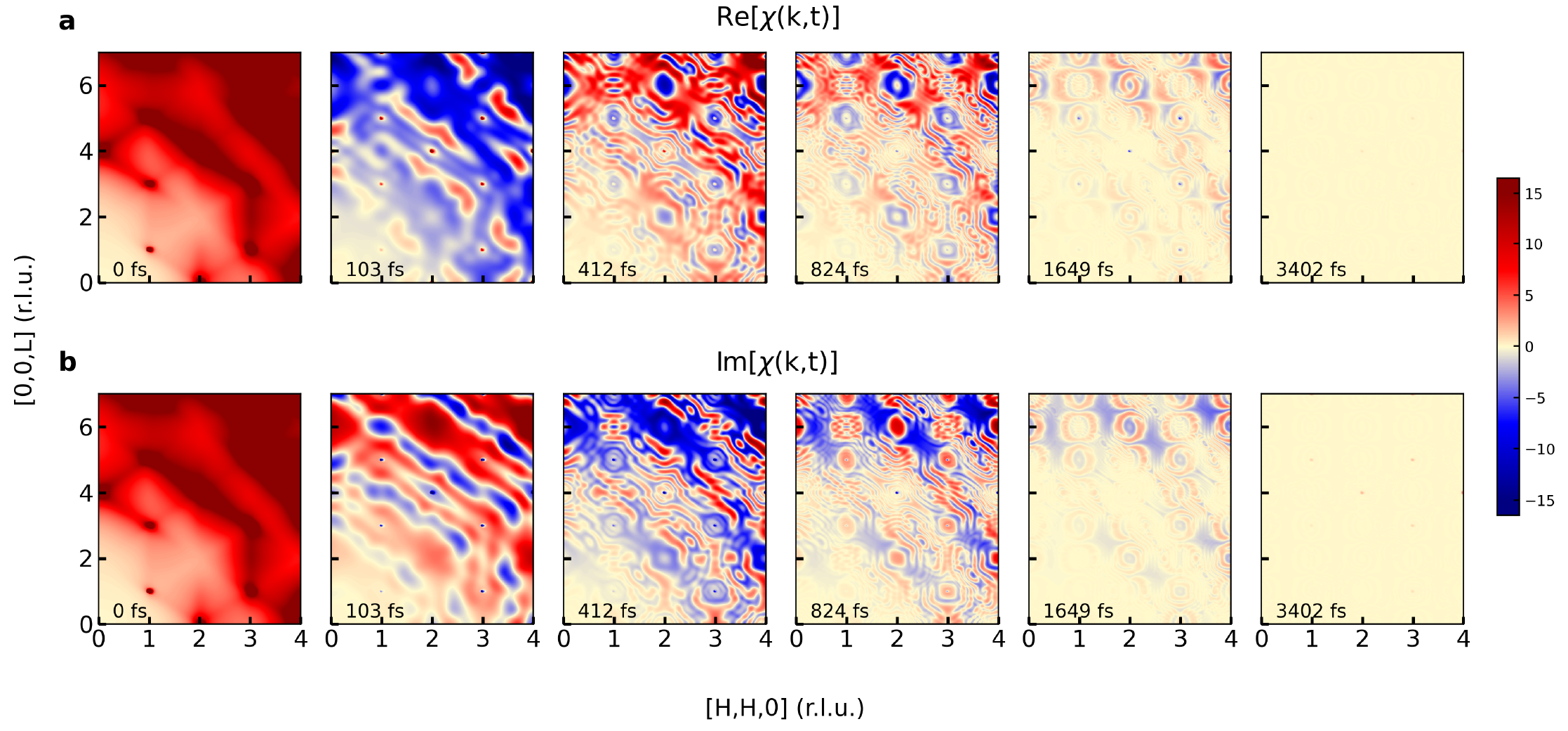}
	\end{center}
	\caption{{Response function	$\chi(\mathbf{k}, t)$ at different instances.} Snapshots of real (in the top row) and  imaginary (in the bottom row) parts  of $\chi(\mathbf{k}, t)$ at 0, 103, 412, 824, 1649 and 
	3402 femtoseconds (fs). $\chi(\mathbf{k}, t)$ is shown in the $(H,H,L)$ reciprocal plane. The contour plots are normalized to their maximum intensity. 
	The lattice dynamics die out at longer time instances.}
	\label{fig1}
\end{figure}

The real and imaginary parts of $\chi(\textbf{k}, \omega)$ at a particular $\mathbf{k} = (0.75, 0.75, 0.75)$ r.l.u.~is presented in Fig.~\ref{fig2}(a). Because of the phonon polarization factor~\citep{squires1996introduction}, at this particular $\mathbf{k}$ value, two phonon modes have finite intensity at energy values $\omega_{1}$ = 28 meV (longitudinal acoustic) 
and $\omega_{2}$ = 60 meV (longitudinal optical). 
The full width at half maximum (FWHM) at these two energy values are $\Gamma_{1} = 0.1$ meV and $\Gamma_{2} = 0.4$ meV [obtained from first-principles simulation of Silicon~\citep{carrete2017almabte}], which are known as decay widths of these modes. To extract the lifetime of these modes, $\chi(\textbf{k}, t)$ is calculated from $\chi(\textbf{k}, \omega)$. The imaginary and real parts of $\chi(\textbf{k}, t)$  are shown in Figs.~\ref{fig2}(b) and ~\ref{fig2}(c), respectively. The lifetime $\tau$ of both the active modes is obtained by fitting the exponentially decaying sinusoidal oscillation at the given $\mathbf{k}$ value, and are $\tau_{1} = 13.7$ and $\tau_{2} = 3.0$\, ps. The fitted $\tau$ values are consistent with the values expected from its inverse relationship with $\Gamma$, i.e., $\Gamma = 1/\pi\tau$, confirming the accuracy of our implementation. Also, the interferences visible in Figs.~\ref{fig2}(b) and \ref{fig2}(c) are due to the presence of two active modes. 

\begin{figure}[h!]
	\begin{center}
	\includegraphics[width=0.8\linewidth]{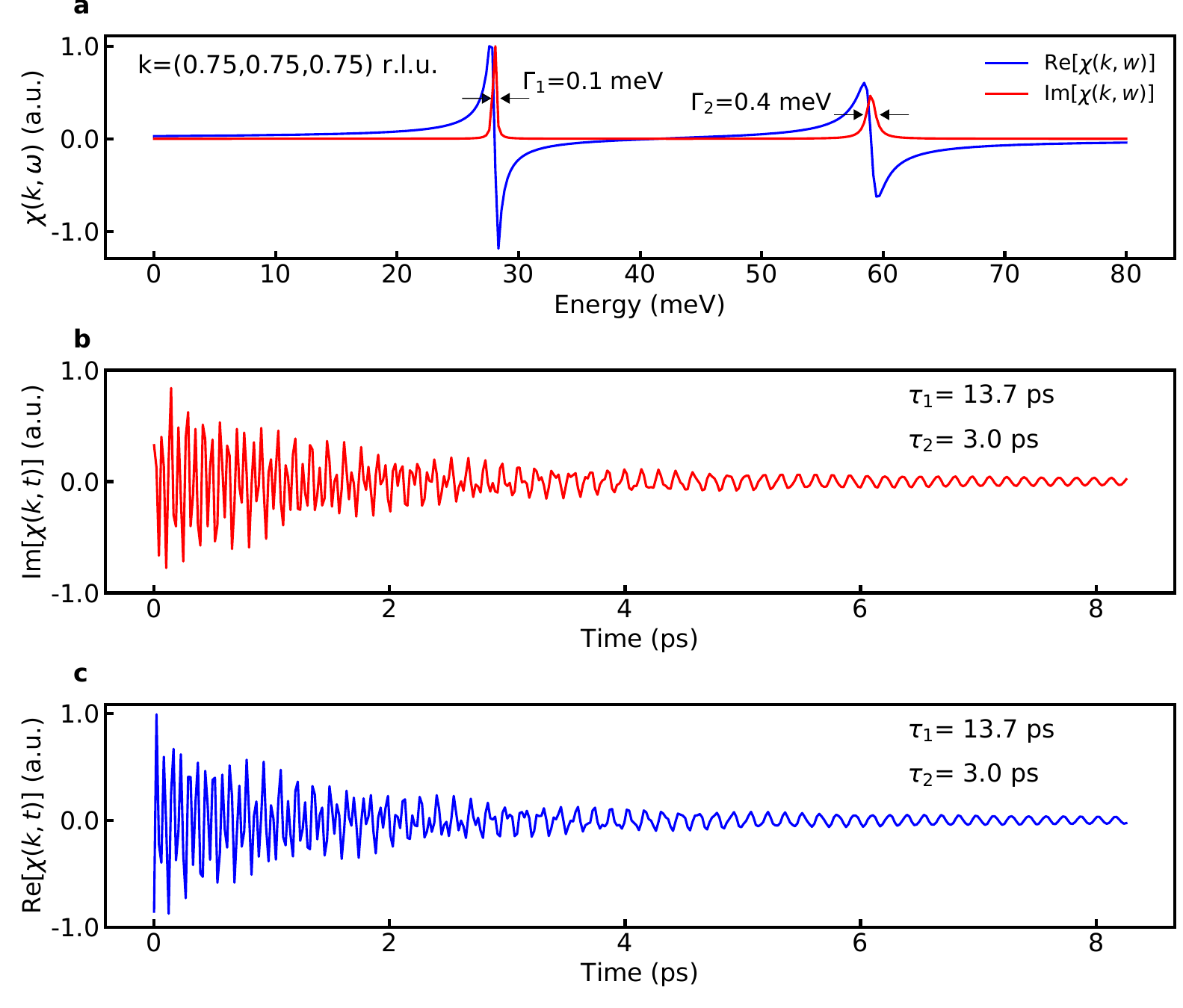}
	\end{center}
	\caption{{Decay width and lifetime of phonon modes at particular $\mathbf{k}$ point.} 
	(a) Real and imaginary parts of $\chi(\textbf{k}, \omega)$ at particular $\mathbf{k} = (0.75, 0.75, 0.75)$\,r.l.u. The decay widths corresponding to two active modes are $\Gamma_{1} = 0.1$\,meV and $\Gamma_{2} = 0.4$\,meV. (b) Imaginary part of $\chi(\textbf{k},t)$ provides the lifetimes of the active phonon modes as $\tau_{1} = 13.7$\,ps  and $\tau_{2} = 3.0$\,ps. (c) The real part of $\chi(\textbf{k},t)$ gives the same values of the lifetimes of the active phonon modes. All the quantities plotted in subplots are normalized.} 
	\label{fig2}
\end{figure}

\begin{figure}[ht]
	\begin{center}
		\includegraphics[width=0.8\linewidth]{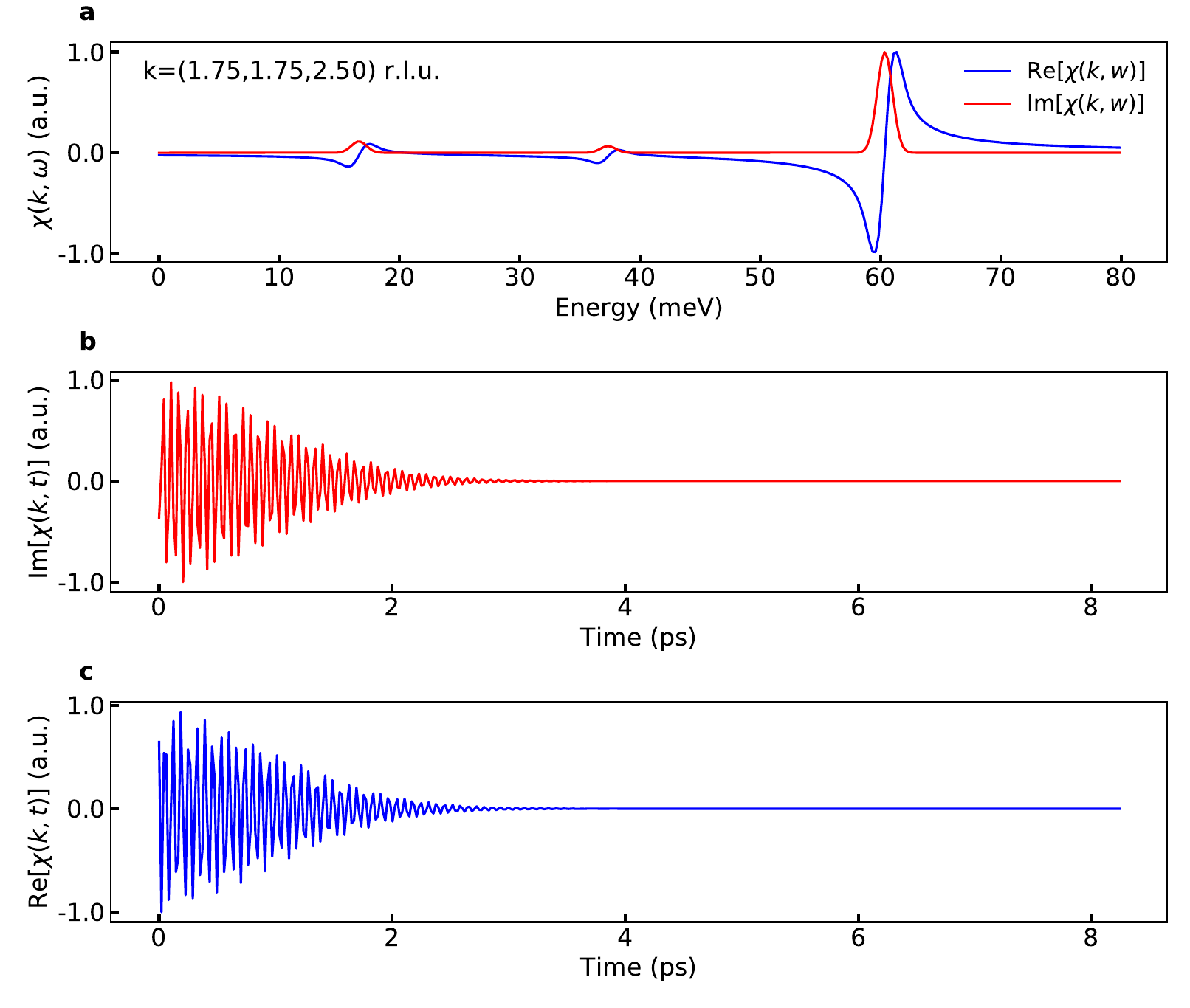}
	\end{center}
	\caption{{Decay width and dynamics of phonon modes excited by an extended source in ${\bf x}$.} 
		(a) Real and imaginary parts of $\chi(\textbf{k}, \omega)$. (b) Imaginary and (c) real parts of $\chi(\textbf{k},t)$. Here, the phonon modes at a specific $\mathbf{k}$ value of $\mathbf{k_0} = (1.75,1.75, 2.50)$\,r.l.u.~are excited by an extended source in ${\bf x}$. All the quantities plotted in subplots are normalized.}
	\label{fig3}
\end{figure}

So far, we have discussed dynamics induced by the point-like disturbance in ${\bf x}$, or first- and second-order states that lead to the excitation of all phonon modes in the reciprocal space. Let us analyze how $\chi(\textbf{k}, t)$ manifests when the extended source in ${\bf x}$ induces the dynamics, which excites the phonon modes active at a single $\mathbf{k}$ value. A general time-dependent extended external source $n_{ext}({\bf x},t)$ can be treated as a Gaussian envelop in the ${\bf k}$-$\omega$ domain with standard deviation $\sigma$ controlling the spatial extent of the source. The induced dynamics $n_{ind}$ can subsequently be modeled as 
\begin{equation}\label{}
n_{ind}(\mathbf{k}, \omega) = \frac{4 \pi}{k^{2}} 
\frac{1}{ \sqrt{2 \pi \sigma^{2}}} ~\textrm{exp}{[{-(\mathbf{k}-\mathbf{k_0})^{2}}/{2 \sigma^{2}}]} 
~\chi(\mathbf{k}, \omega),
\end{equation}
where $\mathbf{k_0}$ is the mean $\mathbf{k}$-value at which phonons are excited. The real and imaginary parts of $\chi(\textbf{k}, \omega)$ at $\mathbf{k_0} = (1.75, 1.75, 2.50)$\,r.l.u.~showing finite intensity for the three phonon modes are shown in Fig.~\ref{fig3}(a). Since momentum transfer ${\bf k}$ and phonon wavevector ${\bf q}$ at $\mathbf{k_0}$ are not entirely parallel or perpendicular to each other, the three modes have mixed transverse and longitudinal character. 
We emphasize that for an infinitely extended source as considered here, dynamics at all other ${\bf k}$ points are zero. As one can observe from Figs.~\ref{fig3}(b) and \ref{fig3}(c), finite linewidths of the phonon modes allow them to decay in a few picoseconds. Note that we do not explicitly include the phonon scattering channels in our approach. When included, these scattering channels will increase the phonon population with time at other ${\bf k}$ points that satisfy the momentum and energy conservation~\citep{trigo2013fourier,stern2018mapping,murphy2019evolution}. 

\subsection{Response Function and Time-Resolved Diffuse X-ray Scattering}\label{5:4.2}
To know how well-grounded the discussed method of extracting $\chi(\mathbf{k}, t)$ from $\mathsf{S}(\mathbf{k}, \omega)$ (obtained from inelastic scattering measurements or simulations) is, we will compare our simulated results with the measured $\chi(\mathbf{k}, t)$. Trigo \emph{et al.} have performed time-resolved diffuse x-ray scattering on germanium~\citep{trigo2013fourier}. In that experiment, an optical pump pulse of 800 nm with a nominal width of 50 fs was used to generate correlated pairs of phonons with equal and opposite momenta at ${\bf k}$ and ${\bf -k}$, i.e., the squeezed states having $\langle { u}(t)\rangle = 0$. Here $u$ is the atomic displacement and $t$ denotes the pump-probe delay time. A 50 fs x-ray pulse having 10 keV photon energy from Linac Coherent Light Source (LCLS) was used to probe diffuse scattering from the squeezed states at various pump-probe delay times. The temporal evolution of the equal-time correlation function was probed~\citep{trigo2013fourier}. Due to the time-resolved nature of the experiment, the time evolution of the squeezed states and the anharmonic decay of phonons is seen in the ${\bf k}$-$t$ domain. Note that for the squeezed states, the diffuse scattering intensity oscillates at twice the phonon frequency~\citep{trigo2013fourier}.

\begin{figure}
	\begin{center}
	\includegraphics[width=0.8\linewidth]{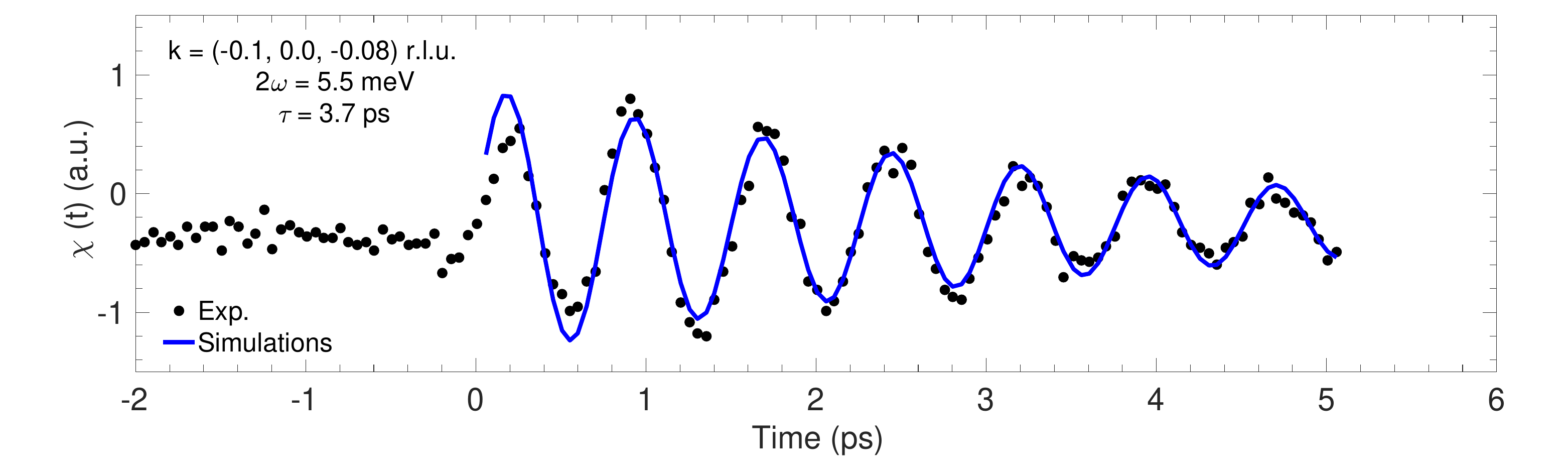}
\end{center}
\caption{Comparison  of experimental data with simulated $\chi(t)$ for germanium. Normalized difference intensity of the diffuse scattering at $\mathbf{k} = (-0.10, 0.00, -0.08)$\,r.l.u.~is from the experiment presented in Ref.~\citep{trigo2013fourier} by black color, and our simulated $\chi(t)$ is shown by blue color.} 
\label{fig4}
\end{figure}
Figure~\ref{fig4} presents the comparison of the experimental data from Ref.~\citep{trigo2013fourier} with our simulated result of $\chi(\mathbf{k}, t)$ for germanium. To demonstrate the merit of our work, we have chosen the data at $\mathbf{k} = (-0.10, 0.00, -0.08)$\,r.l.u.~only. Since $\chi(\mathbf{k}, t)$ is dependent on the phonon mode polarization, the calculated intensity at $\mathbf{k}$ is due to the oscillations at $2\omega\sim$ 5.5\,meV. In the experiment, the normalized difference intensity was shown, which provides the time-resolved value of the equal-time correlation function at $\mathbf{k}$ and $\mathbf{-k}$~\citep{trigo2013fourier}. As reflected from the figure, the present simulated result is in excellent agreement with the experimental data for $\tau\sim$ 3.7\,ps. Such a large value of $\tau$ for dispersive phonon modes, which is straightforward to obtain from ${\bf k}$-$t$ domain measurements, is not so easy to extract from INS or IXS measurements owing to the finite instrument resolution in ${\bf k}$ and $\omega$~\citep{Ehlers, ARCS, HERIX3}.

In spite of the excellent agreement, an important question arises: how do the two different methods -- time-resolved diffuse x-ray scattering and inelastic x-ray scattering, yield the same information. In Ref.~\cite{trigo2013fourier}, it is mentioned that time-resolved diffuse x-ray scattering probes equal-time density-density correlation function: $\langle \hat{n}(\mathbf{-k},t)~\hat{n}(\mathbf{k},t) \rangle$ ~\citep{dixit2014theory, dixit2012imaging}. On the other hand, it is well-established that inelastic x-ray scattering probes density-density correlation function at a different time: $\langle \hat{n}(\mathbf{-k}, t)~\hat{n}(\mathbf{k}, 0) \rangle$, which is related to the Van Hove correlation function~\citep{schulke2007electron,  van1954correlations}. However, there is no contradiction as the experiment was performed without energy resolution and the presented data were energy integrated~\citep{trigo2013fourier}. The density-density correlation function at different times, probed by inelastic x-ray scattering, reduces to an equal-time density-density correlation function in the case where energy resolution is lacking. Therefore, without energy resolution, time-resolved diffuse x-ray scattering and inelastic scattering yield identical information: $\langle \hat{n}(\mathbf{-k},t)~\hat{n}(\mathbf{k},t) \rangle$, apart from a pre-factor as shown in section~\ref{5:2.3}.

\subsection{Response Function in Real-Space}
Not only time-resolved scattering methods in a pump-probe configuration provide the temporal evolution of correlation function, but also help us to visualize lattice dynamics in the ${\bf x}$-$t$ domain~\citep{gaffney2007imaging, sciaini2011femtosecond}, for example, as in the time-resolved electron microscopy experiments~\citep{flannigan20124d, cremons2016femtosecond, cremons2017defect}. In the following, we demonstrate that momentum and energy-resolved inelastic scattering signal also provides the snapshots of the lattice dynamics in the ${\bf x}$-$t$ domain. For this purpose, we need to perform one more Fourier transform from momentum-space to real-space to obtain $\chi(\textbf{x}, t)$ from $\chi(\textbf{k}, t)$. Following Fourier relation, the spatial resolution along the $[H,H,0]$ direction is estimated as $\Delta \mathbf{x} = {2 \pi}/({6.5}$\,\AA$^{-1})$ = 0.96\,\AA\,, whereas along the $[0,0,L]$ direction is $\Delta \mathbf{x} = {2 \pi}/({8.1}$\,\AA$^{-1})$ = 0.78\,\AA. The spatial extent of the dynamics ranges from $(0,0)$ to $(153,108)$\,\AA. 
Similar to the time resolution and duration, the spatial resolution and extent are governed by momentum transfer and momentum resolution of the measurements/simulations.
\begin{figure}[h!]
	\begin{center}
	\includegraphics[width=0.8\linewidth]{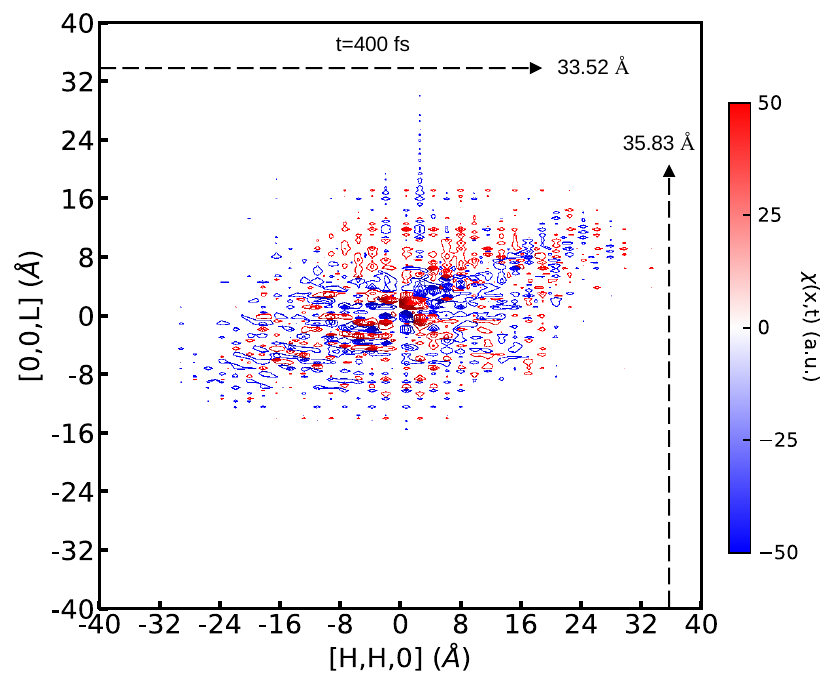}
\end{center}
\caption{{Snapshot of $\chi(\mathbf{x}, t)$ in the $(H,H,L)$ plane at $t$ = 400\,fs for silicon.} 
The dotted lines represent the extent to which there are disturbances in $[1,1,0]$ and $[0,0,1]$ directions, which is given by the dispersion of longitudinal acoustic mode at the zone center.}
\label{fig5}
\end{figure}

The snapshot of $\chi(\textbf{x}, t)$ indicates how and where the disturbance, imparted at $t = 0$, has traveled in real-space. $\chi(\textbf{x}, t)$ at $t$ = 400\,fs in the $(H,H,L)$ plane is shown in Fig.~\ref{fig5}. 
In this case, the dynamics is induced by a point source at ${\bf x} = (0,0,0)$ in silicon. 
As silicon crystal is anisotropic; therefore the phonon propagation is not spherical as generally assumed for the scattering of light waves from a point-like source. Instead, the anisotropic phonon group velocity governs the 3D extent of energy re-distribution after time $t$.
The extent of disturbance along a particular direction at any given time can be estimated from the maximum velocity given by the dispersion of longitudinal acoustic mode at the zone-center. In the present case, the maximum velocity along the $[H,H,0]$ direction is 8958\,ms$^{-1}$, whereas it is 8381\,ms$^{-1}$ along the $[0,0,L]$ direction. 

\begin{figure}
	\begin{center}
	\includegraphics[width=0.8\linewidth]{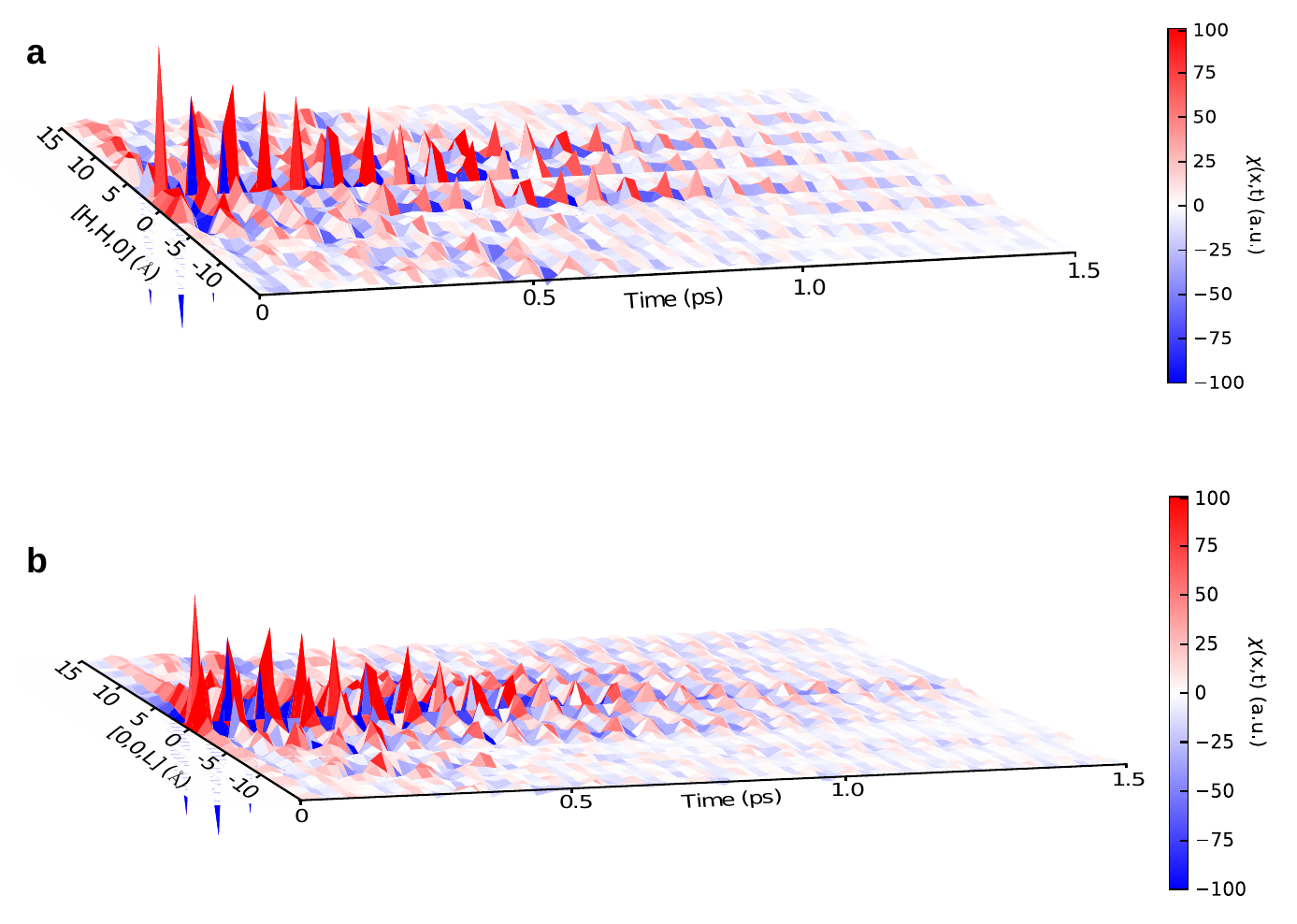}
\end{center}
\caption{{Visualization of the lattice dynamics induced by a point source in silicon in the $(\textbf{x}, t)$ domain.} The dynamics along the $[1,1,0]$ direction are presented in the upper panel, whereas the lower panel shows dynamics along the $[0,0,1]$ direction. The disturbance is still in the system at large instances, but the ripples' height is low. There is no dissipation of energy from the system.}
\label{fig6}
\end{figure}

Figure~\ref{fig6} presents the entire dynamics along $[1,1,0]$ and $[0,0,1]$ directions in upper and lower panels, respectively. Before the disturbance at $ t< 0$, the system is in equilibrium. At $t=0$, the system is struck with a negative disturbance. As a result of this disturbance, a positive recoil at the origin is generated, surrounded by a minimal positive build-up. As visible from the figure, the density-induced disturbance is propagating through the entire system as time evolves. At large-time instances, the disturbance is still in the system, but the order is minimal due to the spread of energy into the system (no dissipation of energy from the system). Such time-resolved images can be captured nowadays using real-space femtosecond electron imaging, as recently demonstrated for the phonon nucleation and launch at a crystal step-edge in the WSe$_2$ flake~\citep{cremons2016femtosecond}. At this point, it is important to mention that Abbamonte and co-workers have employed a similar reconstruction method of $\chi(\mathbf{x}, t)$ from inelastic x-ray scattering data to visualize electron dynamics in various systems~\citep{abbamonte2004imaging, abbamonte2008dynamical, abbamonte2009implicit}.

\subsection{Practical Challenges}

Despite the applicability of our approach to inelastic scattering methods, we should keep experimental and data analysis limitations in mind. For example, \mbox{4D momentum-,} crystallographic direction-, and energy-resolved datasets of lattice dynamics are presently feasible with high-resolution ($\sim$1\,meV, or better) INS and IXS~\citep{Ehlers, ARCS, HERIX3}. In EELS, the energy resolution may always not be suitable for low energy phonons ($<$\, 30\,meV), as intense zero-loss peak masks the spectrum~\citep{krivanek2014vibrational, venkatraman2019vibrational}. Even in INS and IXS measurements, the scattering intensity varies among elements. Since the x-ray scattering cross-section increases with the atomic number $Z$, the phonon intensity of low $Z$ elements is considerably weak. Moreover, acquiring a complete 4D dataset using IXS with an array of monochromators will require a long measurement time (several days, as measurements at a few momentum transfers generally take 15 to 120 minutes). Similarly, in INS, several elements have small coherent scattering cross-sections (i.e., H, V, Co) or high absorption coefficients (i.e., B, Cd). Thus, phonon dispersion measurements remain challenging in materials with these elements. INS measurements also have kinematic constraints (all momentum and energy transfer are not accessible), and energy resolution, instead of being a constant number, depends on the phonon energy~\citep{squires1996introduction, Willis2009}. Deconvolution of energy resolution from the measured 4D dataset to extract the intrinsic details, for example, $\tau$, is not always straightforward~\citep{lin2016mcvine}. Recent advances in experimental techniques and data analysis are overcoming many of the limitations. Together with simulations, they can provide full or complementary information, as demonstrated in this study.

\section{Summary}

In summary, we have established that inelastic scattering methods have the potential to image lattice dynamics with atomic-scale spatiotemporal resolution. Few tens of femtoseconds temporal and $\sim 1$\,\AA~spatial resolutions can be achieved during the reconstruction of lattice dynamics by utilizing the superior energy and momentum resolutions of inelastic scattering measurements. Our proposed  method allows for direct imaging of lattice dynamics and enables us to extract the lifetime of selective phonon modes. Moreover, there is the flexibility to decide whether a single phonon mode or several phonon modes participate in the dynamics by choosing the disturbance source's spatial extent in real-space. The excellent agreement between the present simulated result and the measured lattice dynamics in germanium provides confidence and robustness of the proposed method. We believe that the current approach will be an alternative to time-resolved diffraction methods to image lattice dynamics and beneficial to situations where time-resolved diffraction is not easy to perform, such as neutron scattering and in-situ measurement conditions.

\cleardoublepage
\chapter{Conclusions and Future Directions}

In this thesis, we aim to probe coherent lattice dynamics in real-space and  real-time . 
In the initial segment of the thesis, we have demonstrated  the capability of  high-harmonic spectroscopy to  probe coherent lattice dynamics in solids. 
Using this technique,  we explored the impact of coherent lattice excitations, specifically the in-plane 
longitudinal and transverse optical phonon modes, on the electronic response in monolayer graphene.  
In the later part of the thesis, we delve into imaging the lattice dynamics in both  real-space 
and real-time. For the 4D imagining of coherent lattice dynamics in solids, 
we have proposed a novel approach,  based on the inelastic scattering method.  
We have illustrated the competence of inelastic scattering techniques, when coupled with theoretical analysis, in offering information comparable to what could be obtained through time-resolved diffraction and imaging measurements within pump-probe configurations. 

The strong-field driven nonperturbative light-matter  interaction is numerically simulated by solving  
density matrix-based Semiconductor-Bloch equations.
The electronic structure of graphene is computed within a tight-binding approximation.  
The coherent lattice dynamics in graphene is described classically  by making  hopping parameters time and atomic coordinate-dependent in tight-binding Hamiltonian. 

In \textbf{Chapter 3}, we  have  illustrated that high-harmonic spectroscopy is able  
to investigate the influence of coherent lattice dynamics on the electronic response in graphene on the attosecond timescale.   
It has been assumed that a pump pulse coherently excites  either the $\textsf{iLO}$ or $\textsf{iTO}$  phonon mode, which is probed by a linearly polarized laser pulse to generate the high-harmonic spectrum. 
It has been found that the coherent excitation of the in-plane  phonon modes results in the appearance of sidebands in the spectrum  of the emitted harmonics. 
The spectral positions of the sidebands yield the energy of the excited phonon mode.
When  the $\textsf{iLO}$ phonon mode is excited, the even- (odd-) order sidebands are polarized parallel (perpendicular) to the probe pulse. 
However, upon excitation of $\textsf{iTO}$ phonon mode, all the sidebands are polarized parallel to the probe pulse. 
These observations are explained in terms of a dynamical symmetry analysis within the Floquet framework. Thus, the polarization of the sideband emission offers a sensitive probe of the dynamical symmetries associated with the coherently excited phonon modes. 
We have also explored how HHG is sensitive  to 
various properties of graphene with coherent phonon and various parameters of the probe laser pulse.  

There can be situations when both the phonon modes are coherently excited  while maintaining a specific phase difference.
In \textbf{Chapter 4},  we have demonstrated  that  the phase difference between these phonon modes can be extracted by analyzing the current associated with the sideband of the main harmonic peak.  
Circular 
phonon modes are generated when the phase difference between the phonon modes acquires  90$^{\circ}$. 
We have shown that high-harmonic spectroscopy characterizes  the ``chirality''  of the circular  phonon modes.  
Consequently, the coherent excitation of LCP and  RCP phonon modes results in distinct modulations in the 
high-harmonic spectra.
It has been observed that the excitation of the circular  phonons leads piezo-optic effect  in graphene~\citep{tamaya2021piezo}.  
Therefore, high-harmonic spectroscopy is  capable to characterize not only the energy, polarization, and phase difference but also the ``chirality''  of the phonon modes.

When a  phonon mode is coherently excited in graphene, the  six-fold symmetry undergoes a reduction to a two-fold symmetry. This reduction in symmetry paves the way for the emergence of symmetry-forbidden harmonics. 
This change in symmetry is also explored through high-harmonic spectroscopy. 
Coherently excited in-plane optical phonon mode is probed by the 
circularly polarized  pulse, and the higher-order harmonic spectrum is generated. 
It has been observed that the dynamical symmetry reduction, caused by the coherent lattice dynamics in graphene,  leads to the generation of forbidden harmonics. 

As of now, our discussion has revolved around the probing of lattice dynamics in real-time.
Moving forward, our focus is shifted toward 4D imaging of coherent lattice dynamics in solids.
In \textbf{Chapter 5}, we have demonstrated that inelastic scattering methods, with the aid of theoretical analysis,  are competent to provide similar information as one could obtain from the time-resolved diffraction and imaging measurements. 
To illustrate the robustness of the proposed method, our simulated result of lattice dynamics in germanium is in excellent agreement with the time-resolved diffuse x-ray  scattering measurement performed 
at x-ray free-electron laser in pump-probe setup. 
For a given inelastic scattering data in energy-momentum space, 
the proposed method is useful to image in-situ 
lattice dynamics under different environmental conditions of temperature, pressure, and magnetic field. 
In addition, our approach is suitable to probe first-order states at ${\bf q}\simeq 0$ or disorder-activated continuum,
and second-order `squeezed' states in the momentum-time domain. 

Our work discussed in \textbf{Chapters 3} and  \textbf{4} brings the key advantage of high-harmonic spectroscopy -- the combination of sub-femtosecond to tens of femtoseconds temporal resolution -- to the problem of  probing phonon-driven electronic response and its dependence on the dynamical symmetries in solids. This  opens an avenue in time-resolved probing of phonon-driven dynamical symmetries in solids with sub-cycle temporal resolution. 
Our study can be extended to bilayer graphene, where infrared-active phonon modes can be described in terms of double degenerate in-plane Raman-active phonon modes of monolayer graphene~\citep{gierz2015phonon, rodriguez2021direct}. The investigation of bilayer graphene would provide further insights into the interplay between lattice dynamics and electron behavior.
In a recent development, it has been demonstrated that incorporating phonon deformations into the context of HHG enables the probing of mode-specific electron-phonon coupling and Berry curvature~\citep{hu2023probing}. This innovative all-optical high-harmonic spectroscopy 
offers a promising alternative avenue for characterizing electron-phonon coupling, which complements the insights provided by angle-resolved photoemission spectroscopy.
On the other hand, work discussed in \textbf{Chapter 5}
will profoundly impact where time-resolved diffraction within the pump-probe setup is not feasible, for instance, in inelastic neutron scattering.  
Future investigations can be built upon the mechanisms employed in this thesis and further extend the methodology to study diverse solids, including topological materials, under varying conditions.

\cleardoublepage
\end{spacing}
\begin{spacing}{1.3}
\addcontentsline{toc}{chapter}{Bibliography}
\bibliographystyle{abbrvnat.bst}
\bibliography{solid_HHG}
\cleardoublepage
\end{spacing}
\cleardoublepage \cleardoublepage
\addcontentsline{toc}{chapter}{List of Publications}
\markboth{List of Publications}{List of Publications}
\newpage
\thispagestyle{empty}
\begin{center}
\vspace*{-0.4cm} {\LARGE {\textbf{List of Publications}}}
\end{center}
{\setlength{\baselineskip}{8pt} \setlength{\parskip}{2pt}
\begin{spacing}{1.5}
\vspace*{.7cm} \noindent{\bf \large A. Part of this thesis} \vspace*{.7cm}
\begin{enumerate}
	\item \textbf{Navdeep Rana}, A. P. Roy, Dipanshu Bansal, and Gopal Dixit: Four-dimensional imaging of lattice dynamics using ab-initio simulation, npj Computational Materials   \textbf{7}, 7 (2021).
	\item  \textbf{Navdeep Rana}, and Gopal Dixit: Probing phonon-driven symmetry alterations in graphene via high-harmonic spectroscopy: Physical Review A \textbf{106}, 053116 (2022).
	\item  \textbf{Navdeep Rana}, M. S. Mrudul, Daniil Kartashov, Misha Ivanov, and Gopal Dixit: High-harmonic spectroscopy of coherent lattice dynamics in graphene: Physical Review B \textbf{106}, 064303 (2022).

	\item  \textbf{Navdeep Rana}, and Gopal Dixit: Unveiling phase difference and chirality of circular phonons via high-harmonic spectroscopy, \textbf{under preparation.}
\end{enumerate}

\vspace*{.7cm} \noindent{\bf \large B. Not part of this thesis} \vspace*{.7cm}
\begin{enumerate}
	\item  \textbf{Navdeep Rana}, and Gopal Dixit: All-optical ultrafast valley switching in two-dimensional materials, Physical Review Applied \textbf{19}, 034056 (2023).
	\item  \textbf{Navdeep Rana}, M S Mrudul, and Gopal Dixit: Generation of circularly polarized high-harmonics with identical helicity in two-dimensional materials, Physical Review Applied \textbf{18}, 064049 (2022).
	\item  \textbf{Navdeep Rana}, M S Mrudul, and Gopal Dixit: Sensing strain in graphene via high-harmonic spectroscopy, \textbf{ under preparation.}
	\item Rambabu Rajpoot, Amol R. Holkundkar, \textbf{Navdeep Rana}, and Gopal Dixit: Tailoring polarization of attosecond pulses via co-rotating bicircular laser fields, arXiv:2305.00513, \textbf{ under review.}
	\item  \textbf{Navdeep Rana}, and Gopal Dixit: Tailoring photocurrent in graphene, \textbf{under preparation.}
\end{enumerate}

\newpage
\thispagestyle{empty}
\begin{center}
	\vspace*{-0.4cm} {\LARGE {\textbf{Conferences Attended}}}
\end{center}

\vspace*{.7cm} \noindent{\bf \large A. International}
\vspace*{.7cm}

\begin{enumerate}
	\item  \textbf{Control of Ultrafast (Attosecond and Strong Field) Processes Using Structured Light}, MPIPKS Dresden, Germany (06-14 July 2023), Contributed Talk: ``Sensing strain in graphene via high-harmonic spectroscopy''.
	\item  \textbf{Quantum Battles 2023}, University College London, UK (28-30 June 2023), Contributed Talk: ``All-optical ultrafast valley switching in two-dimensional materials''.
	\item  \textbf{Dynamial Control of Quantum Materials}, MPIPKS Dresden, Germany (22-26 May 2023), Presented Poster: ``High-harmonic spectroscopy of phonon dynamics''.
	\item  \textbf{QUTIF International Conference}, Online (22-25 February 2021), Presented Poster: ``Four-dimensional imaging of lattice dynamics using ab-initio simulation''.
\end{enumerate}

\vspace*{.7cm} \noindent{\bf \large B. National}
\vspace*{.7cm}

\begin{enumerate}
	\item  \textbf{SYMPHY 2023}, IIT Bombay (28-29 January 2023), Contributed Talk: ``Generation of circularly polarized high-harmonics with identical helicity in two-dimensional materials''.
	\item \textbf{Ultrafast Sciences 2022}, IISER Thiruvananthapuram (03-05 November 2022), Presented Poster: ``High-harmonic spectroscopy of lattice dynamics in graphene''. 
	\item  \textbf{Quantum Materials 2022}, IIT Kanpur (18-22 September 2022), Contributed Talk: ``High-harmonic spectroscopy of lattice dynamics in graphene''.
	\item  \textbf{SYMPHY 2021}, IIT Bombay (11-12 December 2021), Contributed Talk: ``High-harmonic spectroscopy of lattice dynamics in graphene''.
	\item \textbf{Quantum Materials 2021}, TIFR Mumbai (08-11 December 2021), Presented Poster: ``Four-dimensional imaging of lattice dynamics using ab-initio simulation''.
\end{enumerate}

\end{spacing}

\cleardoublepage
\end{document}